\documentclass[12pt]{article}

\usepackage{upgreek}
\usepackage{amsfonts}
\usepackage{amsmath}
\usepackage{booktabs}
\usepackage{times}
\usepackage{mathptmx}
\usepackage[T1]{fontenc}	
\usepackage{color}
\usepackage{epsfig}
\usepackage{cite}
\usepackage{colortbl}
\usepackage{xspace}
\usepackage{rotating}
\usepackage{multirow}
\usepackage{ifpdf}

\usepackage{hyperref}

\hypersetup{
    breaklinks=true,
    plainpages=false,
    bookmarksnumbered = true,
    pdfpagemode       = UseNone,
    pdfstartview      = FitH,
    pdfpagelayout     = SinglePage,
    colorlinks        = true,
    linkcolor         = blue,
    urlcolor          = red, 
    citecolor = red
    }

\DeclareMathAlphabet{\mathbfi}{OT1}{ptm}{bx}{it}

\newcommand{\lum}{\ensuremath{\mathcal{L}}}
\newcommand{\jpsi}{\ensuremath{J\mskip -2.5mu/\mskip -1mu\psi\mskip 1mu}}
\newcommand{\tev}{\ensuremath{\,\mathrm{TeV}}}
\newcommand{\gevc}{\ensuremath{\,\mathrm{GeV}\mskip -2mu/\mskip -1mu c}}
\newcommand{\mevc}{\ensuremath{\,\mathrm{MeV}\mskip -2mu/\mskip -1mu c}}
\newcommand{\gevcc}{\ensuremath{\,\mathrm{GeV}\mskip -2mu/\mskip -1mu c^2}}
\newcommand{\mevcc}{\ensuremath{\,\mathrm{MeV}\mskip -2mu/\mskip -1mu c^2}}
\newcommand{\pbinv}{\ensuremath{\,\mathrm{pb}^{-1}}}

\newcommand{\prompt}{\ensuremath{\mathrm{prompt}~\jpsi}}
\newcommand{\fromb}{\ensuremath{\jpsi~\mathrm{from}~b}}
\newcommand{\ptrans}{\ensuremath{p_{\rm T}}}
\newcommand{\microb}{\ensuremath{\,\upmu\mathrm{b}}}

\newcommand{\tpm}{$ & $\,\pm\,$ & $}

\begin{document}

\begin{titlepage}
\pagenumbering{roman}

\vspace*{-1.5cm}
\centerline{\large EUROPEAN ORGANIZATION FOR NUCLEAR RESEARCH (CERN)}
\vspace*{1.5cm}
\hspace*{-0.5cm}
\begin{tabular*}{\linewidth}{lc@{\extracolsep{\fill}}r}
\ifpdf
\vspace*{-1.2cm}\mbox{\!\!\!\includegraphics[width=0.12\textwidth]{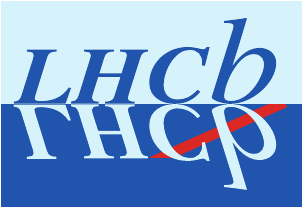}} & & \\
\else
\vspace*{-1.2cm}\mbox{\!\!\!\includegraphics[width=0.12\textwidth]{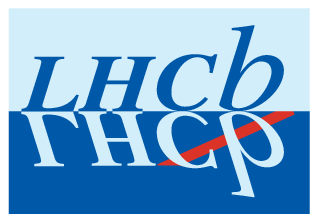}} & & \\
\fi
 & & CERN-PH-EP-2011-018 \\
 & & February 28, 2011 \\ 
 & & \\
\end{tabular*}

\vspace*{3.0cm}

{\bf\huge 
\begin{center}
Measurement of  $\mskip 2mu\mathbfi{\jpsi}\mskip 0.2mu$ production in \\ $\mskip 1.5mu\mathbfi{pp}$ collisions at $\mathbfi{\sqrt{s}}\mathbf{\,=7\,TeV}$
\end{center}
}

\vspace*{0.5cm}

\begin{center}
The LHCb Collaboration\footnote{Authors are listed on the following pages.}
\end{center}

\vspace{\fill}

\begin{abstract}
\noindent
The production of \jpsi\ mesons in proton-proton collisions at  $\sqrt{s}=7\tev$ is studied with the LHCb detector at the LHC.  The  differential cross-section for prompt \jpsi\ production is measured as a function of the \jpsi\ transverse momentum \ptrans\ and rapidity $y$ in the fiducial region  $\ptrans\in[0\,;14]\gevc$  and   $y\in[2.0\,;4.5]$. The differential cross-section and fraction of \jpsi\ from $b$-hadron decays are also measured in the same \ptrans\ and $y$ ranges. The analysis is based on a data sample corresponding to an integrated luminosity of $5.2\pbinv$. 
The measured cross-sections integrated over the fiducial region are $10.52\pm 0.04\pm 1.40^{+1.64}_{-2.20}\,{\microb}$  for \prompt\ production and $1.14 \pm 0.01\pm 0.16\,{\microb}$ for \jpsi\ from $b$-hadron decays, where
the first uncertainty is statistical and the second systematic.
The \prompt\ production cross-section is obtained assuming no \jpsi\ polarisation and the third error indicates the acceptance 
uncertainty due to this assumption.
\end{abstract}

\vspace*{1.5cm}
{\small 
}

\vspace*{1cm}
\centerline{Submitted to Eur. Phys. J. C}
\vspace*{0.5cm}
\vspace{\fill}

\end{titlepage}


\newpage
\setcounter{page}{2}
\mbox{~}
\newpage

\centerline{\large\bf The LHCb Collaboration}
\begin{flushleft}
\small

R.~Aaij$^{23}$, 
B.~Adeva$^{36}$, 
M.~Adinolfi$^{42}$, 
C.~Adrover$^{6}$, 
A.~Affolder$^{48}$, 
Z.~Ajaltouni$^{5}$, 
J.~Albrecht$^{37}$, 
F.~Alessio$^{6,37}$, 
M.~Alexander$^{47}$, 
P.~Alvarez~Cartelle$^{36}$, 
A.A.~Alves~Jr$^{22}$, 
S.~Amato$^{2}$, 
Y.~Amhis$^{38}$, 
J.~Amoraal$^{23}$, 
J.~Anderson$^{39}$, 
R.B.~Appleby$^{50}$, 
O.~Aquines~Gutierrez$^{10}$, 
L.~Arrabito$^{53}$, 
M.~Artuso$^{52}$, 
E.~Aslanides$^{6}$, 
G.~Auriemma$^{22,m}$, 
S.~Bachmann$^{11}$, 
D.S.~Bailey$^{50}$, 
V.~Balagura$^{30,37}$, 
W.~Baldini$^{16}$, 
R.J.~Barlow$^{50}$, 
C.~Barschel$^{37}$, 
S.~Barsuk$^{7}$, 
A.~Bates$^{47}$, 
C.~Bauer$^{10}$, 
Th.~Bauer$^{23}$, 
A.~Bay$^{38}$, 
I.~Bediaga$^{1}$, 
K.~Belous$^{34}$, 
I.~Belyaev$^{30,37}$, 
E.~Ben-Haim$^{8}$, 
M.~Benayoun$^{8}$, 
G.~Bencivenni$^{18}$, 
R.~Bernet$^{39}$, 
M.-O.~Bettler$^{17,37}$, 
M.~van~Beuzekom$^{23}$, 
S.~Bifani$^{12}$, 
A.~Bizzeti$^{17,h}$, 
P.M.~Bj\o rnstad$^{50}$, 
T.~Blake$^{49}$, 
F.~Blanc$^{38}$, 
C.~Blanks$^{49}$, 
J.~Blouw$^{11}$, 
S.~Blusk$^{52}$, 
A.~Bobrov$^{33}$, 
V.~Bocci$^{22}$, 
A.~Bondar$^{33}$, 
N.~Bondar$^{29,37}$, 
W.~Bonivento$^{15}$, 
S.~Borghi$^{47}$, 
A.~Borgia$^{52}$, 
E.~Bos$^{23}$, 
T.J.V.~Bowcock$^{48}$, 
C.~Bozzi$^{16}$, 
T.~Brambach$^{9}$, 
J.~van~den~Brand$^{24}$, 
J.~Bressieux$^{38}$, 
S.~Brisbane$^{51}$, 
M.~Britsch$^{10}$, 
T.~Britton$^{52}$, 
N.H.~Brook$^{42}$, 
H.~Brown$^{48}$, 
A.~B\"{u}chler-Germann$^{39}$, 
A.~Bursche$^{39}$, 
J.~Buytaert$^{37}$, 
S.~Cadeddu$^{15}$, 
J.M.~Caicedo~Carvajal$^{37}$, 
O.~Callot$^{7}$, 
M.~Calvi$^{20,j}$, 
M.~Calvo~Gomez$^{35,n}$, 
A.~Camboni$^{35}$, 
P.~Campana$^{18}$, 
A.~Carbone$^{14}$, 
G.~Carboni$^{21,k}$, 
R.~Cardinale$^{19,i}$, 
A.~Cardini$^{15}$, 
L.~Carson$^{36}$, 
K.~Carvalho~Akiba$^{23}$, 
G.~Casse$^{48}$, 
M.~Cattaneo$^{37}$, 
M.~Charles$^{51}$, 
Ph.~Charpentier$^{37}$, 
N.~Chiapolini$^{39}$, 
X.~Cid~Vidal$^{36}$, 
P.J.~Clark$^{46}$, 
P.E.L.~Clarke$^{46}$, 
M.~Clemencic$^{37}$, 
H.V.~Cliff$^{43}$, 
J.~Closier$^{37}$, 
C.~Coca$^{28}$, 
V.~Coco$^{23}$, 
J.~Cogan$^{6}$, 
P.~Collins$^{37}$, 
F.~Constantin$^{28}$, 
G.~Conti$^{38}$, 
A.~Contu$^{51}$, 
M.~Coombes$^{42}$, 
G.~Corti$^{37}$, 
G.A.~Cowan$^{38}$, 
R.~Currie$^{46}$, 
B.~D'Almagne$^{7}$, 
C.~D'Ambrosio$^{37}$, 
W.~Da~Silva$^{8}$, 
P.~David$^{8}$, 
I.~De~Bonis$^{4}$, 
S.~De~Capua$^{21,k}$, 
M.~De~Cian$^{39}$, 
F.~De~Lorenzi$^{12}$, 
J.M.~De~Miranda$^{1}$, 
L.~De~Paula$^{2}$, 
P.~De~Simone$^{18}$, 
D.~Decamp$^{4}$, 
H.~Degaudenzi$^{38,37}$, 
M.~Deissenroth$^{11}$, 
L.~Del~Buono$^{8}$, 
C.~Deplano$^{15}$, 
O.~Deschamps$^{5}$, 
F.~Dettori$^{15,d}$, 
J.~Dickens$^{43}$, 
H.~Dijkstra$^{37}$, 
M.~Dima$^{28}$, 
P.~Diniz~Batista$^{1}$, 
S.~Donleavy$^{48}$, 
D.~Dossett$^{44}$, 
A.~Dovbnya$^{40}$, 
F.~Dupertuis$^{38}$, 
R.~Dzhelyadin$^{34}$, 
C.~Eames$^{49}$, 
S.~Easo$^{45}$, 
U.~Egede$^{49}$, 
V.~Egorychev$^{30}$, 
S.~Eidelman$^{33}$, 
D.~van~Eijk$^{23}$, 
F.~Eisele$^{11}$, 
S.~Eisenhardt$^{46}$, 
L.~Eklund$^{47}$, 
D.G.~d'Enterria$^{35,o}$, 
D.~Esperante~Pereira$^{36}$, 
L.~Est\`{e}ve$^{43}$, 
E.~Fanchini$^{20,j}$, 
C.~F\"{a}rber$^{11}$, 
G.~Fardell$^{46}$, 
C.~Farinelli$^{23}$, 
S.~Farry$^{12}$, 
V.~Fave$^{38}$, 
V.~Fernandez~Albor$^{36}$, 
M.~Ferro-Luzzi$^{37}$, 
S.~Filippov$^{32}$, 
C.~Fitzpatrick$^{46}$, 
F.~Fontanelli$^{19,i}$, 
R.~Forty$^{37}$, 
M.~Frank$^{37}$, 
C.~Frei$^{37}$, 
M.~Frosini$^{17,f}$, 
J.L.~Fungueirino~Pazos$^{36}$, 
S.~Furcas$^{20}$, 
A.~Gallas~Torreira$^{36}$, 
D.~Galli$^{14,c}$, 
M.~Gandelman$^{2}$, 
P.~Gandini$^{51}$, 
Y.~Gao$^{3}$, 
J-C.~Garnier$^{37}$, 
J.~Garofoli$^{52}$, 
L.~Garrido$^{35}$, 
C.~Gaspar$^{37}$, 
N.~Gauvin$^{38}$, 
M.~Gersabeck$^{37}$, 
T.~Gershon$^{44}$, 
Ph.~Ghez$^{4}$, 
V.~Gibson$^{43}$, 
V.V.~Gligorov$^{37}$, 
C.~G\"{o}bel$^{54}$, 
D.~Golubkov$^{30}$, 
A.~Golutvin$^{49,30,37}$, 
A.~Gomes$^{2}$, 
H.~Gordon$^{51}$, 
M.~Grabalosa~G\'{a}ndara$^{35}$, 
R.~Graciani~Diaz$^{35}$, 
L.A.~Granado~Cardoso$^{37}$, 
E.~Graug\'{e}s$^{35}$, 
G.~Graziani$^{17}$, 
A.~Grecu$^{28}$, 
S.~Gregson$^{43}$, 
B.~Gui$^{52}$, 
E.~Gushchin$^{32}$, 
Yu.~Guz$^{34,37}$, 
T.~Gys$^{37}$, 
G.~Haefeli$^{38}$, 
S.C.~Haines$^{43}$, 
T.~Hampson$^{42}$, 
S.~Hansmann-Menzemer$^{11}$, 
R.~Harji$^{49}$, 
N.~Harnew$^{51}$, 
P.F.~Harrison$^{44}$, 
J.~He$^{7}$, 
K.~Hennessy$^{48}$, 
P.~Henrard$^{5}$, 
J.A.~Hernando~Morata$^{36}$, 
E.~van~Herwijnen$^{37}$, 
A.~Hicheur$^{38}$, 
E.~Hicks$^{48}$, 
W.~Hofmann$^{10}$, 
K.~Holubyev$^{11}$, 
P.~Hopchev$^{4}$, 
W.~Hulsbergen$^{23}$, 
P.~Hunt$^{51}$, 
T.~Huse$^{48}$, 
R.S.~Huston$^{12}$, 
D.~Hutchcroft$^{48}$, 
V.~Iakovenko$^{7,41}$, 
C.~Iglesias~Escudero$^{36}$, 
P.~Ilten$^{12}$, 
J.~Imong$^{42}$, 
R.~Jacobsson$^{37}$, 
M.~Jahjah~Hussein$^{5}$, 
E.~Jans$^{23}$, 
F.~Jansen$^{23}$, 
P.~Jaton$^{38}$, 
B.~Jean-Marie$^{7}$, 
F.~Jing$^{3}$, 
M.~John$^{51}$, 
D.~Johnson$^{51}$, 
C.R.~Jones$^{43}$, 
B.~Jost$^{37}$, 
F.~Kapusta$^{8}$, 
T.M.~Karbach$^{9}$, 
J.~Keaveney$^{12}$, 
U.~Kerzel$^{37}$, 
T.~Ketel$^{24}$, 
A.~Keune$^{38}$, 
B.~Khanji$^{6}$, 
Y.M.~Kim$^{46}$, 
M.~Knecht$^{38}$, 
S.~Koblitz$^{37}$, 
A.~Konoplyannikov$^{30}$, 
P.~Koppenburg$^{23}$, 
A.~Kozlinskiy$^{23}$, 
L.~Kravchuk$^{32}$, 
G.~Krocker$^{11}$, 
P.~Krokovny$^{11}$, 
F.~Kruse$^{9}$, 
K.~Kruzelecki$^{37}$, 
M.~Kucharczyk$^{25}$, 
S.~Kukulak$^{25}$, 
R.~Kumar$^{14,37}$, 
T.~Kvaratskheliya$^{30}$, 
V.N.~La~Thi$^{38}$, 
D.~Lacarrere$^{37}$, 
G.~Lafferty$^{50}$, 
A.~Lai$^{15}$, 
R.W.~Lambert$^{37}$, 
G.~Lanfranchi$^{18}$, 
C.~Langenbruch$^{11}$, 
T.~Latham$^{44}$, 
R.~Le~Gac$^{6}$, 
J.~van~Leerdam$^{23}$, 
J.-P.~Lees$^{4}$, 
R.~Lef\`{e}vre$^{5}$, 
A.~Leflat$^{31,37}$, 
J.~Lefran\c{c}ois$^{7}$, 
O.~Leroy$^{6}$, 
T.~Lesiak$^{25}$, 
L.~Li$^{3}$, 
Y.Y.~Li$^{43}$, 
L.~Li~Gioi$^{5}$, 
M.~Lieng$^{9}$, 
M.~Liles$^{48}$, 
R.~Lindner$^{37}$, 
C.~Linn$^{11}$, 
B.~Liu$^{3}$, 
G.~Liu$^{37}$, 
J.H.~Lopes$^{2}$, 
E.~Lopez~Asamar$^{35}$, 
N.~Lopez-March$^{38}$, 
J.~Luisier$^{38}$, 
B.~M'charek$^{24}$, 
F.~Machefert$^{7}$, 
I.V.~Machikhiliyan$^{4,30}$, 
F.~Maciuc$^{10}$, 
O.~Maev$^{29}$, 
J.~Magnin$^{1}$, 
A.~Maier$^{37}$, 
S.~Malde$^{51}$, 
R.M.D.~Mamunur$^{37}$, 
G.~Manca$^{15,d,37}$, 
G.~Mancinelli$^{6}$, 
N.~Mangiafave$^{43}$, 
U.~Marconi$^{14}$, 
R.~M\"{a}rki$^{38}$, 
J.~Marks$^{11}$, 
G.~Martellotti$^{22}$, 
A.~Martens$^{7}$, 
L.~Martin$^{51}$, 
A.~Mart\'{i}n~S\'{a}nchez$^{7}$, 
D.~Martinez~Santos$^{37}$, 
A.~Massafferri$^{1}$, 
Z.~Mathe$^{12}$, 
C.~Matteuzzi$^{20}$, 
M.~Matveev$^{29}$, 
V.~Matveev$^{34}$, 
E.~Maurice$^{6}$, 
B.~Maynard$^{52}$, 
A.~Mazurov$^{32}$, 
G.~McGregor$^{50}$, 
R.~McNulty$^{12}$, 
C.~Mclean$^{46}$, 
M.~Meissner$^{11}$, 
M.~Merk$^{23}$, 
J.~Merkel$^{9}$, 
M.~Merkin$^{31}$, 
R.~Messi$^{21,k}$, 
S.~Miglioranzi$^{37}$, 
D.A.~Milanes$^{13}$, 
M.-N.~Minard$^{4}$, 
S.~Monteil$^{5}$, 
D.~Moran$^{12}$, 
P.~Morawski$^{25}$, 
J.V.~Morris$^{45}$, 
R.~Mountain$^{52}$, 
I.~Mous$^{23}$, 
F.~Muheim$^{46}$, 
K.~M\"{u}ller$^{39}$, 
R.~Muresan$^{28,38}$, 
F.~Murtas$^{18}$, 
B.~Muryn$^{26}$, 
M.~Musy$^{35}$, 
J.~Mylroie-Smith$^{48}$, 
P.~Naik$^{42}$, 
T.~Nakada$^{38}$, 
R.~Nandakumar$^{45}$, 
J.~Nardulli$^{45}$, 
M.~Nedos$^{9}$, 
M.~Needham$^{46}$, 
N.~Neufeld$^{37}$, 
M.~Nicol$^{7}$, 
S.~Nies$^{9}$, 
V.~Niess$^{5}$, 
N.~Nikitin$^{31}$, 
A.~Oblakowska-Mucha$^{26}$, 
V.~Obraztsov$^{34}$, 
S.~Oggero$^{23}$, 
O.~Okhrimenko$^{41}$, 
R.~Oldeman$^{15,d}$, 
M.~Orlandea$^{28}$, 
A.~Ostankov$^{34}$, 
B.~Pal$^{52}$, 
J.~Palacios$^{39}$, 
M.~Palutan$^{18}$, 
J.~Panman$^{37}$, 
A.~Papanestis$^{45}$, 
M.~Pappagallo$^{13,b}$, 
C.~Parkes$^{47,37}$, 
C.J.~Parkinson$^{49}$, 
G.~Passaleva$^{17}$, 
G.D.~Patel$^{48}$, 
M.~Patel$^{49}$, 
S.K.~Paterson$^{49,37}$, 
G.N.~Patrick$^{45}$, 
C.~Patrignani$^{19,i}$, 
C.~Pavel-Nicorescu$^{28}$, 
A.~Pazos~Alvarez$^{36}$, 
A.~Pellegrino$^{23}$, 
G.~Penso$^{22,l}$, 
M.~Pepe~Altarelli$^{37}$, 
S.~Perazzini$^{14,c}$, 
D.L.~Perego$^{20,j}$, 
E.~Perez~Trigo$^{36}$, 
A.~P\'{e}rez-Calero~Yzquierdo$^{35}$, 
P.~Perret$^{5}$, 
A.~Petrella$^{16,e,37}$, 
A.~Petrolini$^{19,i}$, 
B.~Pie~Valls$^{35}$, 
B.~Pietrzyk$^{4}$, 
D.~Pinci$^{22}$, 
R.~Plackett$^{47}$, 
S.~Playfer$^{46}$, 
M.~Plo~Casasus$^{36}$, 
G.~Polok$^{25}$, 
A.~Poluektov$^{44,33}$, 
E.~Polycarpo$^{2}$, 
D.~Popov$^{10}$, 
B.~Popovici$^{28}$, 
C.~Potterat$^{38}$, 
A.~Powell$^{51}$, 
T.~du~Pree$^{23}$, 
V.~Pugatch$^{41}$, 
A.~Puig~Navarro$^{35}$, 
W.~Qian$^{3}$, 
J.H.~Rademacker$^{42}$, 
B.~Rakotomiaramanana$^{38}$, 
I.~Raniuk$^{40}$, 
G.~Raven$^{24}$, 
S.~Redford$^{51}$, 
W.~Reece$^{49}$, 
A.C.~dos~Reis$^{1}$, 
S.~Ricciardi$^{45}$, 
K.~Rinnert$^{48}$, 
D.A.~Roa~Romero$^{5}$, 
P.~Robbe$^{7,37}$, 
E.~Rodrigues$^{47}$, 
F.~Rodrigues$^{2}$, 
C.~Rodriguez~Cobo$^{36}$, 
P.~Rodriguez~Perez$^{36}$, 
G.J.~Rogers$^{43}$, 
V.~Romanovsky$^{34}$, 
J.~Rouvinet$^{38}$, 
T.~Ruf$^{37}$, 
H.~Ruiz$^{35}$, 
G.~Sabatino$^{21,k}$, 
J.J.~Saborido~Silva$^{36}$, 
N.~Sagidova$^{29}$, 
P.~Sail$^{47}$, 
B.~Saitta$^{15,d}$, 
C.~Salzmann$^{39}$, 
A.~Sambade~Varela$^{37}$, 
M.~Sannino$^{19,i}$, 
R.~Santacesaria$^{22}$, 
R.~Santinelli$^{37}$, 
E.~Santovetti$^{21,k}$, 
M.~Sapunov$^{6}$, 
A.~Sarti$^{18}$, 
C.~Satriano$^{22,m}$, 
A.~Satta$^{21}$, 
M.~Savrie$^{16,e}$, 
D.~Savrina$^{30}$, 
P.~Schaack$^{49}$, 
M.~Schiller$^{11}$, 
S.~Schleich$^{9}$, 
M.~Schmelling$^{10}$, 
B.~Schmidt$^{37}$, 
O.~Schneider$^{38}$, 
A.~Schopper$^{37}$, 
M.-H.~Schune$^{7}$, 
R.~Schwemmer$^{37}$, 
A.~Sciubba$^{18,l}$, 
M.~Seco$^{36}$, 
A.~Semennikov$^{30}$, 
K.~Senderowska$^{26}$, 
N.~Serra$^{23}$, 
J.~Serrano$^{6}$, 
B.~Shao$^{3}$, 
M.~Shapkin$^{34}$, 
I.~Shapoval$^{40,37}$, 
P.~Shatalov$^{30}$, 
Y.~Shcheglov$^{29}$, 
T.~Shears$^{48}$, 
L.~Shekhtman$^{33}$, 
O.~Shevchenko$^{40}$, 
V.~Shevchenko$^{30}$, 
A.~Shires$^{49}$, 
E.~Simioni$^{24}$, 
H.P.~Skottowe$^{43}$, 
T.~Skwarnicki$^{52}$, 
A.C.~Smith$^{37}$, 
K.~Sobczak$^{5}$, 
F.J.P.~Soler$^{47}$, 
A.~Solomin$^{42}$, 
P.~Somogy$^{37}$, 
F.~Soomro$^{49}$, 
B.~Souza~De~Paula$^{2}$, 
B.~Spaan$^{9}$, 
A.~Sparkes$^{46}$, 
E.~Spiridenkov$^{29}$, 
P.~Spradlin$^{51}$, 
F.~Stagni$^{37}$, 
O.~Steinkamp$^{39}$, 
O.~Stenyakin$^{34}$, 
S.~Stoica$^{28}$, 
S.~Stone$^{52}$, 
B.~Storaci$^{23}$, 
U.~Straumann$^{39}$, 
N.~Styles$^{46}$, 
M.~Szczekowski$^{27}$, 
P.~Szczypka$^{38}$, 
T.~Szumlak$^{26}$, 
S.~T'Jampens$^{4}$, 
V.~Talanov$^{34}$, 
E.~Teodorescu$^{28}$, 
H.~Terrier$^{23}$, 
F.~Teubert$^{37}$, 
C.~Thomas$^{51,45}$, 
E.~Thomas$^{37}$, 
J.~van~Tilburg$^{39}$, 
V.~Tisserand$^{4}$, 
M.~Tobin$^{39}$, 
S.~Topp-Joergensen$^{51}$, 
M.T.~Tran$^{38}$, 
A.~Tsaregorodtsev$^{6}$, 
N.~Tuning$^{23}$, 
A.~Ukleja$^{27}$, 
P.~Urquijo$^{52}$, 
U.~Uwer$^{11}$, 
V.~Vagnoni$^{14}$, 
G.~Valenti$^{14}$, 
R.~Vazquez~Gomez$^{35}$, 
P.~Vazquez~Regueiro$^{36}$, 
S.~Vecchi$^{16}$, 
J.J.~Velthuis$^{42}$, 
M.~Veltri$^{17,g}$, 
K.~Vervink$^{37}$, 
B.~Viaud$^{7}$, 
I.~Videau$^{7}$, 
X.~Vilasis-Cardona$^{35,n}$, 
J.~Visniakov$^{36}$, 
A.~Vollhardt$^{39}$, 
D.~Voong$^{42}$, 
A.~Vorobyev$^{29}$, 
An.~Vorobyev$^{29}$, 
H.~Voss$^{10}$, 
K.~Wacker$^{9}$, 
S.~Wandernoth$^{11}$, 
J.~Wang$^{52}$, 
D.R.~Ward$^{43}$, 
A.D.~Webber$^{50}$, 
D.~Websdale$^{49}$, 
M.~Whitehead$^{44}$, 
D.~Wiedner$^{11}$, 
L.~Wiggers$^{23}$, 
G.~Wilkinson$^{51}$, 
M.P.~Williams$^{44,45}$, 
M.~Williams$^{49}$, 
F.F.~Wilson$^{45}$, 
J.~Wishahi$^{9}$, 
M.~Witek$^{25}$, 
W.~Witzeling$^{37}$, 
S.A.~Wotton$^{43}$, 
K.~Wyllie$^{37}$, 
Y.~Xie$^{46}$, 
F.~Xing$^{51}$, 
Z.~Yang$^{3}$, 
G.~Ybeles~Smit$^{23}$, 
R.~Young$^{46}$, 
O.~Yushchenko$^{34}$, 
M.~Zavertyaev$^{10,a}$, 
L.~Zhang$^{52}$, 
W.C.~Zhang$^{12}$, 
Y.~Zhang$^{3}$, 
A.~Zhelezov$^{11}$, 
L.~Zhong$^{3}$, 
E.~Zverev$^{31}$.\bigskip\newline{\it
\footnotesize
$ ^{1}$Centro Brasileiro de Pesquisas F\'{i}sicas (CBPF), Rio de Janeiro, Brazil\\
$ ^{2}$Universidade Federal do Rio de Janeiro (UFRJ), Rio de Janeiro, Brazil\\
$ ^{3}$Center for High Energy Physics, Tsinghua University, Beijing, China\\
$ ^{4}$LAPP, Universit\'{e} de Savoie, CNRS/IN2P3, Annecy-Le-Vieux, France\\
$ ^{5}$Clermont Universit\'{e}, Universit\'{e} Blaise Pascal, CNRS/IN2P3, LPC, Clermont-Ferrand, France\\
$ ^{6}$CPPM, Aix-Marseille Universit\'{e}, CNRS/IN2P3, Marseille, France\\
$ ^{7}$LAL, Universit\'{e} Paris-Sud, CNRS/IN2P3, Orsay, France\\
$ ^{8}$LPNHE, Universit\'{e} Pierre et Marie Curie, Universit\'{e} Paris Diderot, CNRS/IN2P3, Paris, France\\
$ ^{9}$Fakult\"{a}t Physik, Technische Universit\"{a}t Dortmund, Dortmund, Germany\\
$ ^{10}$Max-Planck-Institut f\"{u}r Kernphysik (MPIK), Heidelberg, Germany\\
$ ^{11}$Physikalisches Institut, Ruprecht-Karls-Universit\"{a}t Heidelberg, Heidelberg, Germany\\
$ ^{12}$School of Physics, University College Dublin, Dublin, Ireland\\
$ ^{13}$Sezione INFN di Bari, Bari, Italy\\
$ ^{14}$Sezione INFN di Bologna, Bologna, Italy\\
$ ^{15}$Sezione INFN di Cagliari, Cagliari, Italy\\
$ ^{16}$Sezione INFN di Ferrara, Ferrara, Italy\\
$ ^{17}$Sezione INFN di Firenze, Firenze, Italy\\
$ ^{18}$Laboratori Nazionali dell'INFN di Frascati, Frascati, Italy\\
$ ^{19}$Sezione INFN di Genova, Genova, Italy\\
$ ^{20}$Sezione INFN di Milano Bicocca, Milano, Italy\\
$ ^{21}$Sezione INFN di Roma Tor Vergata, Roma, Italy\\
$ ^{22}$Sezione INFN di Roma Sapienza, Roma, Italy\\
$ ^{23}$Nikhef National Institute for Subatomic Physics, Amsterdam, Netherlands\\
$ ^{24}$Nikhef National Institute for Subatomic Physics and Vrije Universiteit, Amsterdam, Netherlands\\
$ ^{25}$Henryk Niewodniczanski Institute of Nuclear Physics  Polish Academy of Sciences, Cracow, Poland\\
$ ^{26}$Faculty of Physics \& Applied Computer Science, Cracow, Poland\\
$ ^{27}$Soltan Institute for Nuclear Studies, Warsaw, Poland\\
$ ^{28}$Horia Hulubei National Institute of Physics and Nuclear Engineering, Bucharest-Magurele, Romania\\
$ ^{29}$Petersburg Nuclear Physics Institute (PNPI), Gatchina, Russia\\
$ ^{30}$Institute of Theoretical and Experimental Physics (ITEP), Moscow, Russia\\
$ ^{31}$Institute of Nuclear Physics, Moscow State University (SINP MSU), Moscow, Russia\\
$ ^{32}$Institute for Nuclear Research of the Russian Academy of Sciences (INR RAN), Moscow, Russia\\
$ ^{33}$Budker Institute of Nuclear Physics (BINP), Novosibirsk, Russia\\
$ ^{34}$Institute for High Energy Physics (IHEP), Protvino, Russia\\
$ ^{35}$Universitat de Barcelona, Barcelona, Spain\\
$ ^{36}$Universidad de Santiago de Compostela, Santiago de Compostela, Spain\\
$ ^{37}$European Organization for Nuclear Research (CERN), Geneva, Switzerland\\
$ ^{38}$Ecole Polytechnique F\'{e}d\'{e}rale de Lausanne (EPFL), Lausanne, Switzerland\\
$ ^{39}$Physik-Institut, Universit\"{a}t Z\"{u}rich, Z\"{u}rich, Switzerland\\
$ ^{40}$NSC Kharkiv Institute of Physics and Technology (NSC KIPT), Kharkiv, Ukraine\\
$ ^{41}$Institute for Nuclear Research of the National Academy of Sciences (KINR), Kyiv, Ukraine\\
$ ^{42}$H.H. Wills Physics Laboratory, University of Bristol, Bristol, United Kingdom\\
$ ^{43}$Cavendish Laboratory, University of Cambridge, Cambridge, United Kingdom\\
$ ^{44}$Department of Physics, University of Warwick, Coventry, United Kingdom\\
$ ^{45}$STFC Rutherford Appleton Laboratory, Didcot, United Kingdom\\
$ ^{46}$School of Physics and Astronomy, University of Edinburgh, Edinburgh, United Kingdom\\
$ ^{47}$School of Physics and Astronomy, University of Glasgow, Glasgow, United Kingdom\\
$ ^{48}$Oliver Lodge Laboratory, University of Liverpool, Liverpool, United Kingdom\\
$ ^{49}$Imperial College London, London, United Kingdom\\
$ ^{50}$School of Physics and Astronomy, University of Manchester, Manchester, United Kingdom\\
$ ^{51}$Department of Physics, University of Oxford, Oxford, United Kingdom\\
$ ^{52}$Syracuse University, Syracuse, NY, United States of America\\
$ ^{53}$CC-IN2P3, CNRS/IN2P3, Lyon-Villeurbanne, France, associated member\\
$ ^{54}$Pontif\'{i}cia Universidade Cat\'{o}lica do Rio de Janeiro (PUC-Rio), Rio de Janeiro, Brazil, associated to $^2 $\\
\bigskip
$ ^{a}$P.N. Lebedev Physical Institute, Russian Academy of Science (LPI RAS), Moscow, Russia\\
$ ^{b}$Universit\`{a} di Bari, Bari, Italy\\
$ ^{c}$Universit\`{a} di Bologna, Bologna, Italy\\
$ ^{d}$Universit\`{a} di Cagliari, Cagliari, Italy\\
$ ^{e}$Universit\`{a} di Ferrara, Ferrara, Italy\\
$ ^{f}$Universit\`{a} di Firenze, Firenze, Italy\\
$ ^{g}$Universit\`{a} di Urbino, Urbino, Italy\\
$ ^{h}$Universit\`{a} di Modena e Reggio Emilia, Modena, Italy\\
$ ^{i}$Universit\`{a} di Genova, Genova, Italy\\
$ ^{j}$Universit\`{a} di Milano Bicocca, Milano, Italy\\
$ ^{k}$Universit\`{a} di Roma Tor Vergata, Roma, Italy\\
$ ^{l}$Universit\`{a} di Roma La Sapienza, Roma, Italy\\
$ ^{m}$Universit\`{a} della Basilicata, Potenza, Italy\\
$ ^{n}$LIFAELS, La Salle, Universitat Ramon Llull, Barcelona, Spain\\
$ ^{o}$Instituci\'{o} Catalana de Recerca i Estudis Avan\c{c}ats (ICREA), Barcelona, Spain\\
}

\end{flushleft}

\cleardoublepage

\pagestyle{plain}
\setcounter{page}{1}
\pagenumbering{arabic}
\clearpage

\section{Introduction}\label{Introduction}

Understanding  \jpsi\ meson hadroproduction  has been a long-term effort both experimentally and theoretically. 
However, despite the considerable progress made in recent years~\cite{qwgdoc}, none of the existing theoretical
models can successfully describe both the transverse momentum (\ptrans) dependence of the \jpsi\ cross-section
and the \jpsi\ polarisation measured at the Tevatron. The colour-singlet model (CSM) at leading order in 
$\alpha_{\rm s}$~\cite{chang} underestimates \jpsi\ production by two orders of magnitude~\cite{cdf1}, and even
more at high \ptrans. Including additional processes, such as quark and gluon fragmentation~\cite{cacciari} leads to 
a better description of  the \ptrans\ shape at high \ptrans, but still fails to reproduce the measured production rates.
Computations performed in the framework of nonrelativistic quantum chromodynamics (NRQCD), where the 
$c\overline{c}$ pair can be produced in a colour-octet state~\cite{bodwin}, can explain the shape and the magnitude
of the measured \jpsi\ cross-section. However, they predict a substantial transverse component for the polarisation 
of \jpsi\  mesons at large \ptrans. This is in disagreement with the CDF \jpsi\ polarisation measurement~\cite{cdf3}, 
casting doubt on the conclusion that the colour-octet terms dominate \jpsi\ production.
More recent theoretical studies have considered the addition of the $gg\to \jpsi c\overline{c}$ process to the
CSM~\cite{artoisenet,baranov}, or higher order corrections in $\alpha_{\rm s}$: $gg\to \jpsi gg$~\cite{campbell} and
$gg\to \jpsi ggg$~\cite{firstnnlo,stelzer}. With these additions, the discrepancy between theoretical predictions and
experimental measurements significantly decreases. However, the agreement is still not perfect, leaving open the
question of a complete description of \jpsi\ hadroproduction. The large rate of \jpsi\ production at the Large Hadron
Collider (LHC) opens the door to new analyses that extend the phase-space region explored so far, such as that
recently made by the CMS collaboration~\cite{cmsjpsi}. 
In particular, the LHCb detector provides the possibility to extend the measurements to the forward rapidity region.

Three sources  of \jpsi\ production in $pp$ collisions need to be considered when comparing experimental 
observables and theoretical calculations: direct \jpsi\ production, feed-down \jpsi\ from the decay of other heavier
prompt charmonium states like $\chi_{c1}$, $\chi_{c2}$ or $\psi(2S)$, and \jpsi\ from $b$-hadron decay chains. 
The sum of the first two sources will be called ``\prompt'' in the following. The third source will be abbreviated 
as ``\fromb''.

This paper presents the measurement of the differential production cross-section of both prompt \jpsi\  and  \fromb\ 
as a function of the \jpsi\ transverse momentum and rapidity ($y$)  with respect to the beam axis in the fiducial
region  $\ptrans\in[0\,;14]\gevc$ and  $y\in[2.0\,;4.5]$.  The effect due to the unknown \jpsi\ polarisation  is estimated
by providing results for the differential cross-sections for three extreme polarisation cases. The analysis of a larger
data sample is needed to measure the \jpsi\ polarisation over the kinematic range considered.

\section{The LHCb detector, data sample and Monte Carlo simulation}\label{sec:detector}

The LHCb detector is a forward spectrometer described in detail in Ref.~\cite{lhcbdetectorpaper}. The detector
elements are placed along the beam line of the LHC starting with the Vertex Locator (VELO), a silicon strip device
that surrounds the $pp$ interaction region and is positioned with its sensitive area $8\,{\rm mm}$ from the beam 
during collisions. The VELO provides precise measurements of the positions of the primary $pp$ interaction vertices
and decay vertices of long-lived hadrons, and contributes to the measurement of track momenta. Other detectors
used to measure track momenta are a large area silicon strip detector located before a $4\,{\rm Tm}$ dipole magnet
and a combination of silicon strip detectors and straw drift chambers placed after it. Two Ring Imaging Cherenkov
detectors are used to identify charged hadrons. Further downstream an Electromagnetic Calorimeter system (ECAL,
Preshower -- PRS -- and Scintillating Pad Detector -- SPD) is used for photon detection and electron identification,
followed by a Hadron Calorimeter (HCAL). The muon detection consists of five muon stations (MUON) equipped with
multi-wire proportional chambers, with the exception of the centre of the first station, which uses triple-GEM
detectors. 
For the data  included in this analysis all detector components were fully operational and in a stable condition and
the main component of the dipole field was pointing upwards.

The LHCb trigger system consists of two levels. The first level (L0), implemented in hardware, is designed to reduce
the LHC bunch crossing frequency of 40\,MHz to a maximum of 1\,MHz, at which the complete detector is read out.
The ECAL, HCAL and MUON provide the capability of first-level hardware triggering. 
The second level is a software trigger (High Level Trigger, HLT) which runs on an event-filter farm 
and is implemented in two stages.
HLT1 performs a partial event reconstruction to confirm
the L0 trigger decision, and HLT2 performs a full event reconstruction to further discriminate signal events. 

The study reported here uses data corresponding to an integrated luminosity of $5.2\pbinv$ of $pp$ collisions
produced by the LHC at a centre-of-mass energy of 7\tev\ in September 2010, with at maximum 
$1.6\,{\rm MHz}$ collision frequency. The data were collected using two L0 trigger lines: the single-muon line, which
requires one muon candidate with a \ptrans\ larger than $1.4\gevc$, and the  dimuon line,  which requires two 
muon candidates with \ptrans\ larger than $0.56\gevc$ and $0.48\gevc$, respectively. 
They provide the input candidates for the corresponding HLT1 lines: the first one confirms the single-muon
candidates from L0, and applies a harder \ptrans\ selection at $1.8\gevc$; the second line
confirms the dimuon candidates and requires  their combined mass  to be greater than $2.5\gevcc$. 
The HLT2 algorithm selects events having two opposite charged muon candidates with an invariant mass 
greater than $2.9\gevcc$. 
For a fraction of the data,  corresponding to an integrated luminosity of $3.0\pbinv$, the
HLT1 single muon line was pre-scaled by a factor of five. The trigger efficiency is measured independently for the
pre-scaled data set and for the rest of the sample, and the results subsequently combined.

To avoid the possibility that a few events with a high occupancy dominate the HLT CPU time, a set of global event
cuts (GEC) is applied on the hit multiplicities of each sub-detector used by the pattern recognition algorithms. These
cuts were introduced to cope with conditions encountered during the 2010 running period of the LHC, in which the
average number of visible interactions per bunch crossing was equal to 1.8 for the data used for this analysis, a
factor of five above the design value, at a time when only one fifth of the event-filter farm was installed. The GEC
were chosen to reject busy events with a large number of pile-up interactions with minimal loss of luminosity.
The average number of reconstructed primary vertices in selected and triggered events after GEC is
equal to 2.1.

The Monte Carlo samples used for this analysis are based on the {\sc Pythia} 6.4 generator~\cite{pythia} configured
with the parameters detailed in Ref.~\cite{procgenerator}. The EvtGen package~\cite{EvtGen} was used to generate
hadron decays, in particular \jpsi\ and $b$-hadrons, and the GEANT4 package~\cite{Geant4} for the detector
simulation. The prompt charmonium production processes activated in {\sc Pythia} are those from the leading-order
colour-singlet and colour-octet mechanisms.
The $b$-hadron production in {\sc Pythia} is based on leading order 2 $\to$ 2 QCD processes: 
$q\overline{q}\to q'{\overline{q}}'$, $qq'\to qq'$, $q\overline{q}\to gg$, $qg \to qg$, $g\, g\to q\overline{q}$ and 
$gg\to gg$. QED radiative corrections to the decay $\jpsi\ \to \mu^+ \mu^-$ are generated using the PHOTOS 
package~\cite{photos}.

\section{\texorpdfstring{$\mathbfi{\jpsi}$ selection}{J/psi selection}}

The analysis selects events in which at least one primary vertex is reconstructed from at least five charged tracks 
seen in the VELO. \jpsi\ candidates are formed from pairs of opposite sign tracks reconstructed in the full tracking
system. Each track must have \ptrans\ above $0.7\gevc$, have a good quality of the track fit 
($\chi^2\mskip -2.2mu/\rm ndf<4$) and be identified as a muon by ensuring that it penetrates the iron of the MUON 
system.  The two muons are required to originate from a common vertex, and only candidates with a $\chi^2$
probability of the vertex fit larger than $0.5\%$ are kept. Some charged particles can be reconstructed as more than
one track. Duplicate tracks, which share too many hits with another track or are too close to another track, are
removed.

\fromb\ tend to be produced away from the primary vertex and can be separated from \prompt, which are produced
at the primary vertex,  by exploiting the \jpsi\ pseudo-proper time defined as
\begin{equation}\label{eq:tz}
t_z = \frac{(z_{\jpsi}-z_{\rm PV}) \times M_{\jpsi}}{p_z} \,,
\end{equation}
where $z_{\jpsi}$ and $z_{\rm PV}$ are the positions along the $z$-axis (defined along the beam axis, and oriented
from the VELO to the MUON) of the \jpsi\ decay vertex and of the primary vertex; $p_z$ is the measured \jpsi\
momentum in the $z$ direction and $M_{\jpsi}$ the nominal \jpsi\ mass. 
Given that $b$-hadrons are not fully reconstructed, the \jpsi\ momentum is used instead of the exact  $b$-hadron
momentum and the $t_z$ variable provides a good estimate of  the $b$-hadron decay proper time.
For events with several primary vertices ($68\%$ of the events), the one which is closest to the \jpsi\ vertex in the
$z$ direction is selected.

\section{Cross-section determination}

The differential cross-section for \jpsi\ production in a given (\ptrans,$y$) bin is defined as
\begin{equation}
\frac{{\rm d}^2\sigma}{{\rm d}y\,{\rm d}\ptrans} = \frac{N\left(\jpsi\to\mu^+\mu^-\right)}{\lum \times \epsilon_{\rm tot}
\times {\cal B}\left(\jpsi\to\mu^+\mu^-\right)\times \Delta y \times \Delta \ptrans}\label{eq:sigma}\,,
\end{equation}
where $N\left(\jpsi \to \mu^+ \mu^-\right)$ is the number of observed $\jpsi \to \mu^+ \mu^-$ in bin (\ptrans,$y$), 
$\epsilon_{\rm tot}$ the \jpsi\ detection efficiency including acceptance and trigger efficiency in bin $(\ptrans,\, y)$,
\lum\ the integrated luminosity, ${\cal B}\left(\jpsi\to\mu^+\mu^-\right)$  the branching fraction of the 
$\jpsi \to \mu^+\mu^-$ decay ($(5.93\pm0.06)\times 10^{-2}$~\cite{PDG}), and $\Delta y=0.5$ and 
$\Delta \ptrans=1\gevc$  the $y$ and \ptrans\ bin sizes, respectively. The transverse momentum is defined as 
$\ptrans=\sqrt{p_x^2+p_y^2}$ and the rapidity is defined as $y=\displaystyle\frac{1}{2}\ln\frac{E+p_z}{E-p_z}$
\vspace{0.1cm}
where  $(E,\mathbf{p})$  is the \jpsi\ four-momentum in the centre-of-mass frame of the colliding protons.

In each bin of \ptrans\ and $y$, the fraction of signal \jpsi\ from all sources, $f_{\jpsi}$, is estimated from an extended
unbinned maximum likelihood fit to the invariant mass distribution of the reconstructed \jpsi\ candidates in the 
interval $M_{\mu\mu}\in[2.95\,;3.30]\gevcc$, where the signal is described by a Crystal Ball function~\cite{CB} and 
the combinatorial background by an exponential function. The fraction of \fromb\ is then extracted from a fit to the
 $t_z$ distribution. 

\begin{figure}[!tb]
\centering
\ifpdf
\includegraphics[width=7.95cm]{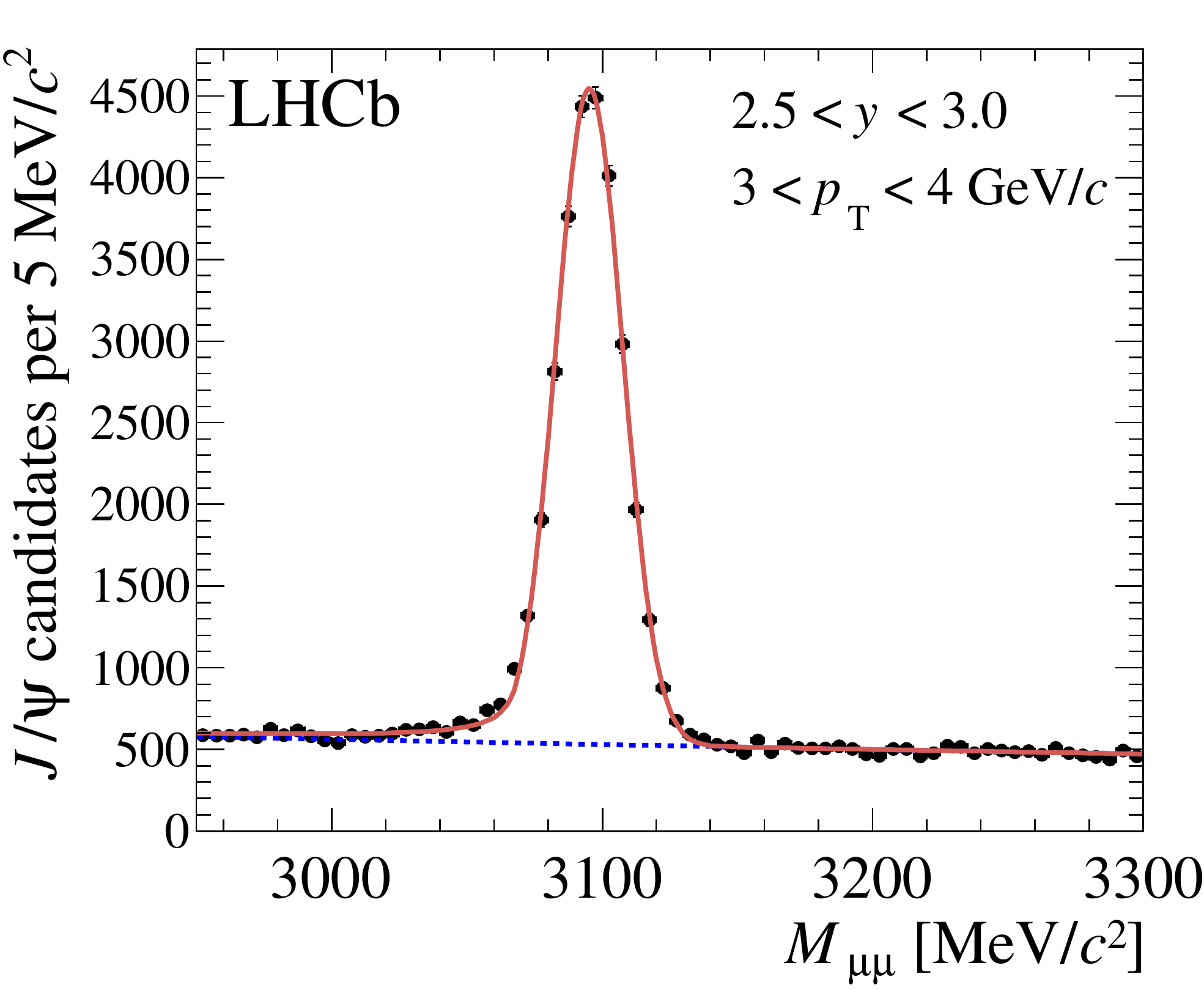}
\includegraphics[width=7.95cm]{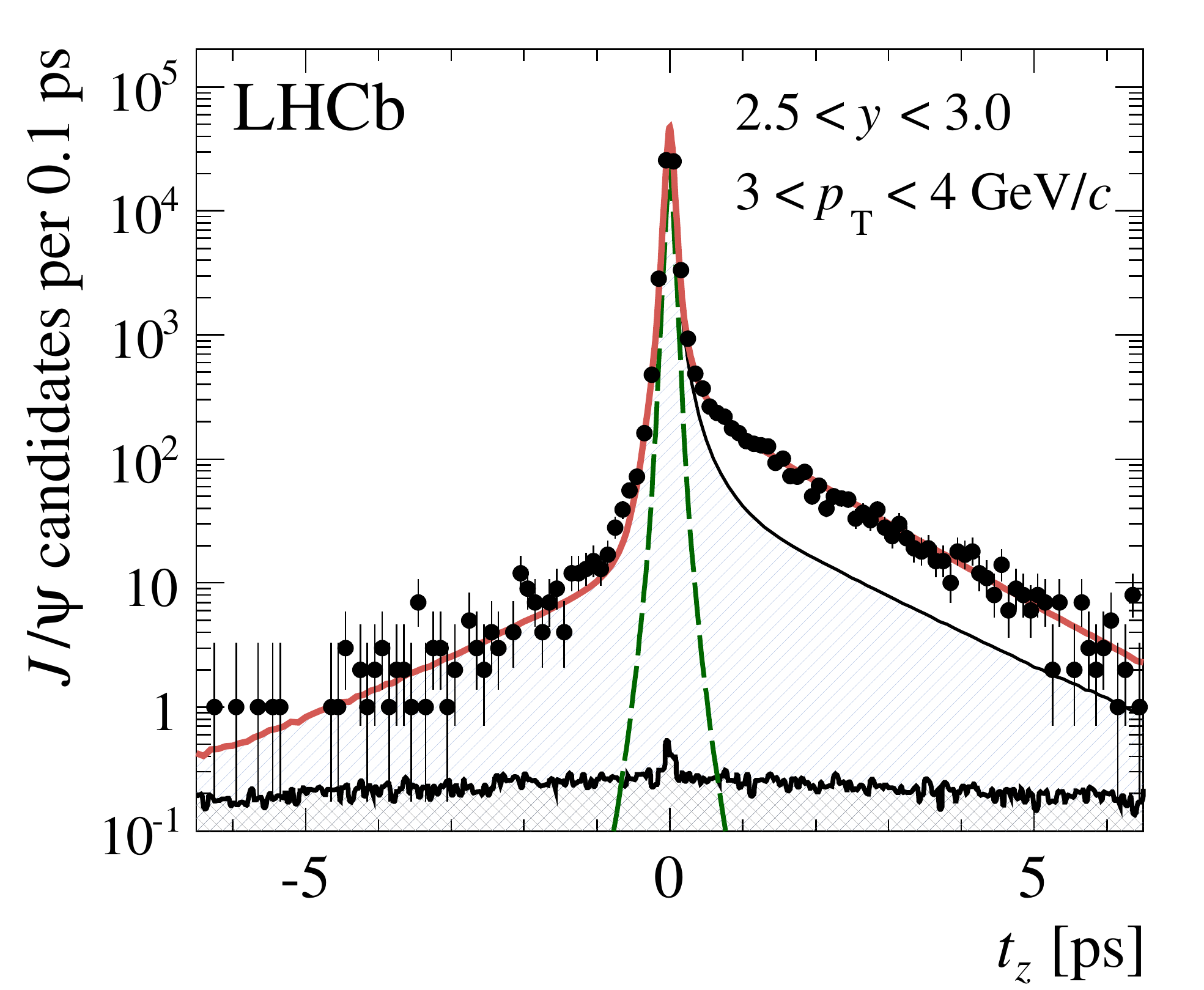}
\else
\includegraphics[width=7.95cm]{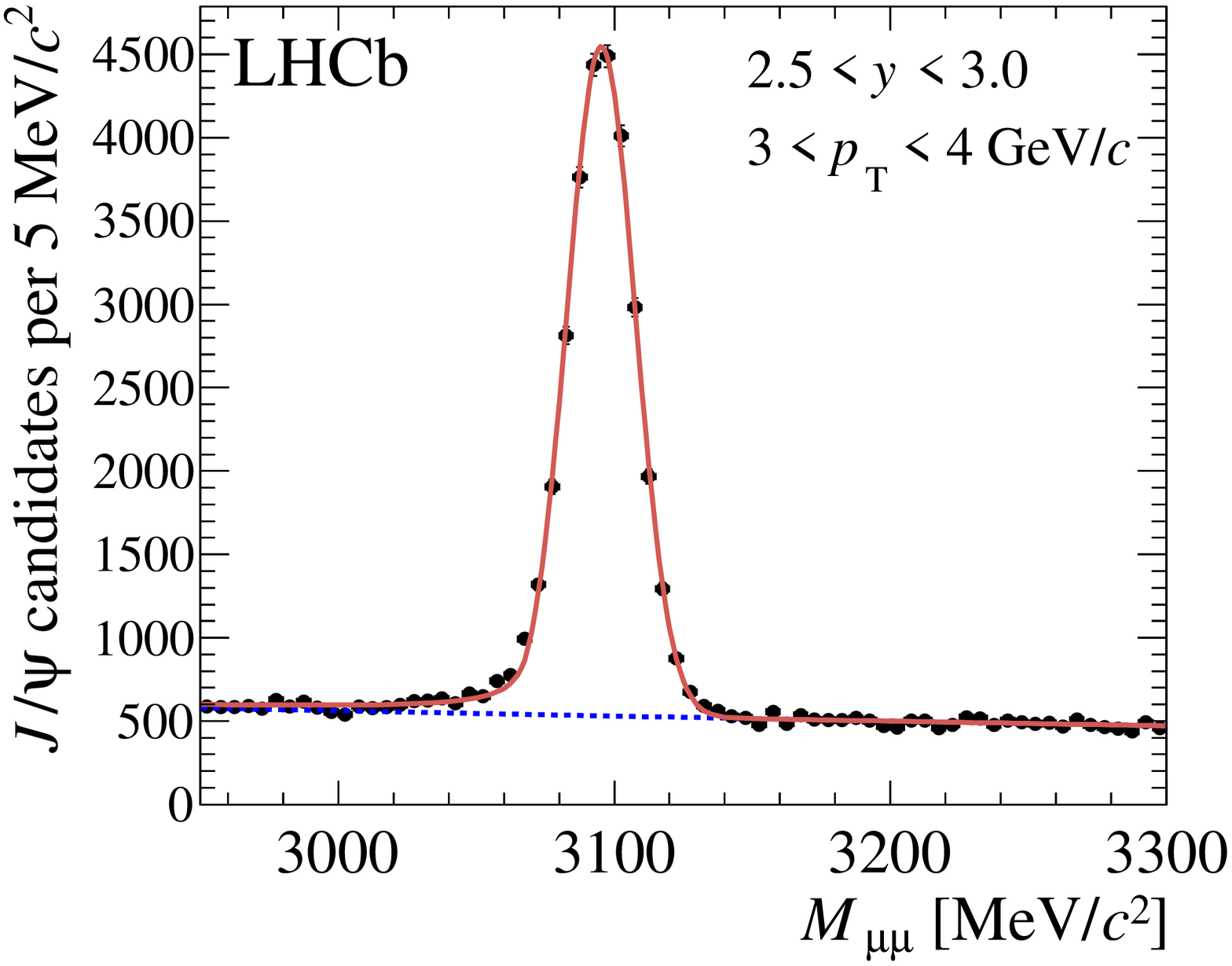}
\includegraphics[width=7.95cm]{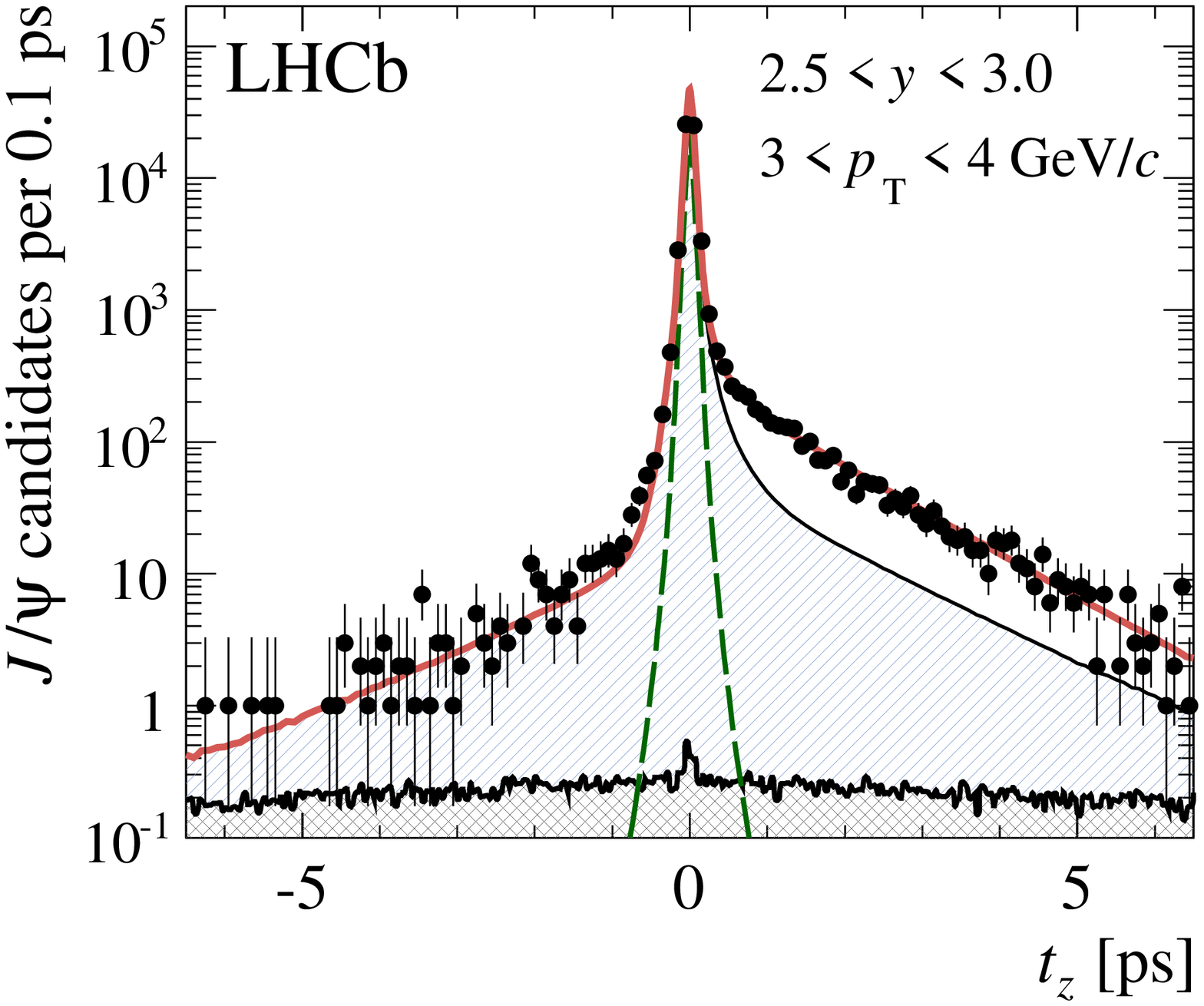}
\fi
\caption{\small Dimuon mass distribution ({\it left}) and $t_z$ distribution ({\it right}), with fit results superimposed, 
for one bin ($3 < \ptrans <4\,\gevc$,\,$2.5<y<3.0$). On the mass distribution, the {\it solid red line} is the total fit 
function, where
the signal is described by a Crystal Ball function, and the {\it dashed blue line} represents the 
exponential background function. On the $t_z$ distribution, the {\it solid red line} is the total fit function 
described in the text, the {\it green dashed line} is the \prompt\ contribution, the {\it single-hatched area} is the
background component and the {\it cross-hatched area} is the tail contribution.}
\label{fig:tzresult}
\end{figure}

As an example, Fig.~\ref{fig:tzresult} (left) shows the mass distribution together with the fit results for one specific bin
($3 < \ptrans <4\,\gevc$, \,$2.5<y<3.0$); the fit gives a mass resolution of $12.3\pm0.1\mevcc$ and a mean of
$3\,095.3\pm0.1\mevcc$, where the errors are statistical only. The mass value is close to the known 
\jpsi\ mass value of $3\,096.916 \pm 0.011\mevcc$~\cite{PDG}, reflecting the current status of the mass-scale
calibration; the difference between the two values has no effect on the results obtained in this
analysis. Summing over all bins, a total signal yield of $565\,000$ events is obtained. 

\subsection{\texorpdfstring{Determination of the fraction of $\mskip 1.5mu\mathbfi{\jpsi}$ from $\mathbfi{b}$}{Determination of the fraction of J/psi from b}}\label{fittz}

The fraction of \fromb, $F_{\rm b}$, is determined from the fits to the pseudo-proper time $t_z$ and the 
$\mu^+\mu^-$ invariant mass in each bin of \ptrans\ and $y$. The signal proper-time distribution is described by a
delta function at $t_z=0$ for the prompt \jpsi\ component, an exponential decay function for the \fromb\ component 
and a long tail arising from the association of the \jpsi\ candidate with the wrong primary vertex.  There are two 
main reasons for the wrong association: 
\begin{enumerate}
\item Two or more primary vertices are close to each other and a primary vertex is reconstructed with tracks 
belonging to the different vertices, at a position that is different from the true primary vertex position.
\item The primary vertex from which the \jpsi\ originates is not found because too few tracks originating from the
vertex are reconstructed, as confirmed by the simulation; the \jpsi\ candidate is then wrongly associated 
with another primary vertex found in the event.
\end{enumerate} 
In the first case, the positions of the reconstructed and of the true primary vertices are correlated. This category of
events is distributed around $t_z = 0$ for the prompt component, with a  width larger than the $t_z$ distribution for
correctly associated primary vertices. The contribution of these events to the $t_z$ distribution is included in the
resolution function described below.

The long tail is predominantly composed of events in the second category. Since the tail distribution affects the
measurement of the \fromb\ component, a method has been developed to extract its shape from data. The method
consists of associating a \jpsi\ from a given event with the primary vertex of the next event in the \jpsi\ sample. This
simulates the position of an uncorrelated primary vertex with which the \jpsi\ is associated. The shape of the tail
contribution to the signal $t_z$ distribution is then obtained from the  distribution of
\begin{equation}
t_z^{\rm next} = \frac{\left(z_{\jpsi}-z_{\rm PV}^{\rm next}\right) \times M_{\jpsi}}{p_z},
\end{equation}
where $z_{\rm PV}^{\rm next}$ is the position along the $z$-axis of the primary vertex of the next event.
The primary vertex reconstruction efficiency is assumed to be equal for \prompt\ and \fromb. Given the high primary
vertex reconstruction efficiency, $99.4\%$, the uncertainty related to this assumption is neglected.

The function describing the $t_z$ distribution of the signal is therefore
\begin{equation}
f_{\rm signal} (t_z;f_{\rm p},f_{\rm b},\tau_{\rm b}) = f_{\rm p}\,\delta(t_z)+f_{\rm b}\frac{ e^{-\frac{t_z}{\tau_{\rm b}}}}{ \tau_{\rm b}}  
+ \left(1-f_{\rm b}-f_{\rm p} \right)h_{\rm tail}(t_z)\,,
\end{equation}
where $f_{\rm p}$ is the fraction of prompt \jpsi\ for which the primary vertex is correctly associated, $f_{\rm b}$ 
the fraction of \fromb\ for which the primary vertex
is correctly associated, $\tau_{\rm b}$ the $b$-hadron pseudo-lifetime and $h_{\rm tail}(t_z)$  the probability 
density function taken as the histogram shape obtained from the ``next event'' method and displayed in 
Fig.~\ref{fig:tzresult} (right). The overall fraction of \fromb\ is  defined as 
$ F_{\rm b} = \displaystyle\frac{f_{\rm b}}{f_{\rm p}+f_{\rm b}}$. This assumes that the fraction of \fromb\ in the 
tail events is equal to the fraction measured with the events for which the primary vertex is correctly reconstructed. 

The prompt and $b$ components of the signal function are convolved with a double-Gaussian resolution function
\begin{equation}
f_{\rm resolution}(t_z;\mu,S_1,S_2,\beta) =
\frac{\beta}{\sqrt{2\pi}S_1\sigma}\,e^{-\frac{(t_z-\mu)^2}{2S_1^2\sigma^2}}+
\frac{1-\beta}{\sqrt{2\pi}S_2\sigma}\,e^{-\frac{(t_z-\mu)^2}{2S_2^2\sigma^2}}\,.
\end{equation}
The widths of the Gaussians are equal to the event-by-event $t_z$ measurement errors $\sigma$, multiplied by 
overall scale factors $S_1$ and $S_2$ to take into account possible mis-calibration effects on $\sigma$.
The parameter $\mu$ is the bias of the $t_z$ measurement and $\beta$ the fraction of the Gaussian with the
smaller scale factor. For bins with low statistics, a single-Gaussian resolution function is used.

The background consists of random combinations of muons from semi-leptonic $b$ and $c$ decays, which tend to
produce positive $t_z$ values, as well as of mis-reconstructed tracks from decays in flight of kaons and pions which
contribute both to positive and negative $t_z$ values. The background distribution is parameterised with an empirical 
function based on the shape of the $t_z$ distribution seen in the \jpsi\ mass sidebands. It is taken as the sum of a 
delta function and five exponential components (three for positive $t_z$ and two for negative $t_z$, 
the negative and positive 
exponentials with the largest lifetimes having their lifetimes $\tau_{\rm L}$ fixed to the same value), convolved with 
the sum of two Gaussian functions of widths $\sigma_1$ and $\sigma_2$ and fractions $\beta'$ and $(1-\beta')$
\begin{equation} 
\begin{split}
f_{\rm background}(t_z) =& {\Biggl [}\left(1-f_1-f_2-f_3-f_4\right)\delta(t_z) + \theta(t_z) 
\left( f_1\, \frac{e^{-\frac{t_z}{\tau_1}}}{\tau_1} + 
f_2 \, \frac{e^{-\frac{t_z}{\tau_2}}}{\tau_2}\right)
+ \\ 
& \theta(-t_z)\, f_3\, \frac{e^{\frac{t_z}{\tau_3}}}{\tau_3} + f_4\, \frac{e^{-\frac{|t_z|}{\tau_{\rm L}}}}{2 \tau_{\rm L}}
{\Biggr] }
\otimes \left(\frac{\beta'}{\sqrt{2\pi}\sigma_1}\, e^{-\frac{t_z^2}{2\sigma_1^2}}+
\frac{1-\beta'}{\sqrt{2\pi}\sigma_2}\, e^{-\frac{t_z^2}{2\sigma_2^2}}\right),
\end{split}
\end{equation}
where $\theta(t_z)$ is the step function. All parameters of the background function are determined independently in 
each bin of \ptrans\ and $y$, but for bins with low statistics the number of exponential components is reduced. 
The parameters are obtained from a fit to the $t_z$ distribution of the \jpsi\ mass sidebands defined as 
$M_{\mu\mu}\in[2.95\,;3.00]\cup[3.20\,;3.25]\gevcc$, and are fixed for the final fit. 

The function used to describe the $t_z$ distribution in the final fit is therefore
\begin{equation}\label{eq:tzfit}
\begin{split}
& f(t_z  ; f_{\rm p} , f_{\rm b} , f_{\jpsi} ,\mu , S_1 , S_2 , \beta ,\tau_{\rm b} ) = \\
& \ \ \ f_{\jpsi} \left[ \left(f_{\rm p}\, \delta(t_z)+f_{\rm b} \, \frac{e^{-\frac{t_z}{\tau_{\rm b}}}}{\tau_{\rm b}}\right) \otimes 
f_{\rm resolution}(t_z;\mu,S_1,S_2,\beta) +   \left(1-f_{\rm b}-f_{\rm p}\right)h_{\rm tail}(t_z) \right] + \\ 
& \ \ \ \ \ \ \left(1-f_{\jpsi} \right) f_{\rm background}( t_z )\,.
\end{split}
\end{equation}

\noindent The total fit function is the sum of the products of the mass and $t_z$ fit functions for the signal and 
background. 
Four bins of \ptrans\ and $y$, which contain less than 150 signal \jpsi\ events as determined from the mass fit, are
excluded from the analysis.

As an example, Fig.~\ref{fig:tzresult} (right) represents the $t_z$ distribution for one specific bin 
($3 < \ptrans <4\,\gevc$, \,$2.5<y<3.0$) with the fit result superimposed. The RMS of the $t_z$ resolution function is 
$53\,{\rm fs}$ and the fraction of tail events to the number of \jpsi\ signal is $(0.40\pm0.01)\%$.
As a measure of the fit quality, a $\chi^2$  is calculated for the fit function using a binned event distribution. The
resulting fit probability for the histogram of Fig.~\ref{fig:tzresult} (right) is equal to $87\%$ and similar good fits 
are seen for the other bins.

\subsection{Luminosity}

The luminosity was measured at specific periods during the data taking using both Van der Meer scans~\cite{vdm} 
and a beam-profile method~\cite{massi}.  Two Van der Meer scans were performed in a single fill. The analysis of 
these scans yields consistent results  for the absolute luminosity scale with a precision of 10$\%$, dominated by the 
uncertainty in the knowledge of the LHC proton beam currents.  In the second approach, six separate periods of 
stable running were chosen, and the beam-profiles measured using beam-gas and beam-beam interactions. Using 
these results, correcting for crossing angle effects, and knowing the beam currents, the luminosity in each period is 
determined following the analysis procedure described in Ref.~\cite{K0}.  Consistent results are found for the 
absolute luminosity scale in each period, with a precision of 10$\%$, also dominated by the beam current 
uncertainty. These results are in good agreement with those of the Van der Meer analysis. The knowledge of the 
absolute luminosity scale is used to calibrate the number of  VELO tracks, which is found to be stable
throughout the data-taking period and can therefore be used to monitor the instantaneous luminosity of the entire
data sample. The integrated luminosity of the runs considered in this analysis is determined to be 
$5.2\pm 0.5\pbinv$.

\subsection{Efficiency calculation}\label{sec_eff}
\begin{figure}[!tb]
\centering
\ifpdf
\includegraphics[width=13.2cm]{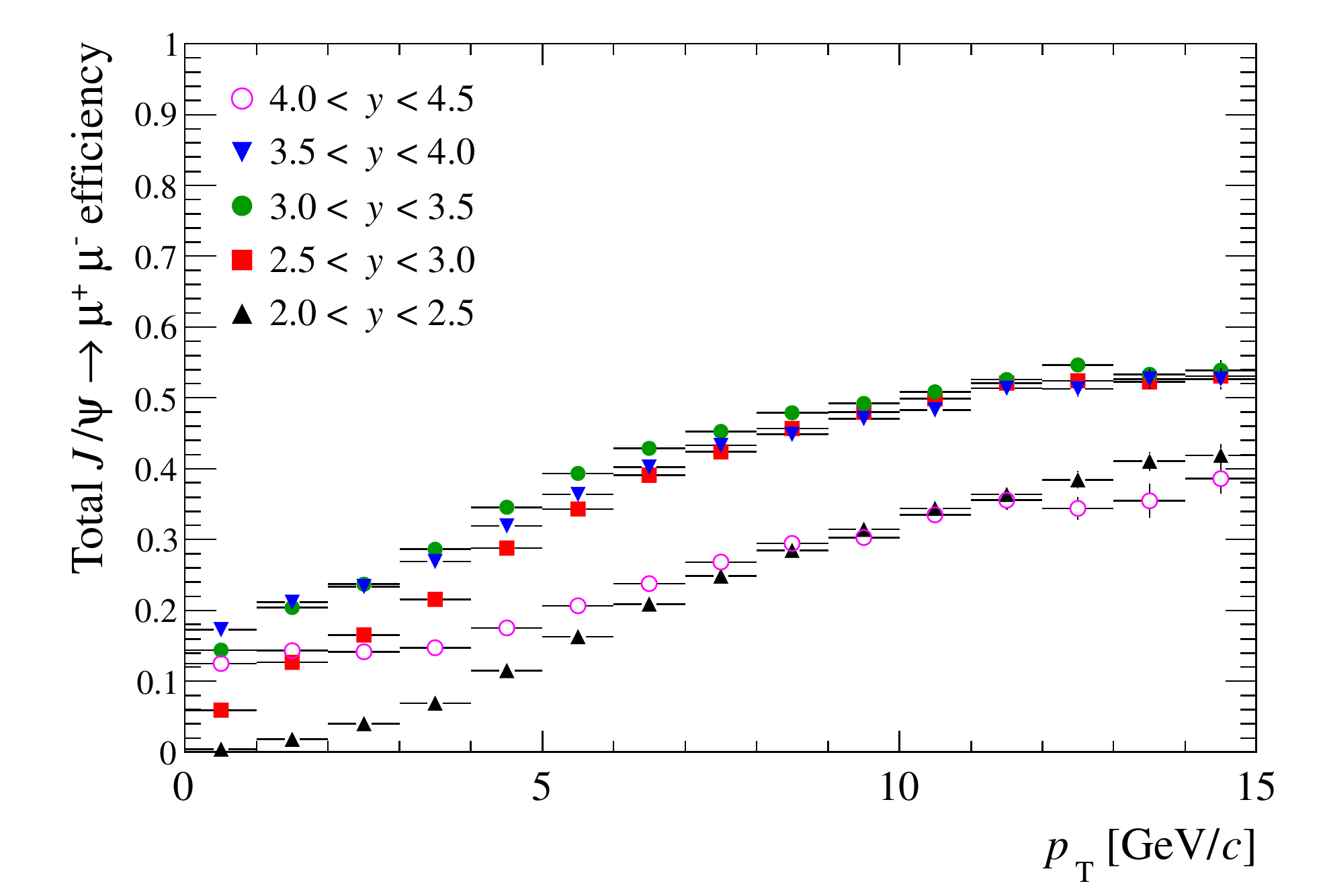}
\else
\includegraphics[width=13.2cm]{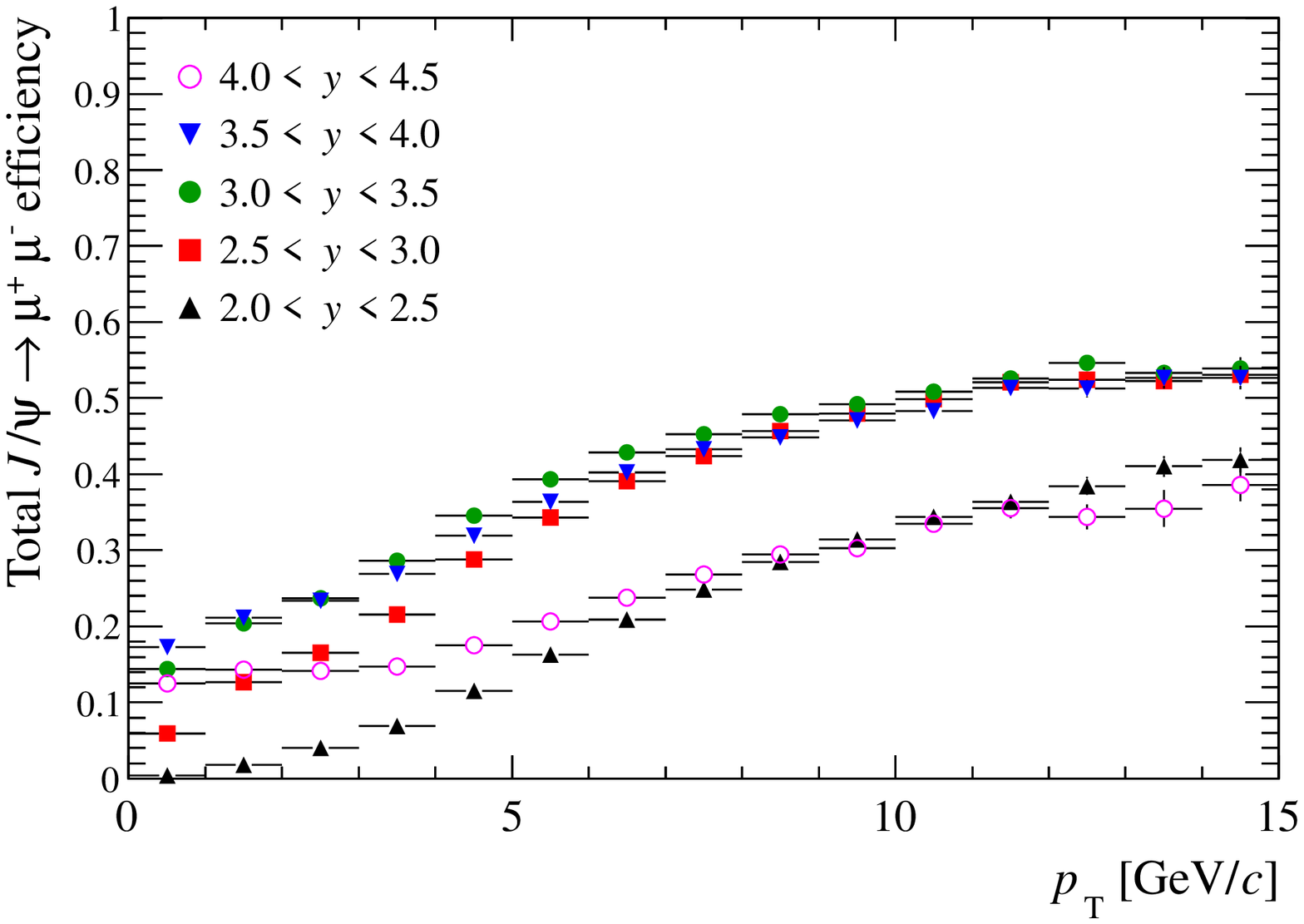}
\fi
\caption{\small Total \jpsi\ efficiency, as a function of \ptrans\ in bins of $y$ assuming that \jpsi\ are produced 
unpolarised. The efficiency is seen to drop somewhat at the edges of the acceptance.
\label{TotalEfficiency}}
\end{figure}

A simulated sample of inclusive, unpolarised \jpsi\ mesons is used to estimate the total efficiency 
$\epsilon_{\rm{tot}}$  in each bin of \ptrans\  and $y$. The total efficiency is the product of the geometrical 
acceptance, the detection, reconstruction and selection efficiencies,  and the trigger efficiency. It is displayed in 
Fig.~\ref{TotalEfficiency}, including both \prompt\ and \fromb.
The efficiencies are assumed to be equal for \prompt\ and \fromb\ in a given 
$(\ptrans,y)$ bin because neither the trigger nor the selection makes use of impact parameter
or decay length information. This assumption is confirmed with studies based on simulation.

A correction to the efficiency is applied to take into account the effect of the global event cuts described in 
Sec.~\ref{sec:detector}, introduced during data taking to remove high multiplicity events. The effect of such cuts on 
events containing a \jpsi\ candidate is not well described by the Monte Carlo simulation; it is therefore evaluated from 
data by using an independent  trigger, which accepts events having at least one track reconstructed in either the 
VELO or the tracking stations.  By comparing the number of such triggered signal \jpsi\ candidates before and after 
GEC, an efficiency of $(93\pm 2)\%$ is determined from data.  

\subsection{\texorpdfstring{Effect of the $\mskip 1.5mu\mathbfi{\jpsi}$ polarisation on the efficiency}{Effect of the J/psi polarisation on the efficiency}}

The efficiency is evaluated from a Monte Carlo simulation in which the \jpsi\  is produced unpolarised. However, 
studies show that non-zero \jpsi\  polarisation may lead to very different total efficiencies. In this analysis, 
the efficiency variation is studied in the helicity frame~\cite{gottfriedjackson}.

The angular distribution of the $\mu^+$ from the \jpsi\ decay is
\begin{equation}
\frac{{\rm d}^2N}{{\rm d}\cos\theta\,{\rm d}\phi} \propto 1 + \lambda_\theta\cos^2\theta 
+ \lambda_{\theta\phi}\sin2\theta\cos\phi 
+\lambda_{\phi}\sin^2\theta\cos2\phi,
\end{equation}
where $\theta$ is defined as the angle between the direction of the $\mu^+$ momentum in the \jpsi\  centre-of-mass 
frame and the direction of the \jpsi\  momentum in the centre-of-mass frame of the colliding protons, and 
$\phi$ is the azimuthal angle measured with respect to the production plane formed by the momenta of the colliding 
protons in the \jpsi\ rest frame. When $\lambda_\phi=0$ and $\lambda_{\theta\phi}=0$, the values 
$\lambda_\theta=+1,\,-1,\,0$ correspond to fully transverse, fully longitudinal, and no polarisation, respectively, 
which are the three default polarisation scenarios considered in this analysis.

The polarisation significantly affects the acceptance and reconstruction efficiencies. The relative efficiency change 
for \prompt\ varies between 3\% and 30\% depending on \ptrans\ and  $y$, when comparing to the unpolarised case. 
Therefore, the measurement of the differential prompt \jpsi\ cross-section will be given for the three default 
polarisations and a separate uncertainty due to the polarisation will be assigned to the integrated cross-section.

Three other polarisation configurations were studied, corresponding to 
$(\lambda_{\theta},\lambda_{\phi},\lambda_{\theta\phi}) = (+1,0,-1)$, $(0,1/\sqrt{2},-1/2)$ and $(0,-1/\sqrt{2},-1/2)$; 
these do not produce variations of the measured prompt cross-sections larger than those obtained with the default 
$(\pm 1,0,0)$ scenarios, except in some of the bins with $4<y<4.5$ where the variations are up to $25\%$ larger. 

The Monte Carlo simulation includes polarisation of \fromb\ as measured at {\sc BaBar} for $B^0$ and $B^+$ 
decays~\cite{babarpol}. The simulation shows that the polarisation that the \jpsi\ acquires in $b$ decays is largely 
diluted when using as helicity quantisation axis the \jpsi\ momentum in the laboratory frame instead of the \jpsi\
momentum in the $b$-hadron rest frame, which is the natural polarisation axis.
The effect of the \fromb\ polarisation on the \jpsi\ acceptance and reconstruction efficiencies is less than $0.5\%$;
therefore, no systematic uncertainty is assigned to the \fromb\ cross-section measurement from the unknown \jpsi\
polarisation.

\section{Systematic uncertainties}\label{sec:systematics}

The different contributions to the systematic uncertainties affecting the cross-section measurement are discussed in 
the following and  summarised in Table~\ref{syst}.

\renewcommand{\arraystretch}{1.2}
\begin{table}[!tb]
\tabcolsep 4mm
\begin{center}
\caption{\small \label{syst}Summary of systematic uncertainties.}
\begin{tabular}{@{}lll@{}}
\toprule
& Source & Systematic uncertainty (\%) \\\midrule\midrule
\multicolumn{3}{l}{\small \it Correlated between bins} \\
& Inter-bin cross-feed & $\phantom{1}0.5$ \\
& Mass fits & $\phantom{1}1.0$ \\
& Radiative tail & $\phantom{1}1.0$ \\
& Muon identification & $\phantom{1}1.1$ \\
& Tracking efficiency & $\phantom{1}8.0$ \\
& Track $\chi^2$ & $\phantom{1}1.0$ \\
& Vertexing & $\phantom{1}0.8$ \\
& GEC & $\phantom{1}2.0$ \\
& ${\cal B}(\jpsi\to\mu^+\mu^-)$ & $\phantom{1}1.0$ \\
& Luminosity & $10.0$ \\
\midrule
\multicolumn{3}{l}{\small \it Uncorrelated between bins} \\
& Bin size & $\phantom{1}0.1$ to $15.0$ \\
& Trigger  & $\phantom{1}1.7$ to $4.5$ \\
\midrule
\multicolumn{3}{l}{\small \it Applied only to \jpsi\ from $b$ cross-sections, correlated between bins} \\
& GEC efficiency on $B$ events & $\phantom{1}2.0$ \\
& $t_z$ fits & $\phantom{1}3.6$ \\
\midrule
\multicolumn{3}{l}{\small \it Applied only to the extrapolation of the $b\overline{b}$ cross-section} \\
& $b$ hadronisation fractions & $\phantom{1}2.0$ \\
& ${\cal B}(b\to\jpsi X)$ &  $\phantom{1}9.0$ \\
\bottomrule
\end{tabular}
\end{center}
\end{table}

Due to the finite \ptrans\ and $y$ resolutions, \jpsi\ candidates can be assigned to a wrong \ptrans\ bin (inter-bin 
cross-feed in Table~\ref{syst}).
According to Monte Carlo simulations, the average \ptrans\ resolution is $12.7\pm0.2\mevc$ and the $y$ resolution 
is $(1.4\pm0.1)\times 10^{-3}$.  The effect of the $y$ resolution is negligible compared to the bin width of 
$\Delta y=0.5$. The effect of the \ptrans\ resolution is estimated by recomputing the efficiency tables after smearing 
the \ptrans\ values with a Gaussian distribution of $\sigma=20\mevc$. The maximum relative deviation observed is 
$0.5\%$ and this is the value used as systematic uncertainty for the differential cross-section measurement. The 
effect on the total cross-section is much smaller and is ignored.

The influence of the choice of the fit function used to describe the shape of the dimuon mass distribution  is 
estimated by fitting the \jpsi\ invariant mass distribution with the sum of two Crystal Ball functions. The relative 
difference of 1\%
in the number of signal events is taken as systematic uncertainty.

A fraction of \jpsi\ events have a lower mass because of the radiative tail. Based on Monte Carlo studies,  $2\%$ of 
the \jpsi\ signal is estimated to be outside the analysis mass window ($M_{\mu\mu}<2.95\gevcc$) and not counted as 
signal. The fitted signal yields are therefore corrected by 2\%, and an uncertainty of 1\% is assigned to the 
cross-section measurements.

To cross-check and assign a systematic uncertainty to the Monte Carlo determination of the muon identification 
efficiency,  the single track muon identification efficiency is measured on data using a tag-and-probe method. This 
method reconstructs \jpsi\ candidates in which one muon is identified by the muon system (``tag'') and the other one 
(``probe'') is identified selecting a track depositing the energy of minimum-ionising particles in the calorimeters. The 
absolute muon identification efficiency is then evaluated on the probe muon, as a function of the muon momentum.
The ratio of the muon identification efficiency measured in data to that obtained in the Monte Carlo simulation is 
convolved with the momentum distribution of muons from \jpsi\ to compute a correction factor to apply on 
simulation-based efficiencies. This factor is found to be $1.024\pm0.011$ and is consistent with being constant 
over the full \jpsi\ transverse momentum and rapidity range; the error on the correction factor is used as a 
systematic uncertainty.
The residual misalignment between the tracking system and the muon detectors is accounted for in this
systematic uncertainty.

Tracking studies have shown that the Monte Carlo simulation reproduces  the track-finding efficiency in data within 
4\%. A systematic uncertainty of 4\% for each muon is therefore assigned, resulting in a total systematic uncertainty 
of 8\% due to the knowledge of the track reconstruction efficiency~\cite{Sheldon}. The effects of the residual 
misalignment of the tracking system are included in this systematic uncertainty.

The selection includes a requirement on the track fit quality, which may not be reliably simulated. A systematic 
uncertainty of 0.5\% is assigned per track, which is the relative difference between the efficiency of this requirement 
in the simulation and data.

Similarly, for the cut on the \jpsi\ vertex $\chi^2$  probability, a difference of $1.6\%$ is measured between the cut 
efficiency computed in data and simulation. The Monte Carlo efficiency is corrected for this difference and a 
systematic uncertainty of 0.8\% (half of the correction) is assigned.

The unknown \jpsi\ transverse momentum and rapidity spectra inside the bins affect the efficiency values used to 
extract the cross-section, because an average value of the efficiency is computed in each bin. This effect is 
important close to the edges of the fiducial region. To take into account possible efficiency variations inside the bins, 
each bin is divided into four sub-bins (two bins in \ptrans\ and two bins in $y$) and the relative deviation between 
the bin efficiency and the average of the 
efficiencies in the sub-bins is  taken as a systematic uncertainty. 

The trigger efficiency can be determined using a trigger-unbiased sample of events that would still be triggered if the
\jpsi\ candidate were removed. The efficiency obtained with this method in each ($\ptrans,y$) bin 
is used to check the efficiencies measured in the simulation. The systematic uncertainty associated with the trigger 
efficiency is the difference between the trigger efficiency measured in the data and in the simulation. 
The largest uncertainties are obtained for the high rapidity bins.

The statistical error on the GEC efficiency ($2\%$) is taken as an additional syste\-matic uncertainty associated with 
the trigger. This efficiency is extracted from data as explained in Sec.~\ref{sec_eff}; it is essentially the efficiency of 
the GEC on prompt \jpsi. In the simulation, a 
$2\%$ difference is seen between the \prompt\ and the \fromb\ efficiency, which is used 
as an additional systematic uncertainty, applied only to the \fromb\ cross-section measurement.

Uncertainties related to the $t_z$ fit procedure are taken into account by varying the central value of the prompt  
\jpsi\ component, $\mu$, which is found to be different from zero. This shift could be due to an improper description 
of the background for events close to $t_z=0$. The impact of such a  shift  is studied by fixing $\mu$ at two extreme 
values, $\mu=-3\,{\rm fs}$ and $\mu=3\,{\rm fs}$ and repeating the $t_z$ fit. The relative variation of the number of 
\fromb, $3.6\%$, is used as a systematic uncertainty.

The extrapolation to the full polar angle to obtain the $b\overline{b}$ cross-section uses the average branching 
fraction of inclusive $b$-hadron decays to \jpsi\ measured at LEP, i.e., 
${\cal B}(b\to\jpsi X)=(1.16\pm0.10)\%$~\cite{delphibtojpsi}.  
The underlying assumption is that the $b$-hadron fractions in $pp$ collisions at $\sqrt{s}=7\,{\rm TeV}$ are 
identical to those seen in $Z\to b\overline{b}$ decays. However, the $b$ hadronisation fractions may differ at 
hadronic machines. 
To estimate the systematic uncertainty due to possibly different fractions, the ${\cal B}(b\to\jpsi X)$ is computed by 
taking as input for the calculation the fractions
measured at the Tevatron~\cite{CDFbfractions,HFAG} and
assuming the partial widths of $B_{\rm u}$, $B_{\rm d}$, $B_{\rm s}$ and $\Lambda_b$ to $\jpsi X$ to be equal.
The relative difference between the estimates of the branching fractions based on the fragmentation functions 
measured at LEP and at the Tevatron, 2\%, is taken as systematic uncertainty, which only affects the
extrapolation of the $b\overline{b}$ cross-section.

\section{Results}

\begin{figure}[!tb]
\centering
\ifpdf
\includegraphics[width=14.1cm]{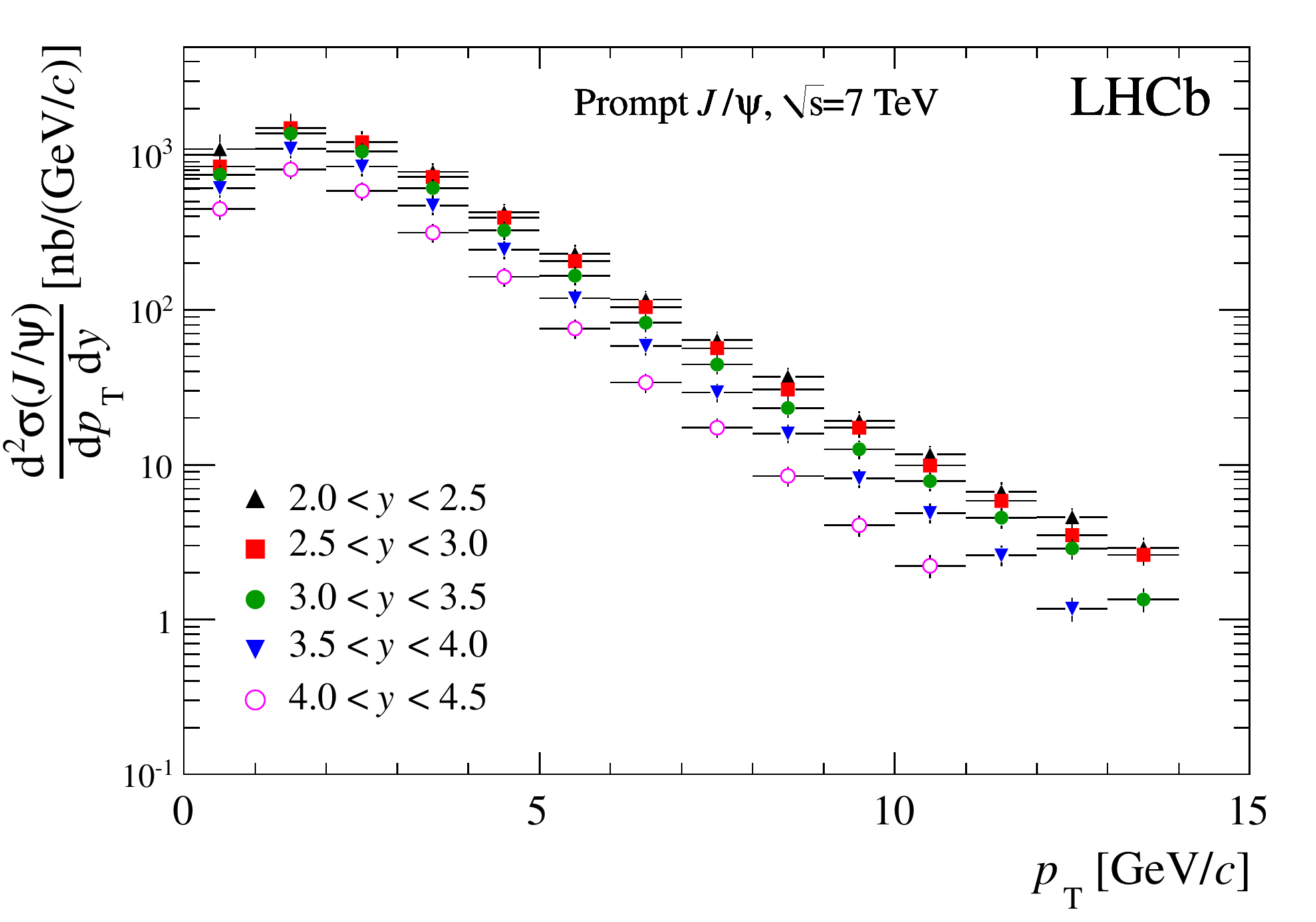}
\else
\includegraphics[width=14.1cm]{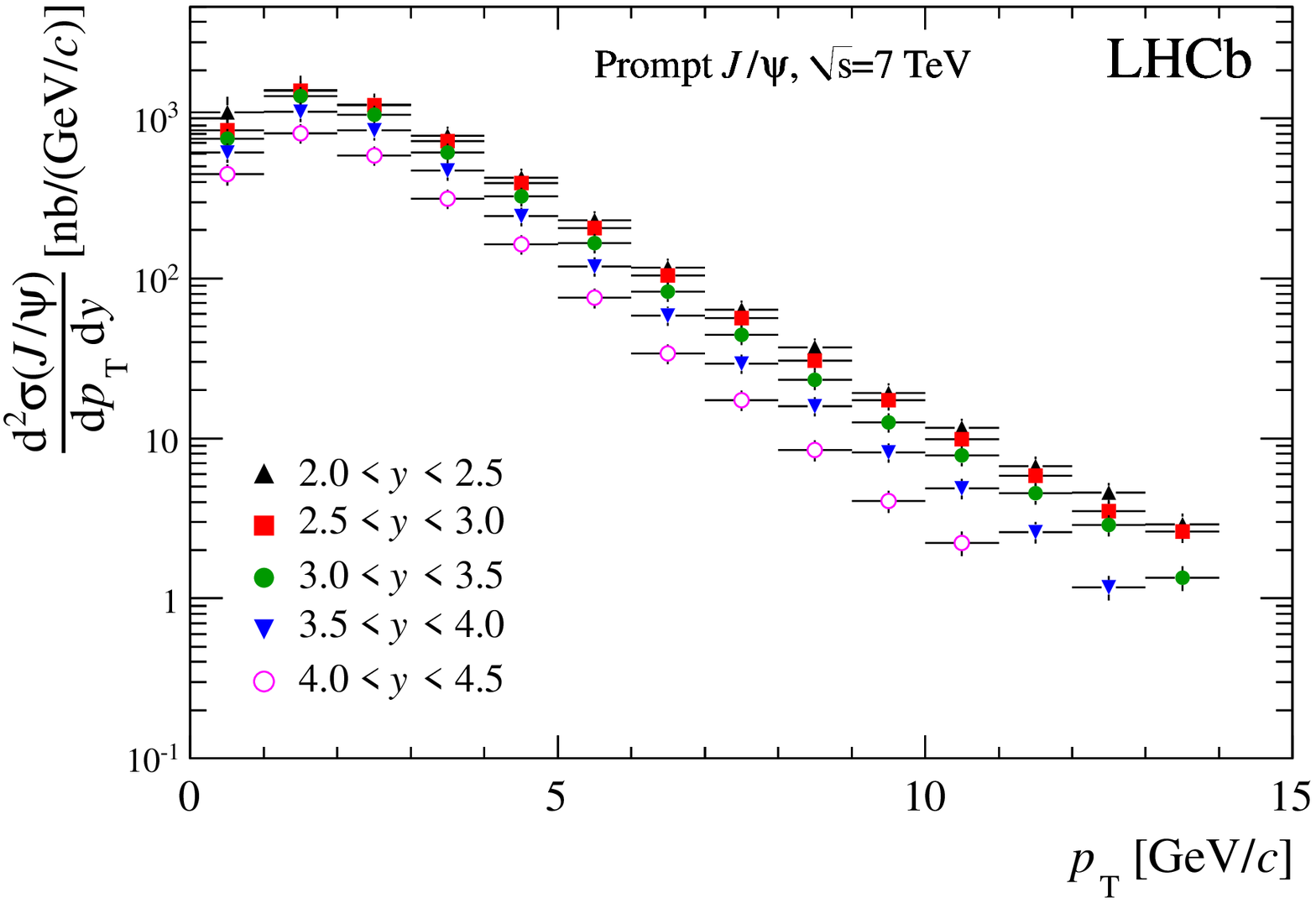}
\fi
\vspace{-0.4cm}
\caption{\small Differential production cross-section for prompt \jpsi\ as a function of \ptrans\ in bins of $y$ , 
assuming that \prompt\ are produced unpolarised. The errors are the quadratic sums of the statistical and 
systematic uncertainties. } \label{sigmaresults_prompt}
\end{figure}

\begin{figure}[!bt]
\centering
\ifpdf
\includegraphics[width=14.1cm]{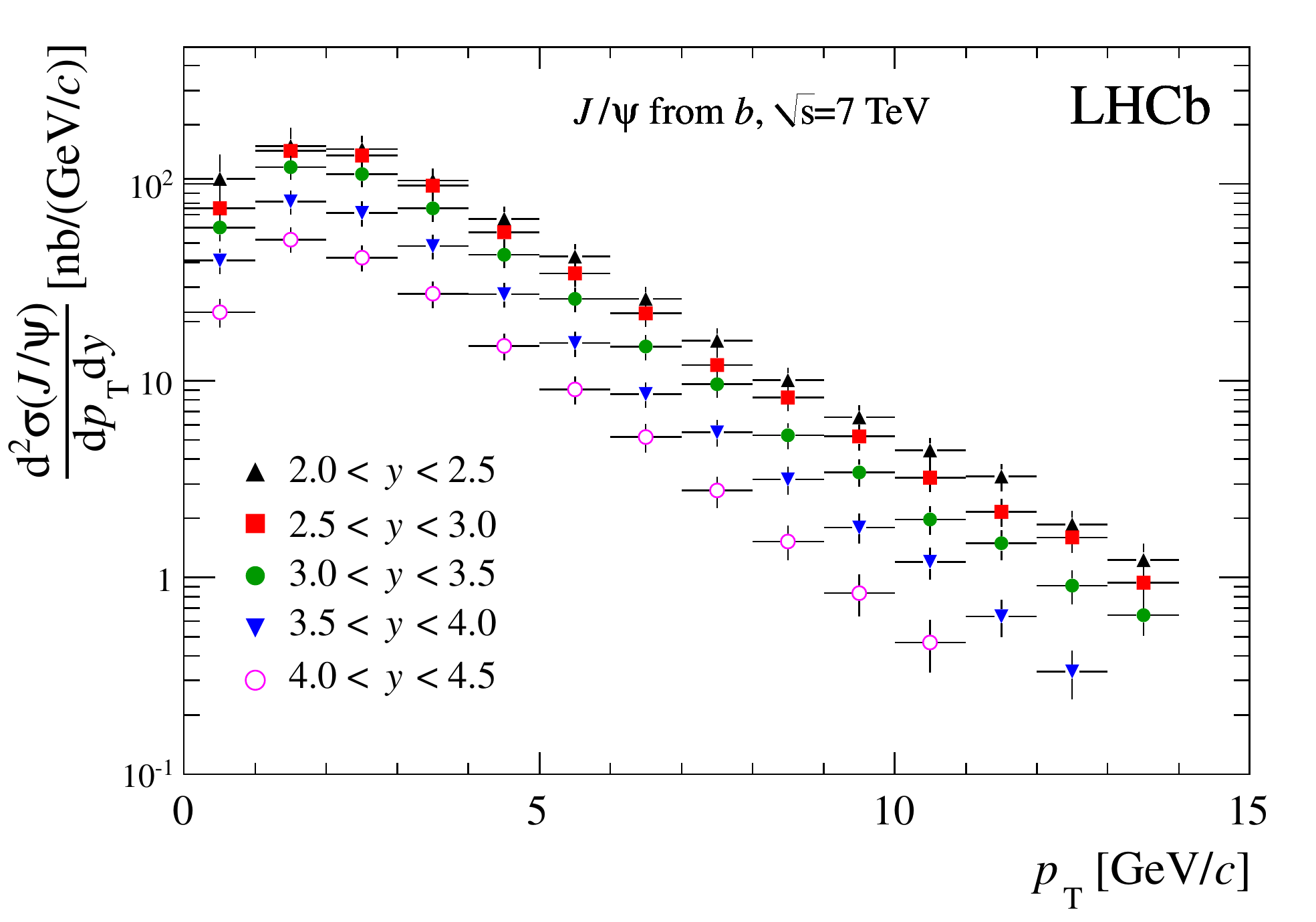}
\else
\includegraphics[width=14.1cm]{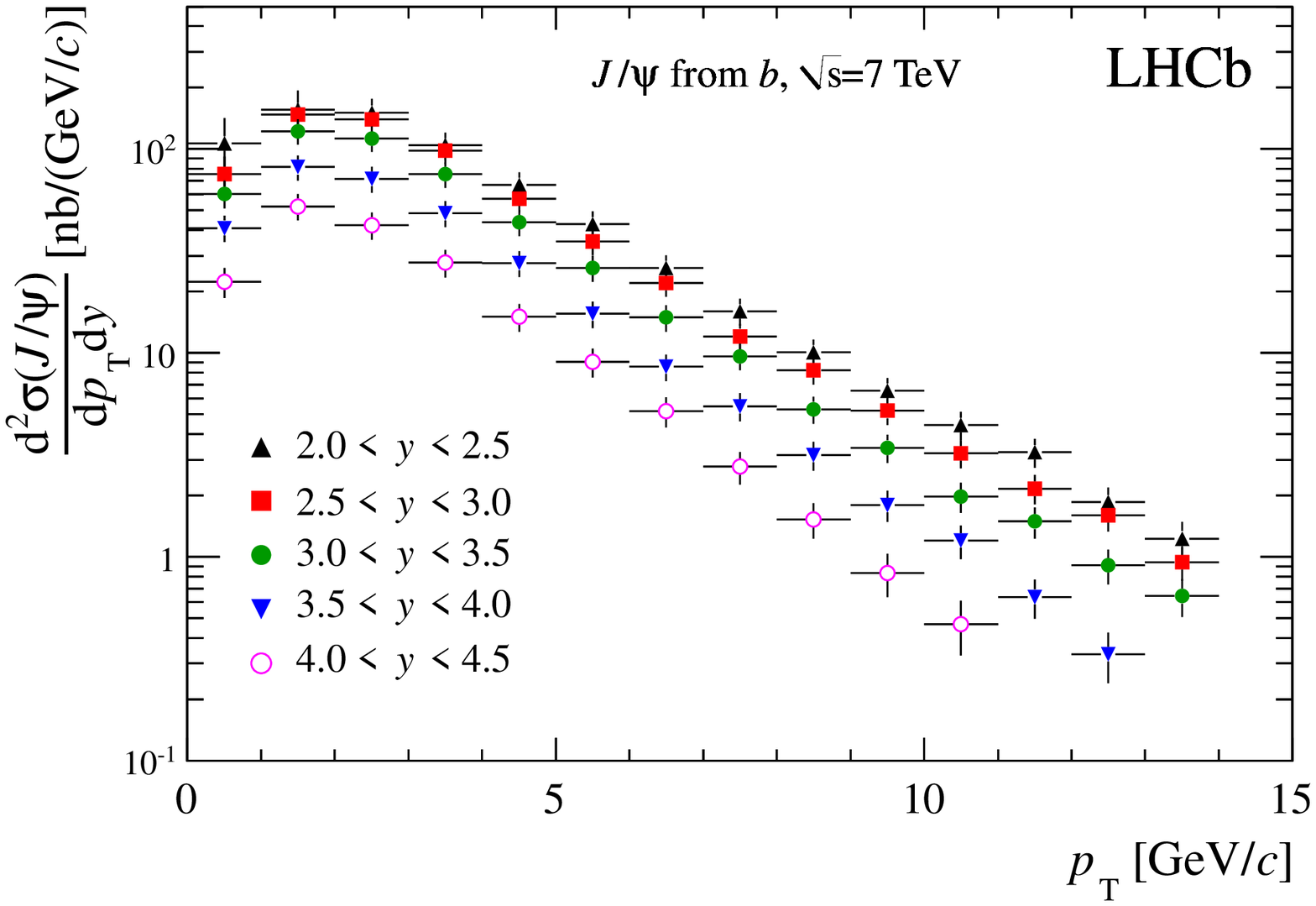}
\fi
\vspace{-0.4cm}
\caption{\small Differential production cross-section for \fromb\ as a function of \ptrans\  in bins of $y$. The errors are
the quadratic sums of the statistical and systematic uncertainties. } \label{sigmaresults_fromb}
\end{figure}

\begin{figure}[!b]
\centering
\ifpdf
\includegraphics[width=7.95cm]{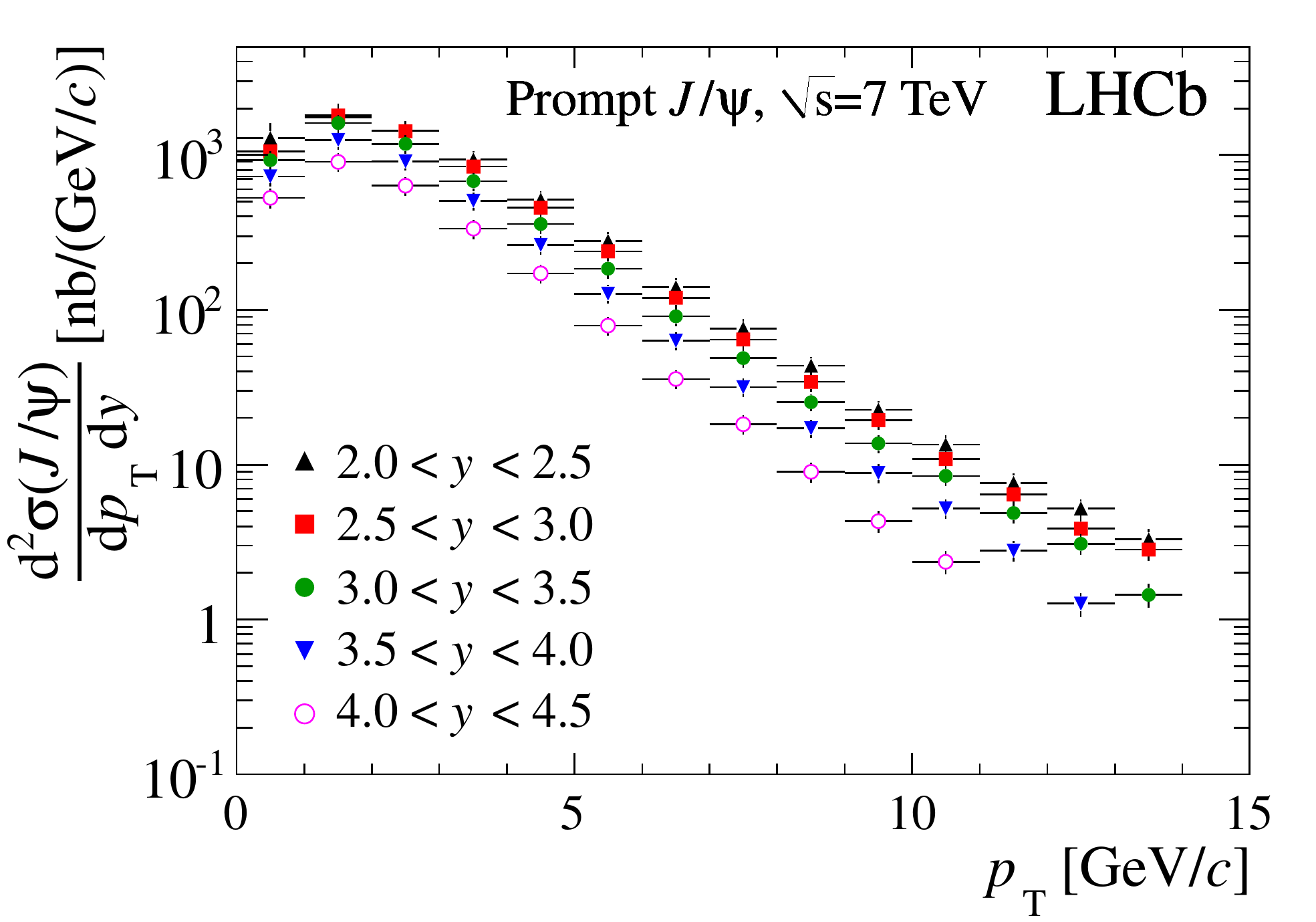}
\includegraphics[width=7.95cm]{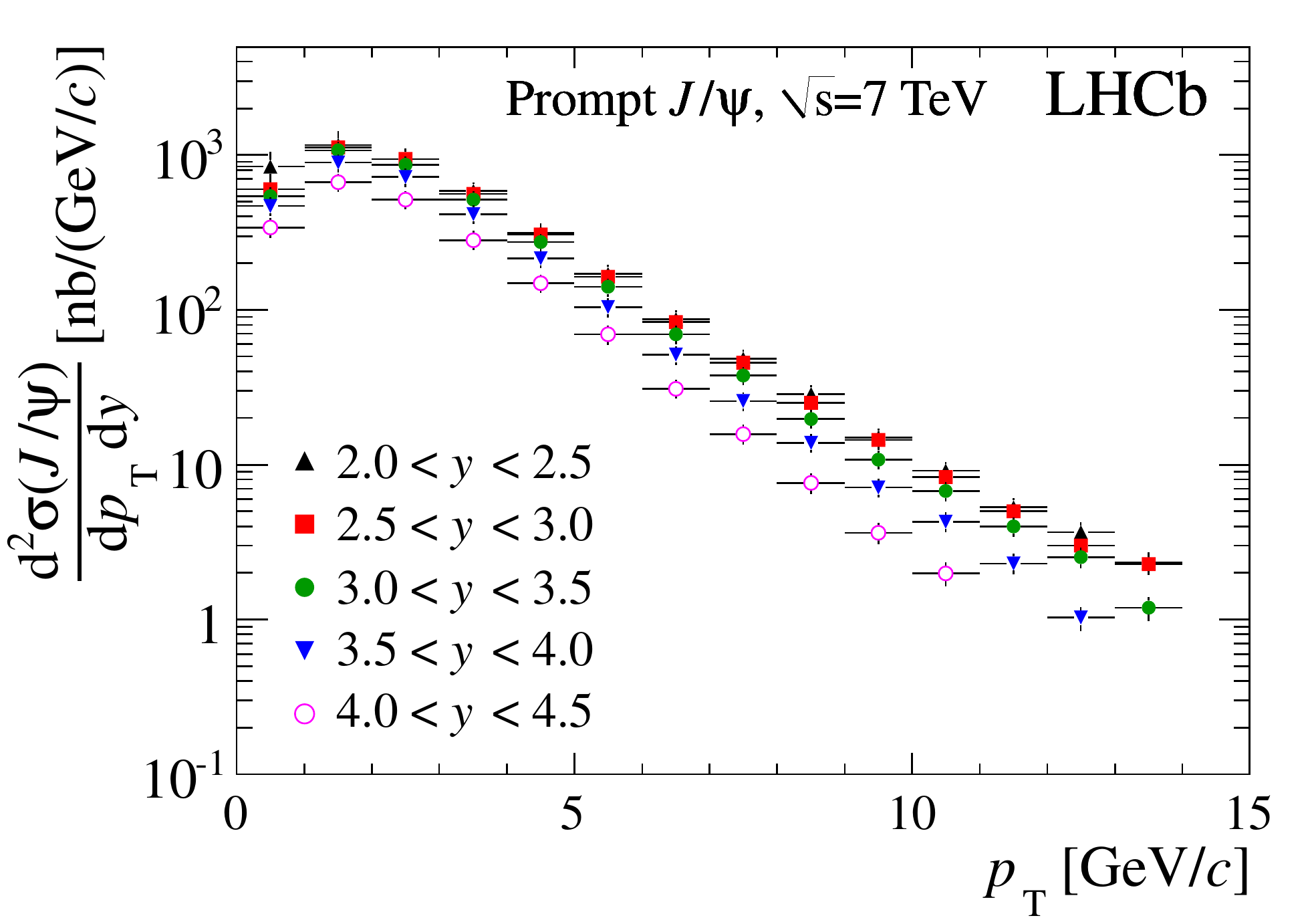}
\else
\includegraphics[width=7.95cm]{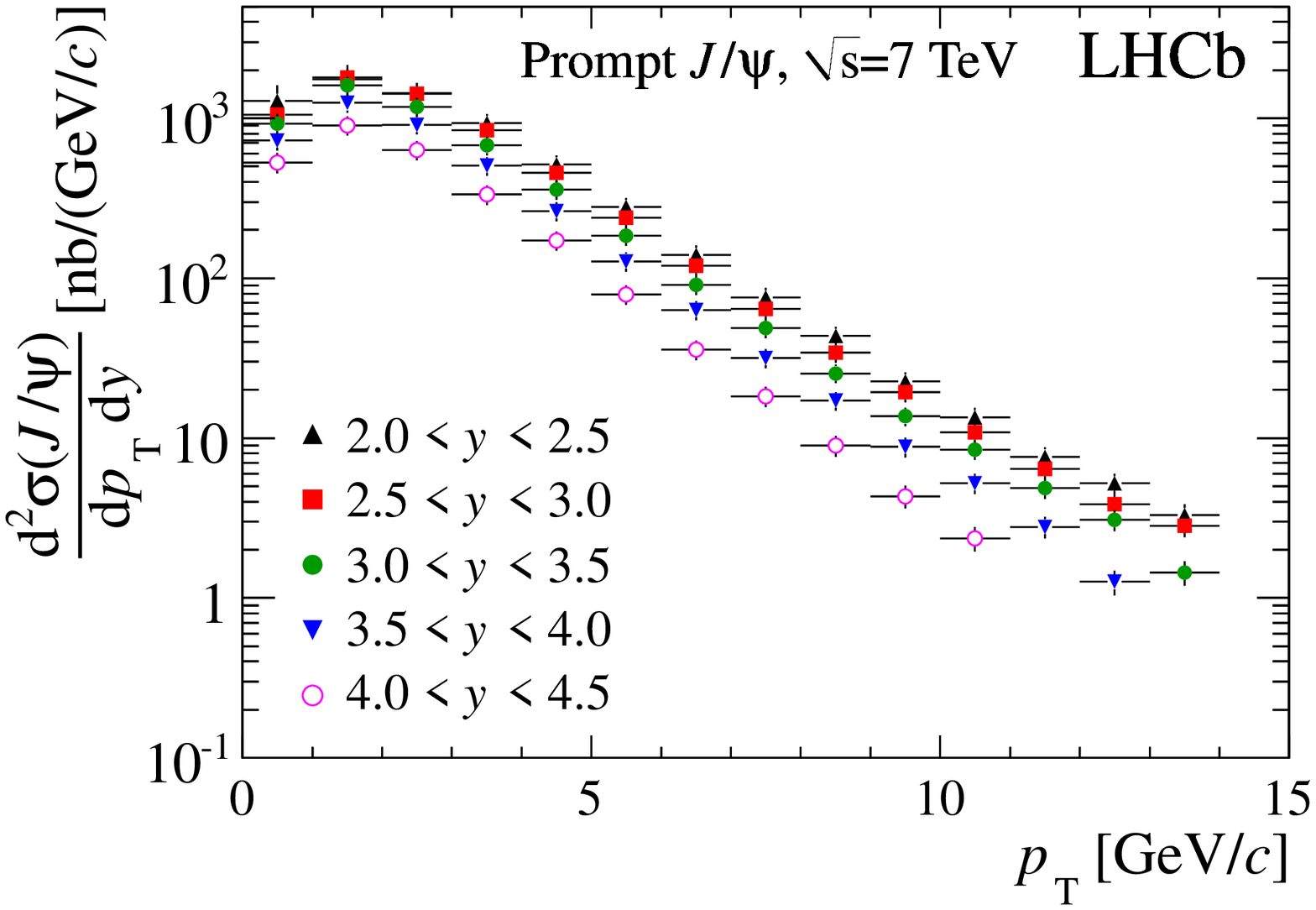}
\includegraphics[width=7.95cm]{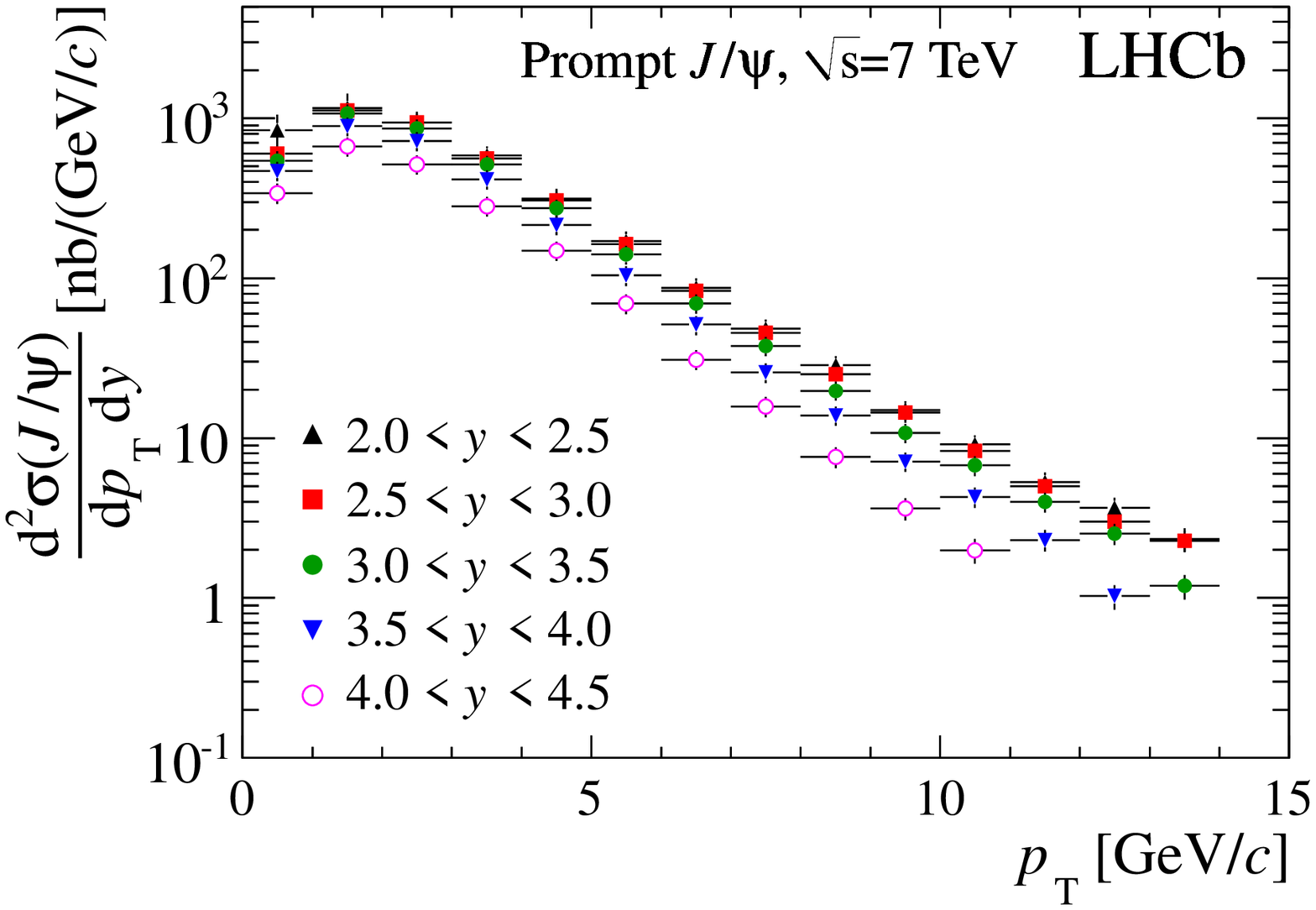}
\fi
\caption{\small Differential production cross-section for \prompt\ as a function of \ptrans\  in bins of $y$, assuming full 
transverse ({\it left}) or full longitudinal ({\it right}) \jpsi\ polarisation. The errors are the quadratic sums of the 
statistical and systematic uncertainties.} \label{xsecallpol}
\end{figure}

The measured  double-differential cross-sections for prompt \jpsi\ and \fromb\  in the various ($\ptrans,y$) bins, after 
all corrections and assuming no polarisation, are given in Tables~\ref{promptresult} and \ref{bresult},
and displayed in Figs.~\ref{sigmaresults_prompt} and~\ref{sigmaresults_fromb}. The results for full transverse and 
full longitudinal polarisation of the \jpsi\ in the helicity frame are given in Tables~\ref{promptresulttransverse} and 
\ref{promptresultlongitudinal}, and displayed in Fig.~\ref{xsecallpol}.


The integrated cross-section for prompt \jpsi\ production in the
defined fiducial region, summing over all bins of the analysis, is
\begin{equation}
\sigma\left({\rm prompt}\,\, \jpsi, \ptrans <14\,\gevc,\,2.0<y<4.5\right)  \, = \, 
10.52\pm 0.04\pm 1.40^{+1.64}_{-2.20}\,\,\upmu{\rm b},
\end{equation}
where the first uncertainty is statistical and the second systematic. The result is quoted assuming unpolarised \jpsi\
and the last error indicates the uncertainty related to this assumption.
The integrated cross-section for the production of \fromb\ in the same fiducial region is
\begin{equation}\label{sigma_from_b}
\sigma\left(\fromb,\, \ptrans <14\,\gevc,\,2.0<y<4.5\right)  \, =  \, 1.14 \pm 0.01\pm 0.16\,\upmu{\rm b},
\end{equation}
where the first uncertainty is statistical and the second systematic.

The mean and RMS of the \ptrans\ spectrum in each $y$ bin 
are displayed in Table~\ref{tab:meanpt}. The \jpsi\ mesons from $b$-hadron
decays have a mean \ptrans\ and RMS which
are approximately 20\% larger than those
of prompt \jpsi\ mesons. For each \jpsi\ source,
the mean \ptrans\ and RMS are observed to decrease with 
increasing $y$. 

\renewcommand{\arraystretch}{1.5}
\begin{table}[!t]
\caption{\small Mean \ptrans\ and RMS for \prompt\ (assumed unpolarised) and \fromb. The first uncertainty is 
statistical, the second systematic and the third for \prompt\ the uncertainty due to the unknown polarisation.}
\vspace{-0.1cm}
\begin{center}
\scalebox{0.77}{
\begin{tabular}{@{}lllllll@{}}
\toprule
\ptrans\ range& & \multicolumn{2}{c}{Prompt $J/\psi$} & & \multicolumn{2}{c}{\fromb} \\ \cmidrule{3-4}\cmidrule{6-7}
(\gevc) & $y$ bin  & $\langle\ptrans\rangle$ (\gevc) & RMS \ptrans\ (\gevc) & & $\langle\ptrans\rangle$ (\gevc) & 
RMS \ptrans\ (\gevc) \\
\midrule
$0-14$ & $2.0-2.5\ \ \ $ & $2.51 \pm 0.03 \pm 0.10_{-0.01}^{+0.02}\ \ \ $ & 
$1.80 \pm 0.01 \pm 0.04_{-0.02}^{+0.00}$ & 
&  $3.06 \pm 0.09 \pm 0.11$ & $2.22 \pm 0.02 \pm 0.04$ \\
$0-14$ & $2.5-3.0$ & $2.53 \pm 0.01 \pm 0.06_{-0.04}^{+0.06}$ & $1.74 \pm 0.01 \pm 0.01_{-0.02}^{+0.02}$ & 
&  $3.04 \pm 0.02 \pm 0.05$ & $2.12 \pm 0.01 \pm 0.01$ \\
$0-14$ & $3.0-3.5$ & $2.46 \pm 0.01 \pm 0.02_{-0.05}^{+0.07}$ & $1.68 \pm 0.01 \pm 0.01_{-0.01}^{+0.02}$ & 
&  $2.93 \pm 0.02 \pm 0.02$ & $2.03 \pm 0.01 \pm 0.01$ \\
$0-13$ & $3.5-4.0$ & $2.38 \pm 0.01 \pm 0.02_{-0.05}^{+0.07}$ & $1.61 \pm 0.01 \pm 0.01_{-0.01}^{+0.01}$ & 
&  $2.82 \pm 0.02 \pm 0.02$ & $1.92 \pm 0.02 \pm 0.01$ \\
$0-11$ & $4.0-4.5$ & $2.29 \pm 0.01 \pm 0.02_{-0.05}^{+0.08}$ & $1.50 \pm 0.01 \pm 0.01_{-0.01}^{+0.01}$ & 
&  $2.73 \pm 0.03 \pm 0.03$ & $1.77 \pm 0.03 \pm 0.01$ \\
\bottomrule 
\end{tabular}
}
\end{center}
\label{tab:meanpt} 
\end{table}
\renewcommand{\arraystretch}{1.}

\begin{figure}[!ht]
\centering
\ifpdf
\includegraphics[width=7.95cm]{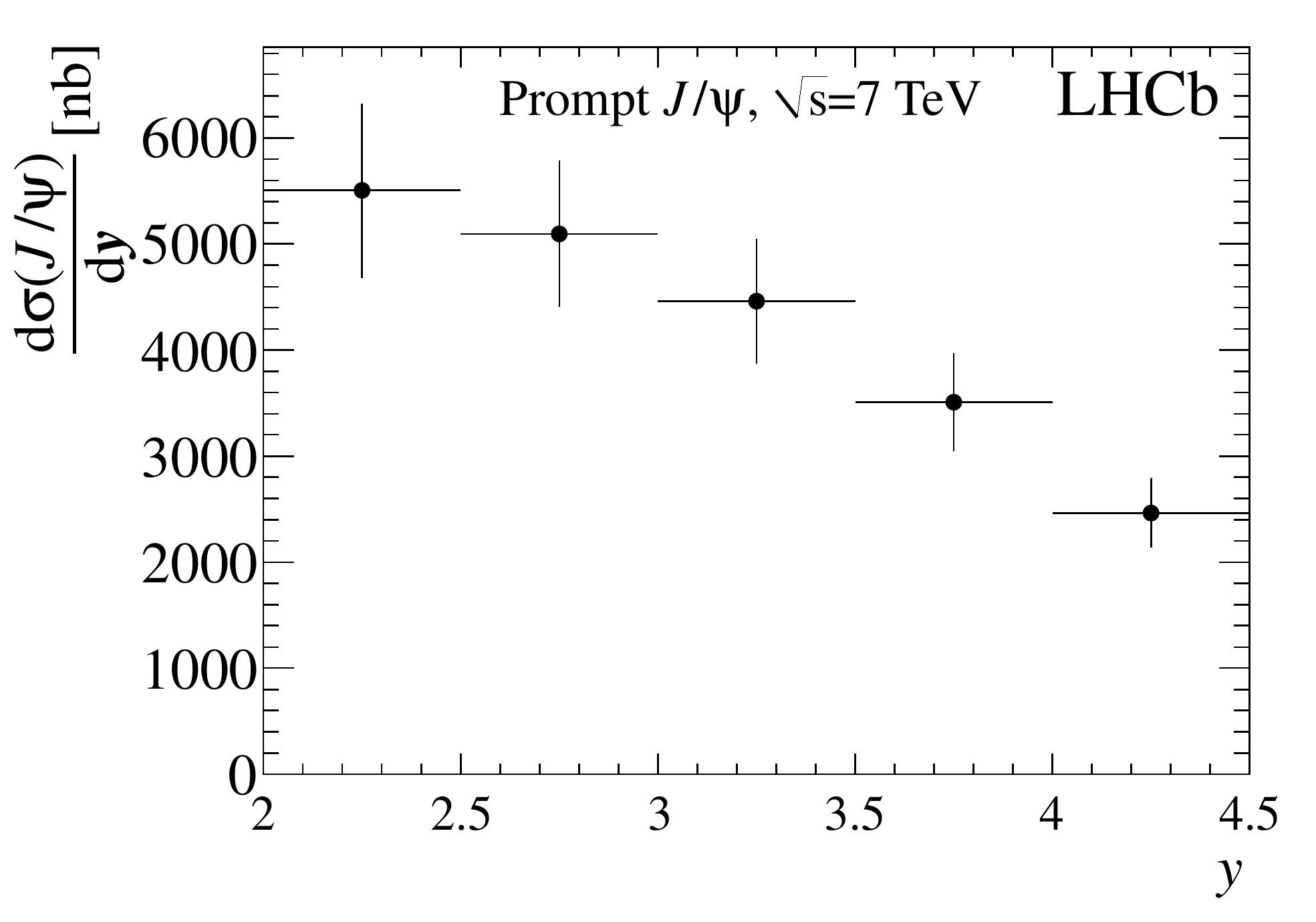}
\includegraphics[width=7.95cm]{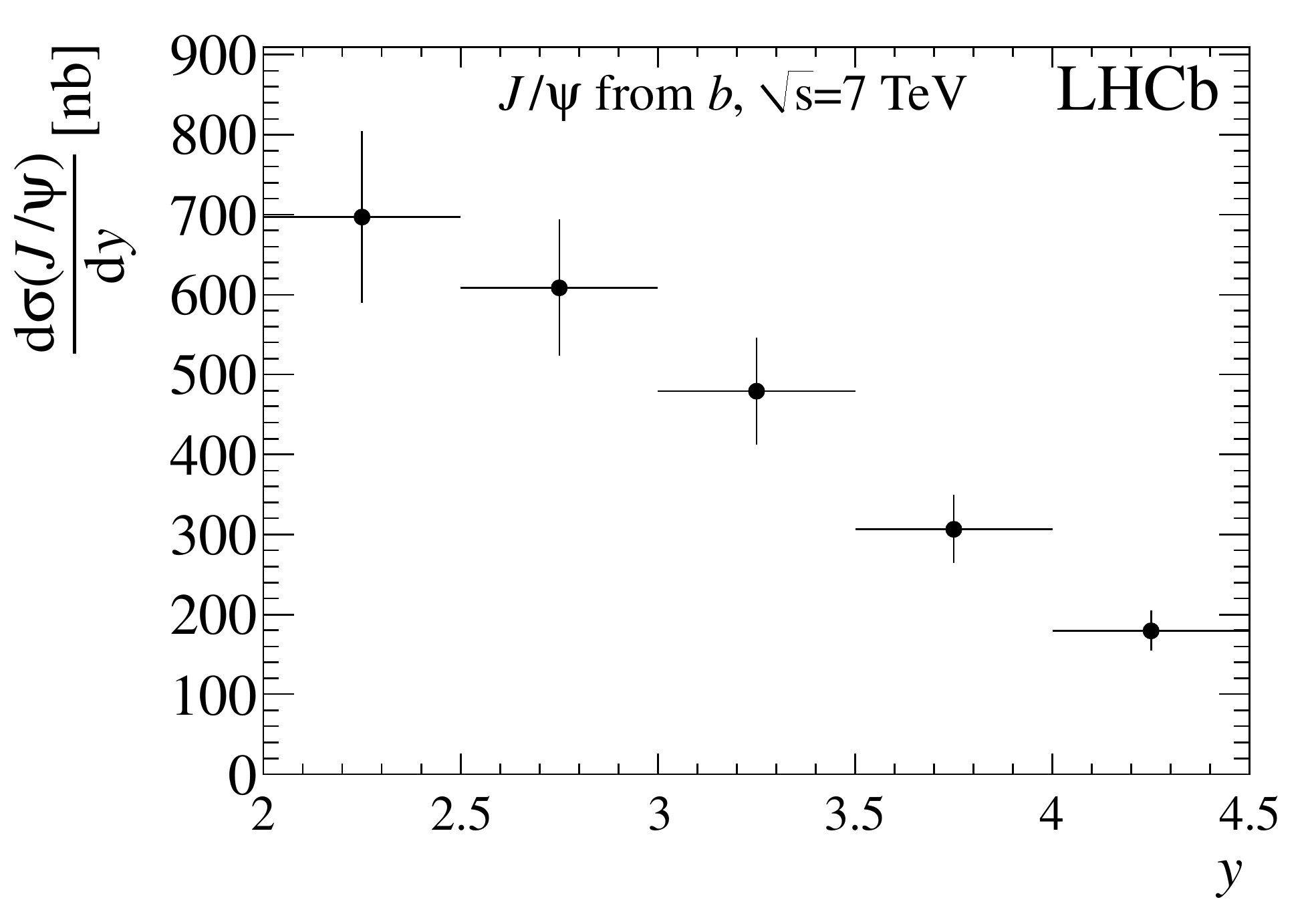}
\else
\includegraphics[width=7.95cm]{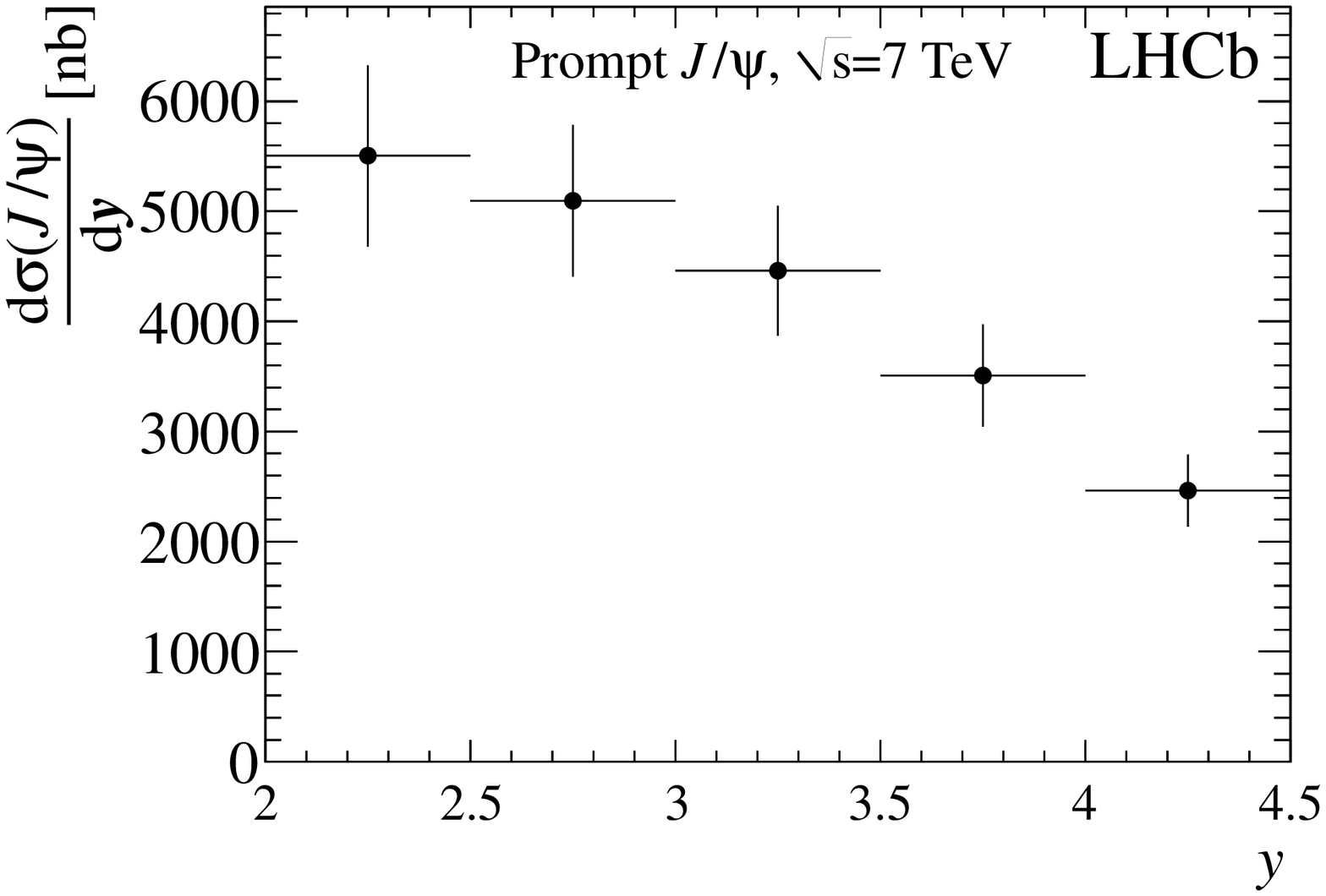}
\includegraphics[width=7.95cm]{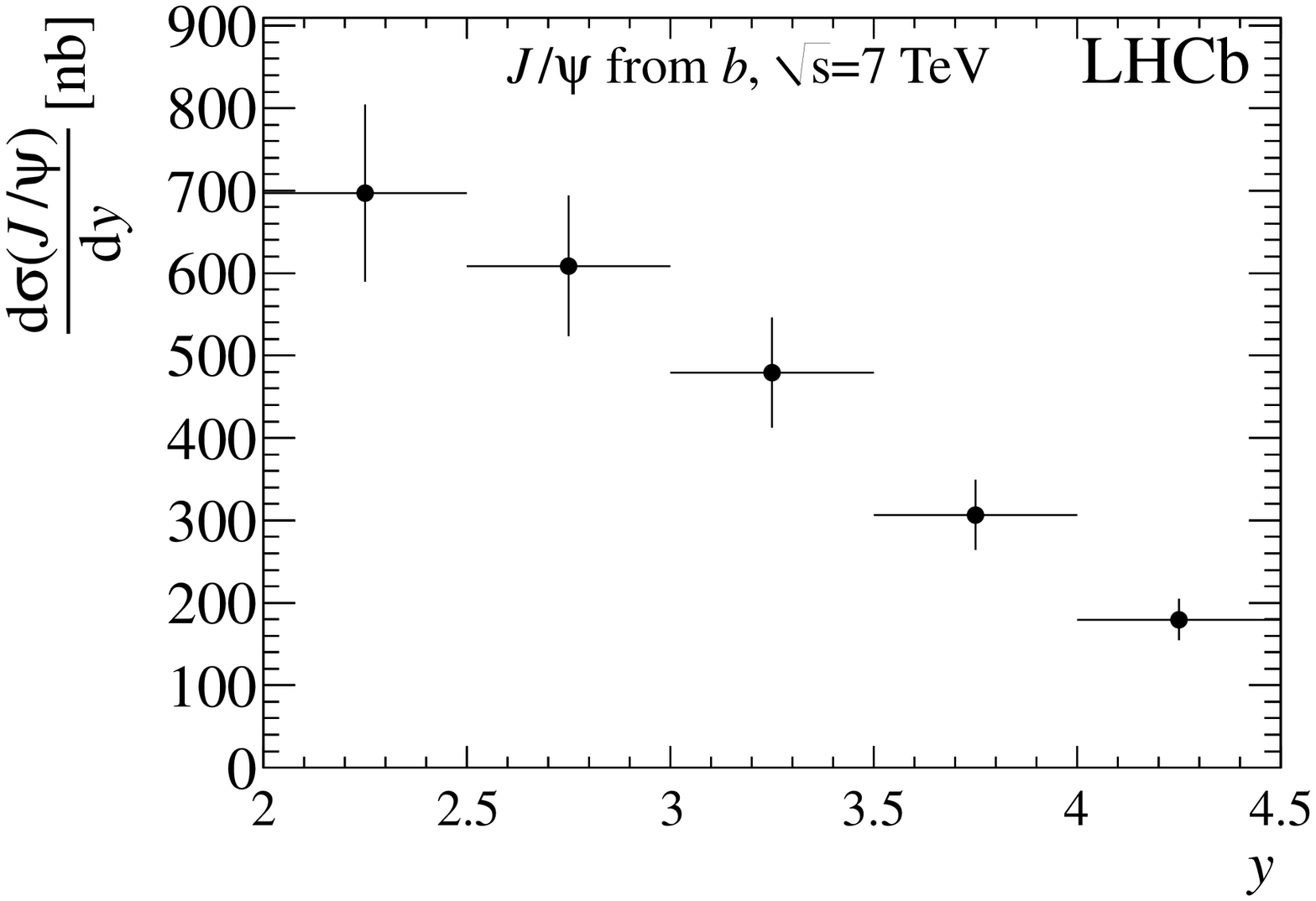}
\fi
\caption{\small Differential production cross-section as a function of $y$ integrated over \ptrans, for unpolarised
\prompt\ ({\it left}) and \fromb\ ({\it right}). The errors are the quadratic sums of the statistical and 
systematic uncertainties.} \label{fig:rapidity}
\end{figure}

Table~\ref{tab:rapidity} and Fig.~\ref{fig:rapidity} show the differential cross-sections $\frac{{\rm d}\sigma}{{\rm d}y}$ 
integrated over \ptrans, both for unpolarised \prompt\ and \fromb. For the two production sources, the cross-sections 
decrease significantly between the central and forward regions of the LHCb acceptance.

\renewcommand{\arraystretch}{1.3}
\begin{table}[!t]
\caption{\small $\frac{{\rm d}\sigma}{{\rm d}y}$ in nb for \prompt\ (assumed unpolarised) and \fromb, integrated 
over \ptrans. The first uncertainty is statistical, the second is the component of the systematic uncertainty that is 
uncorrelated between bins and the third is the correlated component.}
\begin{center}
\scalebox{1.0}{
\begin{tabular}{@{}llr@{}c@{}r@{}c@{}r@{}c@{}rrr@{}c@{}r@{}c@{}r@{}c@{}r@{}}
\toprule
\ptrans\ range (\gevc)  & \multicolumn{1}{c}{$y$ bin} & \multicolumn{7}{c}{Prompt $J/\psi$} & 
& \multicolumn{7}{c}{\fromb} \\ \midrule
$0-14$ & \ $2.0-2.5\  \ $ & $5504 \tpm 83 \tpm 381 \tpm 726$ & & $697\tpm27\tpm40\tpm96$ \\
$0-14$ & \ $2.5-3.0\ \ $       & $5096 \tpm 21  \tpm 142 \tpm 672$ & & $608\tpm7\tpm13\tpm84$  \\
$0-14$ & \ $3.0-3.5\ \ $       & $4460 \tpm 14 \tpm 59 \tpm 589  $ & & $479\tpm5\tpm5\tpm66$ \\
$0-13$ & \ $3.5-4.0\ \ $       & $3508 \tpm 12 \tpm 40 \tpm 463$   & & $307\tpm4\tpm3\tpm42$ \\
$0-11$ & \ $4.0-4.5\ \ $       & $2462 \tpm 12 \tpm 48 \tpm 325$   & & $180\tpm4\tpm3\tpm25$ \\
\bottomrule 
\end{tabular}
}
\end{center}
\label{tab:rapidity} 
\end{table}
\renewcommand{\arraystretch}{1.}

\subsection{\texorpdfstring{Fraction of $\mskip 1.5mu\mathbfi{\jpsi}$ from $\mathbfi{b}$}{Fraction of J/psi from b}}

\begin{figure}[!t]
\centering
\ifpdf
\includegraphics[width=14.1cm]{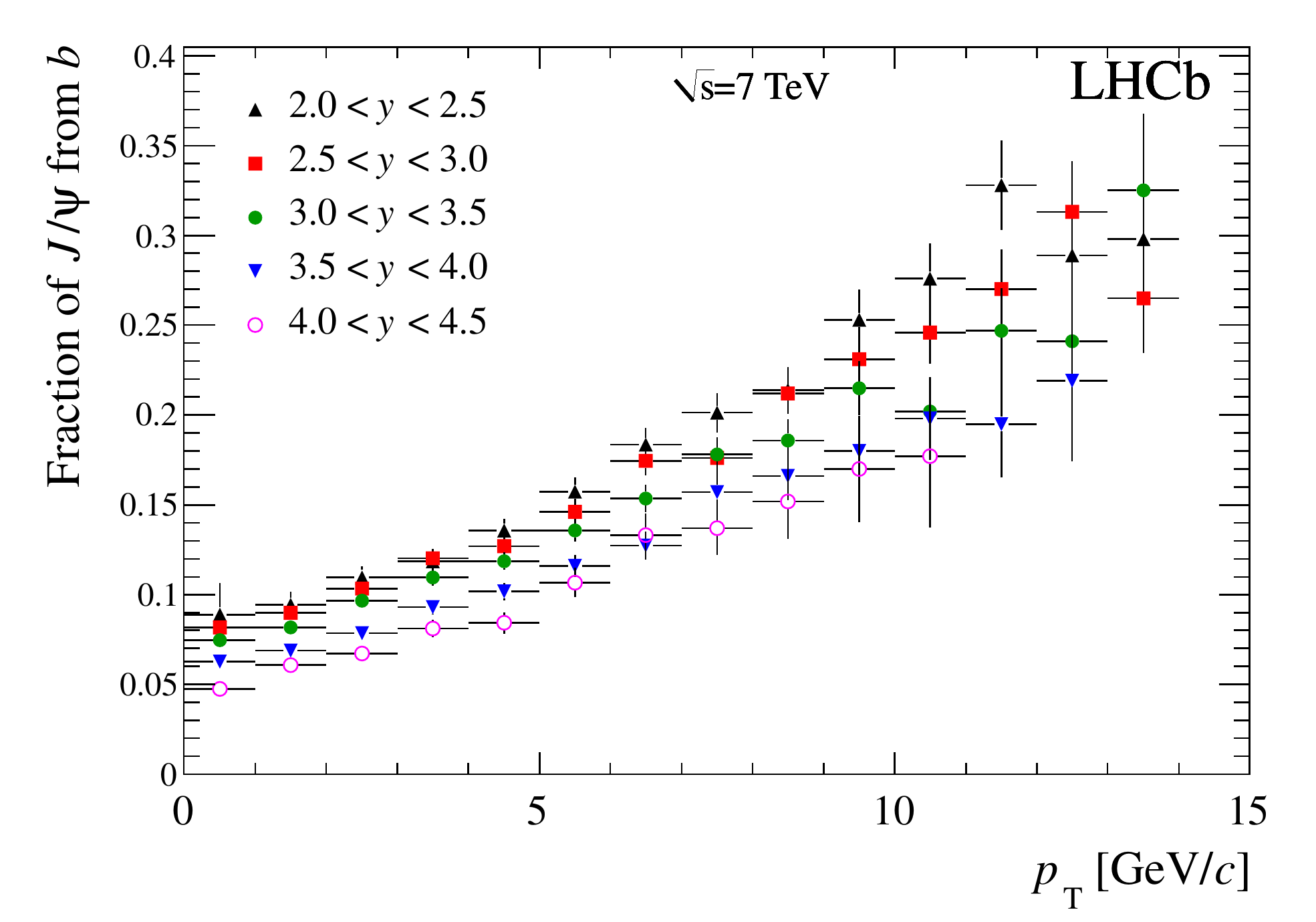}
\else
\includegraphics[width=14.1cm]{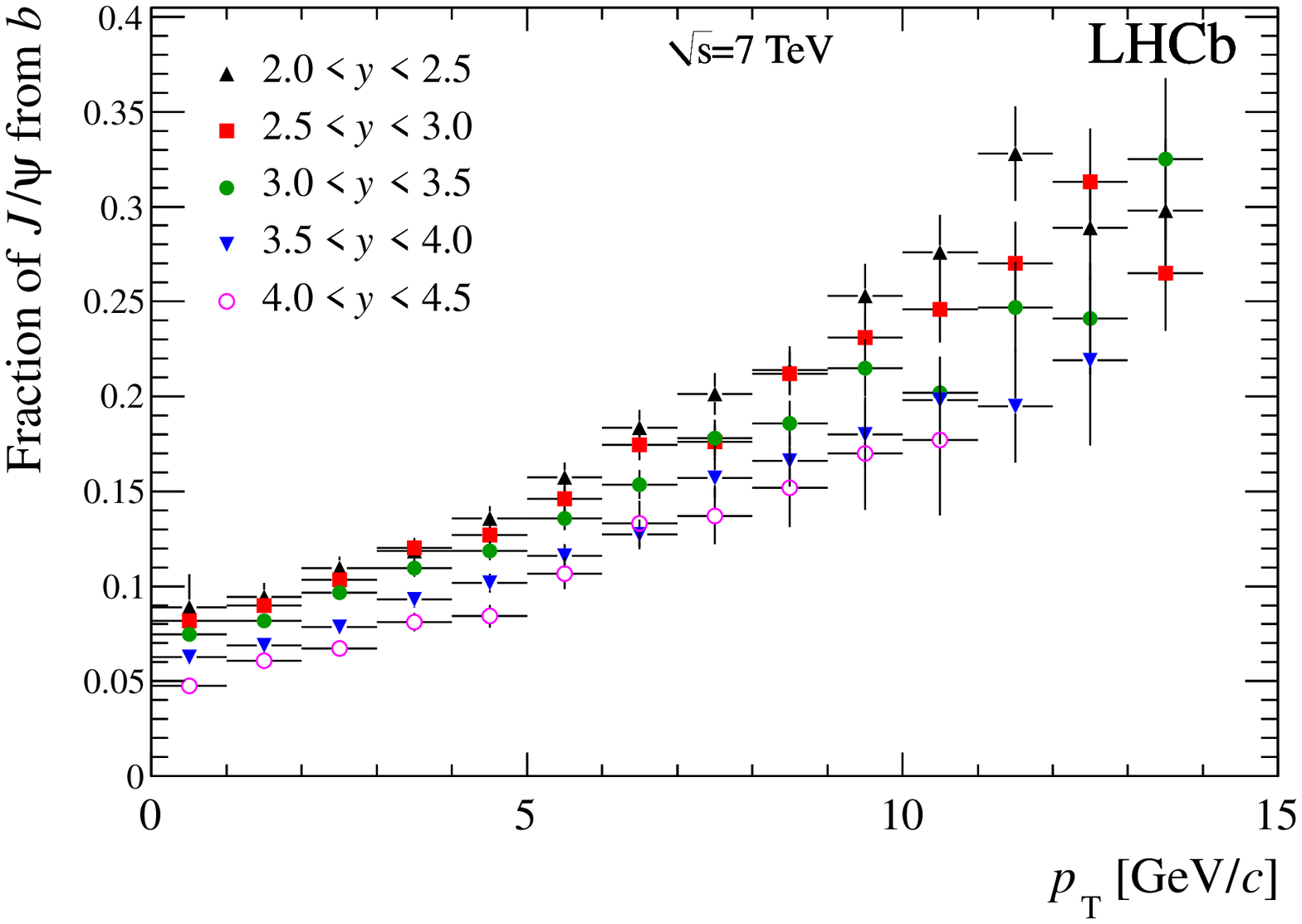}
\fi
\caption{\small Fraction of \fromb\ as a function of \ptrans, in bins of $y$. } \label{fig:bfraction}
\end{figure}

Table~\ref{tab:bfraction} and Fig.~\ref{fig:bfraction} give the values of the fraction of \fromb\ in the different bins
assuming that the \prompt\ are produced unpolarised. The third uncertainty in Table~\ref{tab:bfraction} gives the 
deviation from the central value when the \prompt\ are fully transversely or fully longitudinally polarised in the helicity 
frame.

In Fig.~\ref{fig:bfraction}, only the statistical and systematic uncertainties are displayed, added quadratically, but not 
the uncertainties associated with the \prompt\ polarisation. The fraction of \fromb\ increases as a function of \ptrans. 
For a constant \ptrans, the fraction of \fromb\ decreases with increasing $y$, indicating that $b$-hadrons are 
produced more centrally than  prompt \jpsi.

\subsection {Cross-section extrapolation}\label{sec:extrapolation}
Using the LHCb Monte Carlo simulation based on {\sc Pythia} 6.4~\cite{pythia} and EvtGen~\cite{EvtGen},  the 
result quoted in Eq.~\eqref{sigma_from_b}
is extrapolated  to the full polar angle range
\begin{equation}
\sigma(pp \to b\overline{b} X) = 
\alpha_{4\pi} \,\frac{\sigma\left(\fromb,\, \ptrans <14\,\gevc,\,2.0<y<4.5\right)}{2 \, {\cal B}(b\to\jpsi X)},
\end{equation}
where $\alpha_{4\pi}=5.88$ is the ratio of \fromb\ events in the full range to the number of events  in the region 
$2.0<y<4.5$ and ${\cal B}(b\to\jpsi X)=(1.16\pm0.10)\%$ is the average branching fraction of inclusive $b$-hadron 
decays to \jpsi\ measured at LEP\cite{delphibtojpsi}. 
The result is
\begin{equation}
\sigma(pp \to b\overline{b} X) = 288\pm 4 \pm 48 \, \upmu{\rm b}\,,
\end{equation}
where the first uncertainty is statistical and the second systematic. 
The systematic uncertainty includes the uncertainties on the $b$ fractions ($2\%$) and on ${\cal B}(b \to \jpsi X)$. 
No additional uncertainty has been included for the extrapolation factor $\alpha_{4\pi}$ estimated from the 
simulation.
The above result is in excellent agreement with 
$\sigma(pp \to b\overline{b} X) = 284 \pm 20 \pm 49 \, \upmu {\rm b}$ obtained from $b$ decays into 
$D^0 \mu\nu X$~\cite{Sheldon}. The extrapolation factor $\alpha_{4\pi}$ has also been estimated using predictions 
made in the framework of fixed-order next-to-leading log (FONLL) computations~\cite{cacciari1}, and found to be 
equal to $\alpha_{4\pi}^{\rm FONLL}=5.21$.

\section{Comparison with theoretical models}

\begin{figure}[!t]
\centering
\ifpdf
\includegraphics[width=16.5cm]{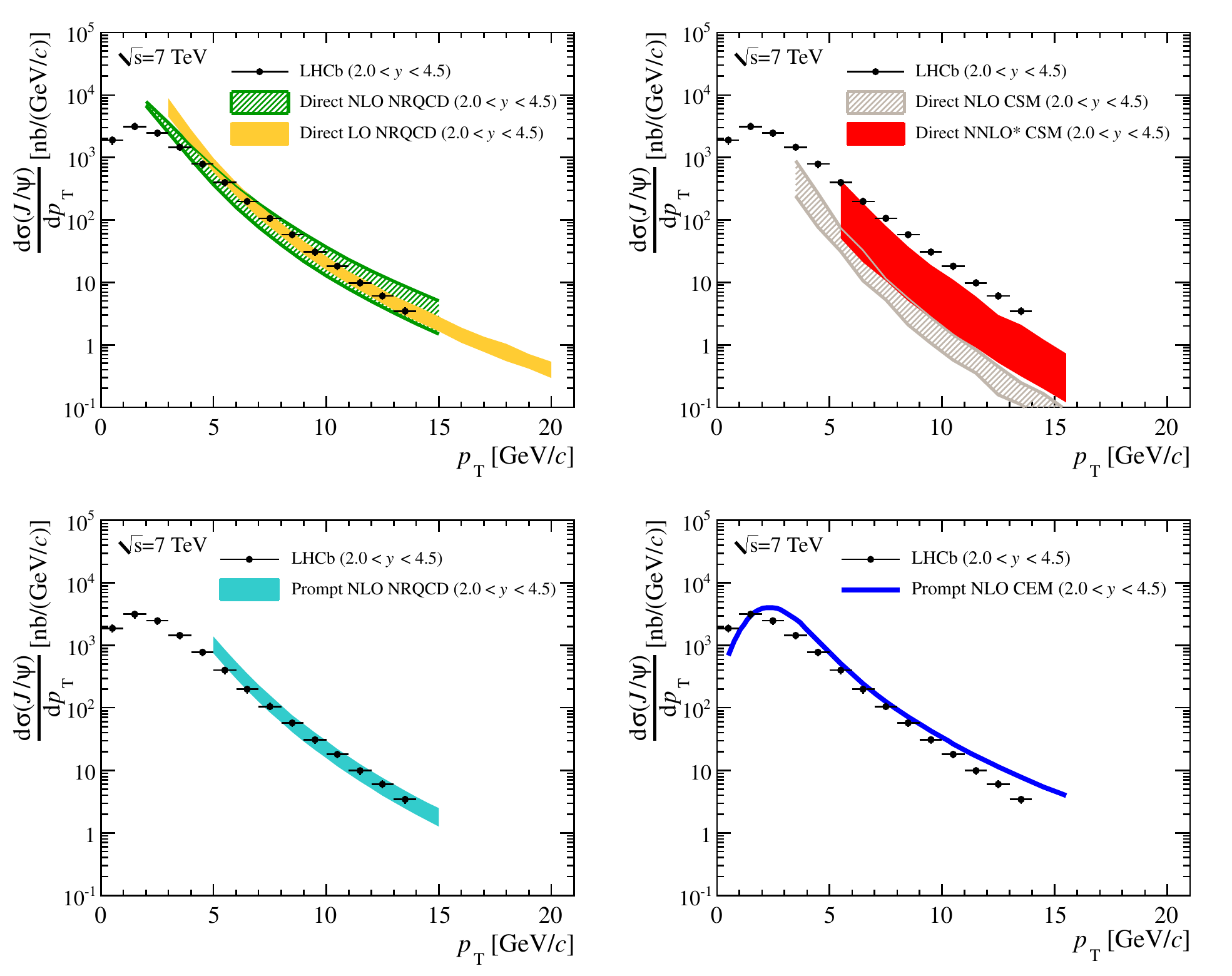}
\else
\includegraphics[width=16.5cm]{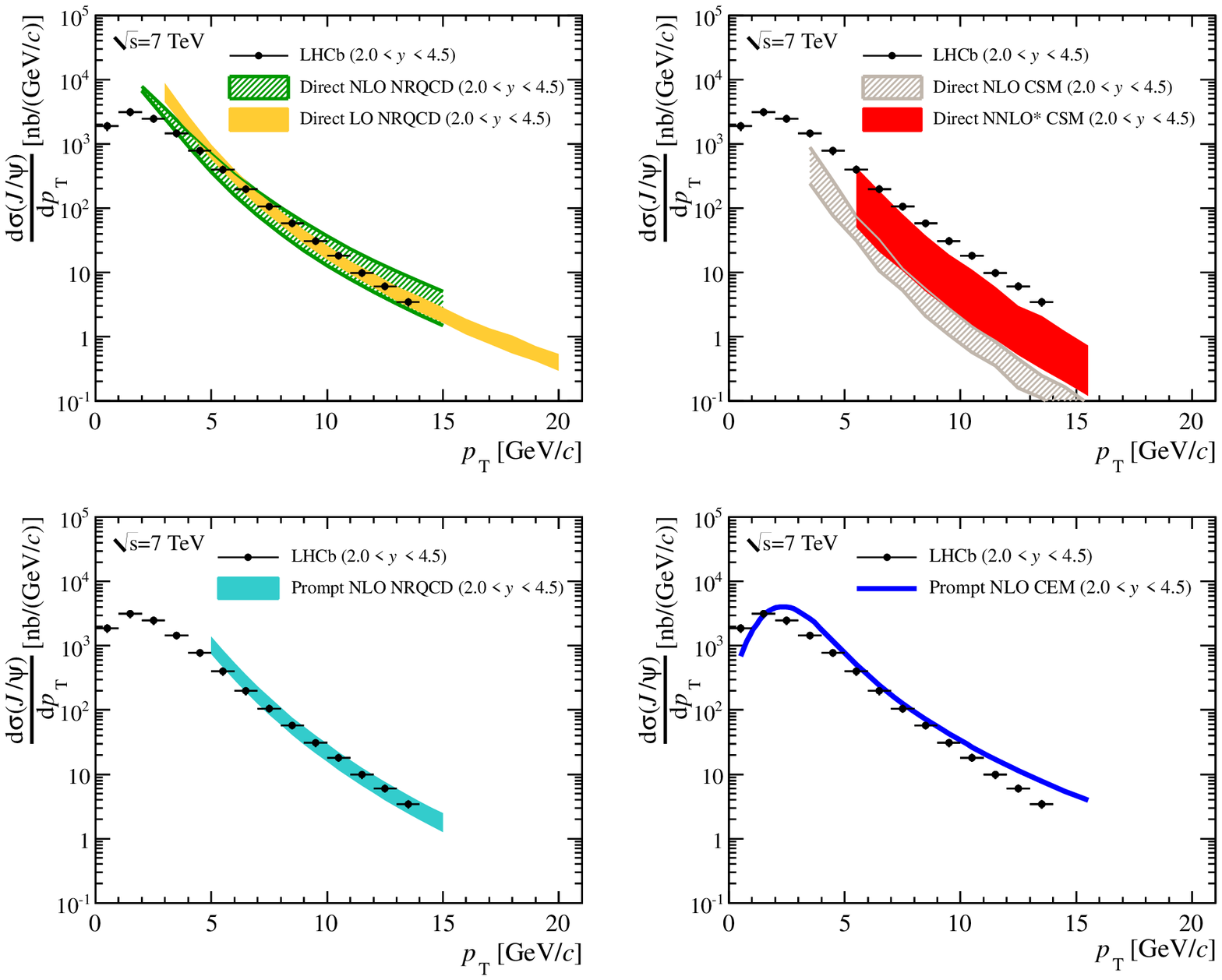}
\fi
\caption{\small Comparison of the LHCb results for the differential prompt \jpsi\ production for unpolarised \jpsi\ 
({\it circles with error bars}) with: ({\it top, left}) {\it direct} \jpsi\ production as predicted by LO and NLO NRQCD;
({\it top, right}) {\it direct} \jpsi\ production as predicted by NLO and NNLO$\star$ CSM; 
({\it bottom, left}) {\it prompt} \jpsi\ production as predicted by NLO NRQCD;
({\it bottom, right}) {\it prompt} \jpsi\ production as predicted by NLO CEM. A more detailed description of the models and 
their references is given in the text.} \label{fig:theorycomparison}
\end{figure}

Figure~\ref{fig:theorycomparison} compares the LHCb measurement of the differential prompt \jpsi\ production with 
several recent theory predictions in the LHCb acceptance region:
\begin{itemize}
\item top, left: direct \jpsi\ production as calculated from NRQCD at leading-order in $\alpha_s$ (LO, filled orange 
uncertainty band)~\cite{artoisenet2} and next-to-leading order (NLO), with colour-octet long distance matrix elements 
determined from HERA and Tevatron data (hatched green uncertainty band)~\cite{nlonrqcd}, summing the 
colour-singlet and colour-octet contributions.
\item top, right: direct production as calculated from a NNLO\raisebox{0.5mm}{$\star$} colour-singlet model (CSM, 
filled red uncertainty band)~\cite{stelzer,nlostar2}. The notation NNLO\raisebox{0.5mm}{$\star$} denotes an 
evaluation that is not a complete next-to-next leading order computation and that can be affected by logarithmic 
corrections, which are however not easily quantifiable. Direct production as calculated from NLO CSM (hatched grey 
uncertainty band)~\cite{artoisenet,campbell} is also represented.
\item bottom, left: \prompt\ production as calculated from NRQCD at NLO, including contributions from $\chi_c$ and
$\psi(2S)$ decays, summing the colour-singlet and colour-octet contributions~\cite{chao1}.
\item bottom, right: \prompt\ production as calculated from a NLO colour-evaporation model (CEM), including 
contributions from $\chi_c$ and $\psi(2S)$ decays~\cite{cem}.
\end{itemize}

\noindent
It should be noted that some of the theoretical models compute the direct \jpsi\ production, whereas the \prompt\ 
measurement includes \jpsi\ from $\chi_c$ decays and,
to a smaller extent, $\psi(2S)$ decays. However, if one takes into account the feed-down contribution, which has 
been estimated to be of the order of $30\%$ averaging over several experimental measurements at lower 
energies~\cite{faccioli}, a satisfactory agreement is found with the theoretical predictions.

\begin{figure}[!t]
\centering
\ifpdf
\includegraphics[width=12.55cm]{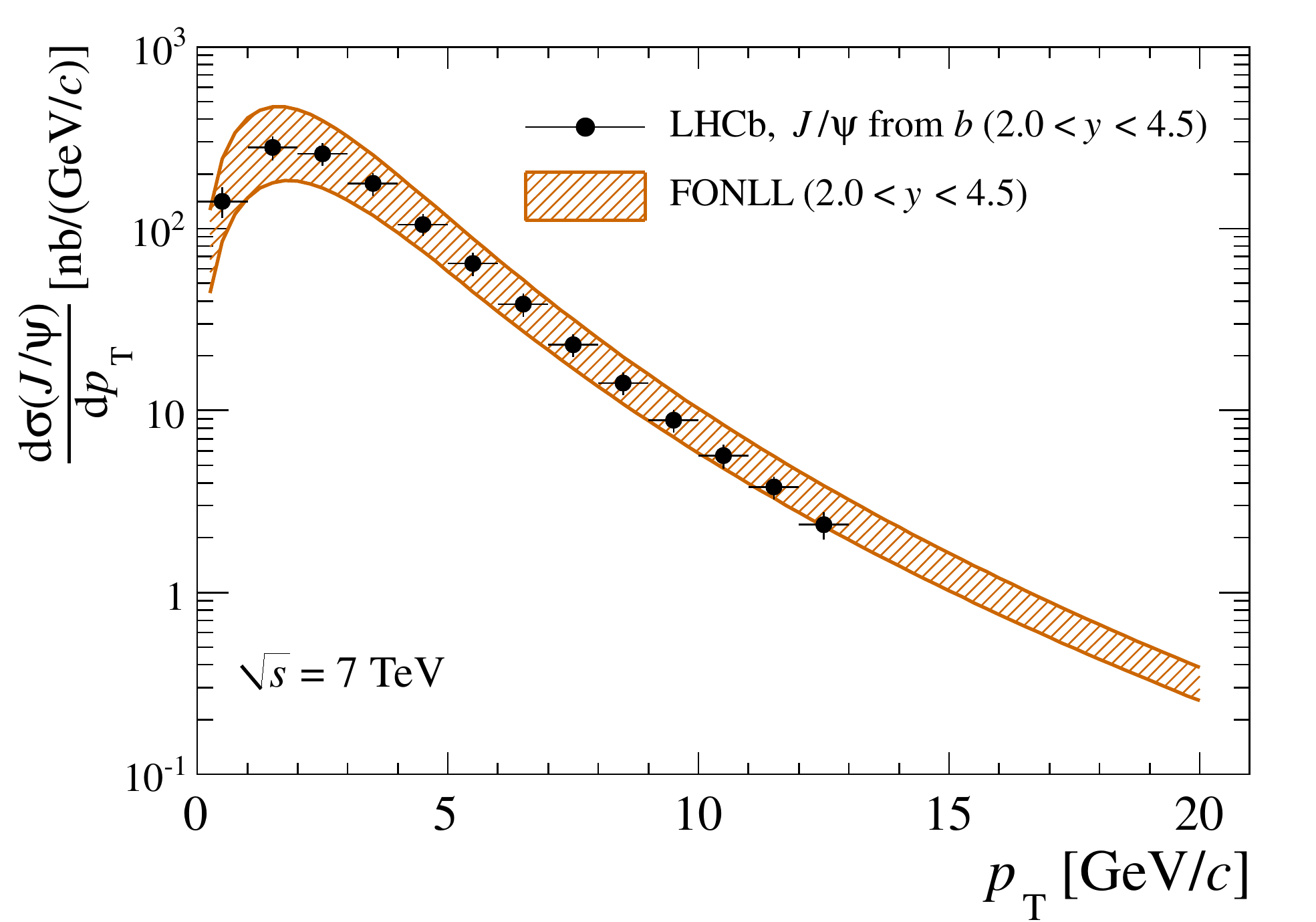}
\else
\includegraphics[width=12.55cm]{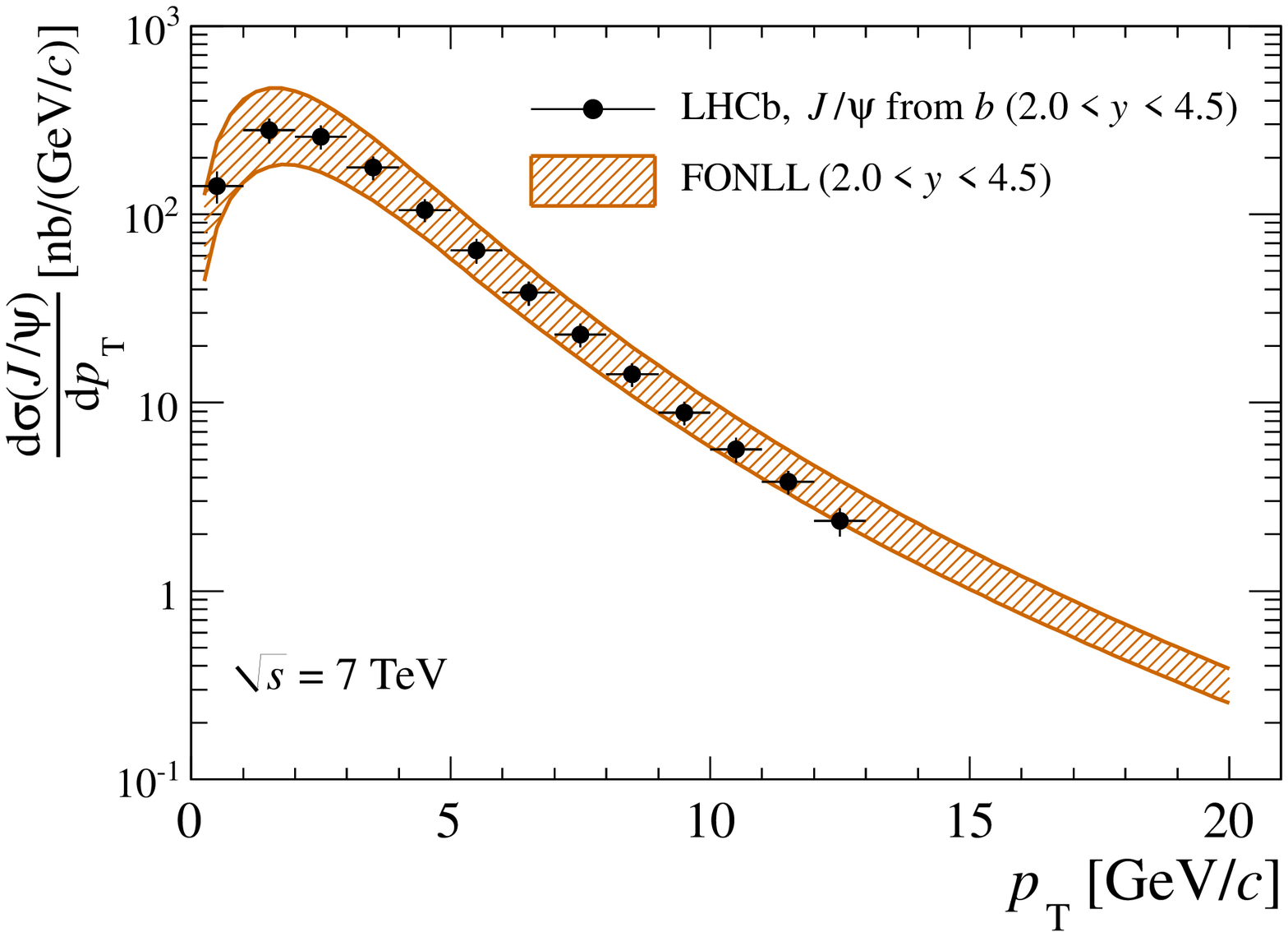}
\fi
\caption{\small Comparison of the LHCb results for the differential \fromb\ production for unpolarised \jpsi\ ({\it circles 
with error bars}) with \fromb\ production as predicted by FONLL ({\it hatched orange uncertainty band}). 
A more detailed description of the model and its references is given in the text.} \label{fig:bspectrumtheory}
\end{figure}

Figure~\ref{fig:bspectrumtheory} shows a comparison of the LHCb measurement of the differential \fromb\ 
cross-section with a calculation based on the FONLL formalism~\cite{cacciari1}.  This model predicts
the $b$-quark production cross-section, and includes the fragmentation of the $b$-quark into $b$-hadrons and their 
decay into \jpsi\ mesons. The measurements show a very good agreement with the calculation.

\section{Conclusions}

The differential cross-section for \jpsi\ production is measured as a function of the \jpsi\ transverse momentum and 
rapidity in the forward region, $2.0<y<4.5$. The analysis is based on a data sample corresponding to an
integrated luminosity of $5.2\pbinv$ collected at the Large Hadron Collider at a centre-of-mass energy of 
$\sqrt{s}=7\tev$, and the contributions of \prompt\ and \fromb\ production are individually measured. 
The results obtained are in good agreement with earlier measurements of the \jpsi\ production cross-section in $pp$
collisions at the same centre-of-mass energy, performed by CMS in a region corresponding to the low rapidity part of
the LHCb acceptance~\cite{cmsjpsi}.  
This measurement is the first measurement of \prompt\ and \fromb\ production in the forward region at 
$\sqrt{s}=7\tev$.

A comparison with recent theoretical models shows good general agreement with the measured \prompt\ 
cross-section in the LHCb acceptance at high \ptrans. This confirms the progress in the theoretical 
calculations of \jpsi\ 
hadroproduction, even if the uncertainties on the predictions are still large. However, the measurement of the 
differential cross-section alone is not sufficient to be able to discriminate amongst the various models, and studies of 
other observables such as the \jpsi\ polarisation will be necessary. The measurement of the cross-section for \fromb\ 
is found to agree very well with FONLL predictions.
An estimate of the $b\overline{b}$ cross-section in $pp$ collisions at $\sqrt{s}=7\tev$ is also obtained, which is in 
excellent agreement with measurements performed analysing
different $b$ decay modes~\cite{Sheldon}.

\section*{Acknowledgments}
We express our gratitude to our colleagues in the CERN accelerator departments for the excellent performance of 
the LHC. We thank the technical and administrative staff at CERN and at the LHCb institutes, and acknowledge 
support from the National Agencies: CAPES, CNPq, FAPERJ and FINEP (Brazil); CERN; NSFC (China); CNRS/IN2P3 
(France); BMBF, DFG, HGF and MPG (Germany); SFI (Ireland); INFN (Italy); FOM and NWO (Netherlands); SCSR 
(Poland); ANCS (Romania); MinES of Russia and Rosatom (Russia); MICINN, XUNGAL and GENCAT (Spain); SNSF 
and SER (Switzerland); NAS Ukraine (Ukraine); STFC (United Kingdom); NSF (USA). We also acknowledge the 
support received from the ERC under FP7 and the Region Auvergne. 

We thank P.~Artoisenet, M.~Butensch\"{o}n, M.~Cacciari, K.~T.~Chao, B.~Kniehl, J.-P.~Lansberg and R.~Vogt for 
providing theoretical predictions of \jpsi\ cross-sections in the LHCb acceptance range.

\renewcommand{\arraystretch}{1.35}
\begin{table}[!ht]
\begin{center}
\caption{\small \label{promptresult}$\frac{{\rm d}^2\sigma}{{\rm d}p_{\rm T}{\rm d}y}$ in nb/(\gevc) for prompt \jpsi\ in 
bins of the \jpsi\ transverse momentum and rapidity, assuming no polarisation. The first error is statistical, the second  
is the component of the systematic uncertainty that is uncorrelated between bins and the third is the correlated 
component.}
\vskip 0.5cm
\scalebox{0.98}{
\begin{tabular}{@{}r@{}c@{}rrr@{}c@{}r@{}c@{}r@{}c@{}rr@{}c@{}r@{}c@{}r@{}c@{}rr@{}c@{}r@{}c@{}r@{}c@{}r@{}}
\toprule
\multicolumn{4}{c}{$\ptrans\,(\gevc)$} & \multicolumn{7}{c}{$2.0<y<2.5$} & \multicolumn{7}{c}{$2.5<y<3.0$} & 
\multicolumn{7}{c}{$3.0<y<3.5$} \\
\midrule
$0$&$-$&$1$     && $1091\tpm70\tpm226\tpm144$    & $844\tpm 13\tpm 133\tpm 111$ & $749\tpm 7\tpm 46\tpm 99 $ \\
$1$&$-$&$2$     && $1495\tpm38\tpm282\tpm 197$   & $1490\tpm 12\tpm 39\tpm 197$ & $1376\tpm 8\tpm 26\tpm 182$ \\
$2$&$-$&$3$     && $1225\tpm 20\tpm109\tpm 162$  & $1214\tpm 9\tpm 24\tpm 160$   & $1053\tpm 7\tpm 19\tpm 139$ \\
$3$&$-$&$4$     && $777\tpm 11\tpm 44\tpm 103$     & $719\tpm 6\tpm 18\tpm 95$       & $611\tpm 5\tpm 14\tpm 81$ \\
$4$&$-$&$5$     && $424\tpm 6\tpm 22\tpm 56$         & $392\tpm 3\tpm 12\tpm 52$       & $325\tpm 3\tpm 9\tpm 43$ \\
$5$&$-$&$6$     && $230\tpm 4\tpm 12\tpm 30$         & $206\tpm 2\tpm 8\tpm 27$         & $167\tpm 2\tpm 5\tpm 22$ \\
$6$&$-$&$7$     && $116\tpm 2\tpm 6\tpm 15$           & $104\tpm 1\tpm 4\tpm 14$         & $82\tpm 1\tpm 3\tpm 11$ \\
$7$&$-$&$8$     && $64\tpm 1\tpm 3\tpm 8$               & $57\tpm 1\tpm 3\tpm 7$             & $44\tpm 1\tpm 1\tpm 6$ \\
$8$&$-$&$9$     && $37\tpm 1\tpm 1\tpm 5$               & $31\tpm 1\tpm 1\tpm 4$             & $23\tpm 1\tpm 1\tpm 3$ \\
$9$&$-$&$10$   && $19.3\tpm 0.7\tpm 0.5\tpm 2.6$   & $17.4\tpm 0.5\tpm 0.2\tpm 2.3$ & $12.6\tpm 0.4\tpm 0.1\tpm 1.7$ \\
$10$&$-$&$11$ && $11.6\tpm 0.5\tpm 0.3\tpm 1.5$   & $9.8\tpm 0.4\tpm 0.1\tpm 1.3$   & $7.8\tpm 0.3\tpm 0.1\tpm 1.0$ \\
$11$&$-$&$12$ && $6.7\tpm 0.4\tpm 0.2\tpm 0.9$     & $5.9\tpm 0.3\tpm 0.1\tpm 0.8$   & $4.5\tpm 0.3\tpm 0.1\tpm 0.6$ \\
$12$&$-$&$13$ && $4.6\tpm 0.3\tpm 0.2\tpm 0.6$     & $3.5\tpm 0.2\tpm 0.1\tpm 0.5$   & $2.9\tpm 0.2\tpm 0.1\tpm 0.4$ \\
$13$&$-$&$14$ && $2.9\tpm 0.3\tpm 0.1\tpm 0.4$     & $2.6\tpm 0.2\tpm 0.1\tpm 0.3$   & $1.3\tpm 0.2\tpm 0.1\tpm 0.2$ \\
\bottomrule
& & & & \multicolumn{7}{c}{$3.5<y<4.0$} & \multicolumn{7}{c}{$4.0<y<4.5$} \\ \midrule
$0$&$-$&$1$     && $614\tpm 6\tpm 23\tpm 81$         & $447\tpm 5\tpm 28\tpm 59$ \\
$1$&$-$&$2$     && $1101\tpm 7\tpm 23\tpm 145$     & $807\tpm 7\tpm 28\tpm 107$ \\
$2$&$-$&$3$     && $839\tpm 6\tpm 19\tpm 111$       & $588\tpm 6\tpm 22\tpm 78$ \\
$3$&$-$&$4$     && $471\tpm 4\tpm 13\tpm 62$         & $315\tpm 4\tpm 14\tpm 42$ \\
$4$&$-$&$5$     && $244\tpm 3\tpm 7\tpm 32$           & $163\tpm 3\tpm 6\tpm 22$  \\
$5$&$-$&$6$     && $119\tpm 2\tpm 5\tpm 16$           & $76\tpm 2\tpm 3\tpm 10$ \\
$6$&$-$&$7$     && $59\tpm 1\tpm 2\tpm 8$               & $34\tpm 1.1\tpm 1.4\tpm 4.5$ \\
$7$&$-$&$8$     && $29\tpm 1\tpm 1\tpm 4$               & $17\tpm 0.7\tpm 0.8\tpm 2.3$ \\
$8$&$-$&$9$     && $15.9\tpm 0.5\tpm 0.1\tpm 2.1$   & $8.5\tpm 0.5\tpm 0.4\tpm 1.1$ \\
$9$&$-$&$10$   && $8.2\tpm 0.4\tpm 0.1\tpm 1.1$     & $4.1\tpm 0.3\tpm 0.2\tpm 0.5$ \\
$10$&$-$&$11$ && $4.9\tpm 0.3\tpm 0.1\tpm 0.6$     & $2.2\tpm 0.2\tpm 0.1\tpm 0.3$ \\
$11$&$-$&$12$ && $2.6\tpm 0.2\tpm 0.1\tpm 0.3$     & \\
$12$&$-$&$13$ && $1.2\tpm 0.1\tpm 0.1\tpm 0.2$     & \\
\bottomrule
\end{tabular}
}
\end{center}
\end{table}

\renewcommand{\arraystretch}{1.35}
\begin{table}[!ht]
\begin{center}
\caption{\small \label{bresult}$\frac{{\rm d}^2\sigma}{{\rm d}p_{\rm T}{\rm d}y}$ in nb/(\gevc) for  \fromb\ in bins of the \jpsi\ transverse momentum and rapidity. The first error is statistical, the second is the component of the systematic uncertainty that is uncorrelated between bins and the third is the correlated component.}
\vskip 0.5cm
\scalebox{0.98}{
\begin{tabular}{@{}r@{}c@{}rrr@{}c@{}r@{}c@{}r@{}c@{}rr@{}c@{}r@{}c@{}r@{}c@{}rr@{}c@{}r@{}c@{}r@{}c@{}r@{}}
\toprule
\multicolumn{4}{c}{$\ptrans\,(\gevc)$} & \multicolumn{7}{c}{$2.0<y<2.5$} & \multicolumn{7}{c}{$2.5<y<3.0$} & \multicolumn{7}{c}{$3.0<y<3.5$} \\
\midrule
$0$&$-$&$1$     && $107\tpm23\tpm22\tpm15$    & $75\tpm 4\tpm 12\tpm 10$     & $60\tpm 2\tpm 4\tpm 8 $ \\
$1$&$-$&$2$     && $156\tpm11\tpm30\tpm 22$   & $147\tpm 4\tpm 4\tpm 20$     & $123\tpm 3\tpm 2\tpm 17$ \\
$2$&$-$&$3$     && $151\tpm 6\tpm14\tpm 21$    & $140\tpm 3\tpm 3\tpm 19$     & $113\tpm 2\tpm 2\tpm 16$ \\
$3$&$-$&$4$     && $105\tpm 4\tpm 6\tpm 15$     & $98\tpm 2\tpm 2\tpm 14$         & $75\tpm 2\tpm 2\tpm 10$ \\
$4$&$-$&$5$     && $67\tpm 2\tpm 3\tpm 9$         & $57\tpm 1\tpm 2\tpm 8$         & $44\tpm 1\tpm 1\tpm 6$ \\
$5$&$-$&$6$     && $43\tpm 2\tpm 2\tpm 6$         & $35\tpm 1\tpm 1\tpm 5$           & $26\tpm 1\tpm 1\tpm 4$ \\
$6$&$-$&$7$     && $26\tpm 1\tpm 1\tpm 4$           & $22\tpm 1\tpm 1\tpm 3$           & $14.9\tpm 0.6\tpm 0.5\tpm 2.1$ \\
$7$&$-$&$8$     && $16.1\tpm 0.7\tpm 0.8\tpm 2.2$       & $12.1\tpm 0.5\tpm 0.6\tpm 1.7$       & $9.4\tpm 0.4\tpm 0.3\tpm 1.3$ \\
$8$&$-$&$9$     && $10.1\tpm 0.6\tpm 0.3\tpm 1.4$       & $8.2\tpm 0.4\tpm 0.8\tpm 1.1$         & $5.3\tpm 0.3\tpm 0.1\tpm 0.7$ \\
$9$&$-$&$10$   && $6.5\tpm 0.4\tpm 0.2\tpm 0.9$   & $5.2\tpm 0.3\tpm 0.1\tpm 0.7$   & $3.4\tpm 0.2\tpm 0.1\tpm 0.5$ \\
$10$&$-$&$11$ && $4.4\tpm 0.3\tpm 0.1\tpm 0.6$   & $3.2\tpm 0.2\tpm 0.1\tpm 0.4$   & $2.0\tpm 0.2\tpm 0.1\tpm 0.3$ \\
$11$&$-$&$12$ && $3.3\tpm 0.3\tpm 0.1\tpm 0.4$   & $2.2\tpm 0.2\tpm 0.1\tpm 0.3$   & $1.5\tpm 0.2\tpm 0.1\tpm 0.2$ \\
$12$&$-$&$13$ && $1.9\tpm 0.2\tpm 0.1\tpm 0.3$   & $1.6\tpm 0.2\tpm 0.1\tpm 0.2$   & $0.9\tpm 0.1\tpm 0.1\tpm 0.1$ \\
$13$&$-$&$14$ && $1.2\tpm 0.2\tpm 0.1\tpm 0.2$   & $0.9\tpm 0.1\tpm 0.1\tpm 0.1$   & $0.6\tpm 0.1\tpm 0.1\tpm 0.1$ \\
\bottomrule
&&&& \multicolumn{7}{c}{$3.5<y<4.0$} & \multicolumn{7}{c}{$4.0<y<4.5$} \\ \midrule
$0$&$-$&$1$     && $41\tpm 2\tpm 2\tpm 6$                 & $22\tpm 2\tpm 1\tpm 3$ \\
$1$&$-$&$2$     && $82\tpm 2\tpm 2\tpm 11$               & $52\tpm 2\tpm 2\tpm 7$ \\
$2$&$-$&$3$     && $71\tpm 2\tpm 2\tpm 10$               & $42\tpm 2\tpm 2\tpm 6$ \\
$3$&$-$&$4$     && $48\tpm 1\tpm 1\tpm 7$                 & $28\tpm 1\tpm 1\tpm 4$ \\
$4$&$-$&$5$     && $28\tpm 1\tpm 1\tpm 4$                 & $15.0\tpm 1.0\tpm 0.6\tpm 2.1$  \\
$5$&$-$&$6$     && $15.6\tpm 0.7\tpm 0.7\tpm 2.2$     & $9.0\tpm 0.7\tpm 0.3\tpm 1.3$ \\
$6$&$-$&$7$     && $8.6\tpm 0.4\tpm 0.3\tpm 1.2$       & $5.2\tpm 0.5\tpm 0.2\tpm 0.7$ \\
$7$&$-$&$8$     && $5.5\tpm 0.3\tpm 0.2\tpm 0.8$       & $2.8\tpm 0.3\tpm 0.1\tpm 0.4$ \\
$8$&$-$&$9$     && $3.2\tpm 0.3\tpm 0.1\tpm 0.4$       & $1.5\tpm 0.2\tpm 0.1\tpm 0.2$ \\
$9$&$-$&$10$   && $1.8\tpm 0.2\tpm 0.1\tpm 0.2$       & $0.8\tpm 0.2\tpm 0.1\tpm 0.1$ \\
$10$&$-$&$11$ && $1.2\tpm 0.2\tpm 0.1\tpm 0.2$       & $0.5\tpm 0.1\tpm 0.1\tpm 0.1$ \\
$11$&$-$&$12$ && $0.6\tpm 0.1\tpm 0.1\tpm 0.1$   & \\
$12$&$-$&$13$ && $0.3\tpm 0.1\tpm 0.1\tpm 0.1$   & \\
\bottomrule
\end{tabular}
}
\end{center}
\end{table}
\renewcommand{\arraystretch}{1.2}

\renewcommand{\arraystretch}{1.35}
\begin{table}[!ht]
\begin{center}
\caption{\small \label{promptresulttransverse}$\frac{{\rm d}^2\sigma}{{\rm d}p_{\rm T}{\rm d}y}$ in nb/(\gevc) for prompt \jpsi\ in bins of the \jpsi\ transverse momentum and rapidity, assuming fully  transversely polarised \jpsi. The first error is statistical, the second is the component of the systematic uncertainty that is uncorrelated between bins and the third is the correlated component.}
\vskip 0.5cm
\scalebox{0.98}{
\begin{tabular}{@{}r@{}c@{}rrr@{}c@{}r@{}c@{}r@{}c@{}rr@{}c@{}r@{}c@{}r@{}c@{}rr@{}c@{}r@{}c@{}r@{}c@{}r@{}}
\toprule
\multicolumn{4}{c}{$\ptrans\,(\gevc)$} & \multicolumn{7}{c}{$2.0<y<2.5$} & \multicolumn{7}{c}{$2.5<y<3.0$} & \multicolumn{7}{c}{$3.0<y<3.5$} \\
\midrule
$0$&$-$&$1$     && $1282\tpm83\tpm266\tpm169$    & $1058\tpm 16\tpm 166\tpm 140$ & $924\tpm 9\tpm 56\tpm 122 $ \\
$1$&$-$&$2$     && $1751\tpm44\tpm331\tpm 231$   & $1791\tpm 15\tpm 47\tpm 236$   & $1603\tpm 10\tpm 31\tpm 212$ \\
$2$&$-$&$3$     && $1438\tpm24\tpm129\tpm 190$   & $1423\tpm 11\tpm 28\tpm 188$   & $1182\tpm 7\tpm 21\tpm 156$ \\
$3$&$-$&$4$     && $932\tpm 13\tpm 53\tpm 123$     & $839\tpm 7\tpm 21\tpm 111$       & $675\tpm 5\tpm 15\tpm 89$ \\
$4$&$-$&$5$     && $513\tpm 7\tpm 27\tpm 68$         & $455\tpm 4\tpm 14\tpm 60$         & $358\tpm 3\tpm 10\tpm 47$ \\
$5$&$-$&$6$     && $278\tpm 4\tpm 15\tpm 37$         & $238\tpm 3\tpm 9\tpm 32$           & $184\tpm 2\tpm 6\tpm 24$ \\
$6$&$-$&$7$     && $140\tpm 3\tpm 7\tpm 19$           & $120\tpm 2\tpm 5\tpm 16$           & $91\tpm 1\tpm 3\tpm 12$ \\
$7$&$-$&$8$     && $76\tpm 2\tpm 4\tpm 10$             & $64\tpm 1\tpm 3\tpm 8$               & $49\tpm 1\tpm 2\tpm 6$ \\
$8$&$-$&$9$     && $44\tpm 1\tpm 1\tpm 6$               & $34\tpm 1\tpm 1\tpm 5$               & $25\tpm 1\tpm 1\tpm 3$ \\
$9$&$-$&$10$   && $23\tpm 1\tpm 1\tpm 3$               & $19.3\tpm 0.6\tpm 0.2\tpm 2.6$   & $13.7\tpm 0.5\tpm 0.1\tpm 1.8$ \\
$10$&$-$&$11$ && $13.5\tpm 0.6\tpm 0.4\tpm 1.8$   & $10.9\tpm 0.4\tpm 0.1\tpm 1.4$   & $8.5\tpm 0.4\tpm 0.1\tpm 1.1$ \\
$11$&$-$&$12$ && $7.7\tpm 0.4\tpm 0.3\tpm 1.0$     & $6.4\tpm 0.3\tpm 0.1\tpm 0.8$     & $4.9\tpm 0.3\tpm 0.1\tpm 0.6$ \\
$12$&$-$&$13$ && $5.2\tpm 0.3\tpm 0.2\tpm 0.7$     & $3.8\tpm 0.3\tpm 0.1\tpm 0.5$     & $3.1\tpm 0.2\tpm 0.1\tpm 0.4$ \\
$13$&$-$&$14$ && $3.3\tpm 0.3\tpm 0.1\tpm 0.4$     & $2.8\tpm 0.2\tpm 0.1\tpm 0.4$     & $1.4\tpm 0.2\tpm 0.1\tpm 0.2$ \\
\bottomrule
&&&& \multicolumn{7}{c}{$3.5<y<4.0$} & \multicolumn{7}{c}{$4.0<y<4.5$} \\ \midrule
$0$&$-$&$1$     && $728\tpm 7\tpm 27\tpm 96$         & $530\tpm 6\tpm 33\tpm 70$ \\
$1$&$-$&$2$     && $1246\tpm 8\tpm 26\tpm 164$     & $902\tpm 7\tpm 31\tpm 119$ \\
$2$&$-$&$3$     && $913\tpm 6\tpm 21\tpm 120$       & $631\tpm 6\tpm 24\tpm 83$ \\
$3$&$-$&$4$     && $505\tpm 4\tpm 14\tpm 67$         & $334\tpm 4\tpm 15\tpm 44$ \\
$4$&$-$&$5$     && $262\tpm 3\tpm 8\tpm 35$           & $172\tpm 3\tpm 7\tpm 23$  \\
$5$&$-$&$6$     && $128\tpm 2\tpm 5\tpm 17$           & $79\tpm 2\tpm 3\tpm 11$ \\
$6$&$-$&$7$     && $63\tpm 1\tpm 2\tpm 8$               & $36\tpm 1\tpm 2\tpm 5$ \\
$7$&$-$&$8$     && $32\tpm 1\tpm 1\tpm 4$               & $18.3\tpm 0.7\tpm 0.8\tpm 2.4$ \\
$8$&$-$&$9$     && $17.1\tpm 0.6\tpm 0.2\tpm 2.3$   & $8.9\tpm 0.5\tpm 0.4\tpm 1.2$ \\
$9$&$-$&$10$   && $8.8\tpm 0.4\tpm 0.1\tpm 1.2$     & $4.3\tpm 0.3\tpm 0.2\tpm 0.5$ \\
$10$&$-$&$11$ && $5.2\tpm 0.3\tpm 0.1\tpm 0.7$     & $2.4\tpm 0.2\tpm 0.1\tpm 0.3$ \\
$11$&$-$&$12$ && $2.8\tpm 0.2\tpm 0.1\tpm 0.4$     & \\
$12$&$-$&$13$ && $1.3\tpm 0.1\tpm 0.1\tpm 0.2$     & \\
\bottomrule
\end{tabular}
}
\end{center}
\end{table}
\renewcommand{\arraystretch}{1.2}

\renewcommand{\arraystretch}{1.35}
\begin{table}[!ht]
\begin{center}
\caption{\small \label{promptresultlongitudinal}$\frac{{\rm d}^2\sigma}{{\rm d}p_{\rm T}{\rm d}y}$ in nb/(\gevc) for 
prompt \jpsi\ in bins of the \jpsi\ transverse momentum and rapidity, assuming fully longitudinally polarised \jpsi. The 
first error is statistical, the second is the component of the systematic uncertainty that is uncorrelated between bins 
and the third is the correlated component.}
\vskip 0.5cm
\scalebox{0.98}{
\begin{tabular}{@{}r@{}c@{}rrr@{}c@{}r@{}c@{}r@{}c@{}rr@{}c@{}r@{}c@{}r@{}c@{}rr@{}c@{}r@{}c@{}r@{}c@{}r@{}}
\toprule
\multicolumn{4}{c}{$\ptrans\,(\gevc)$} & \multicolumn{7}{c}{$2.0<y<2.5$} & \multicolumn{7}{c}{$2.5<y<3.0$} & \multicolumn{7}{c}{$3.0<y<3.5$} \\
\midrule
$0$&$-$&$1$     && $839\tpm54\tpm174\tpm111$      & $601\tpm 9\tpm 94\tpm 79$         & $543\tpm 5\tpm 33\tpm 72 $ \\
$1$&$-$&$2$     && $1157\tpm29\tpm219\tpm 153$   & $1114\tpm 9\tpm 29\tpm 147$     & $1073\tpm 7\tpm 21\tpm 142$ \\
$2$&$-$&$3$     && $945\tpm16\tpm84\tpm 125$       & $938\tpm 7\tpm 19\tpm 124$       & $865\tpm 5\tpm 16\tpm 114$ \\
$3$&$-$&$4$     && $583\tpm 8\tpm 33\tpm 77$         & $559\tpm 4\tpm 14\tpm 74$         & $514\tpm 4\tpm 11\tpm 68$ \\
$4$&$-$&$5$     && $315\tpm 4\tpm 16\tpm 42$         & $307\tpm 3\tpm 9\tpm 41$           & $274\tpm 2\tpm 8\tpm 36$ \\
$5$&$-$&$6$     && $171\tpm 3\tpm 9\tpm 23$           & $163\tpm 2\tpm 6\tpm 22$           & $140\tpm 2\tpm 4\tpm 19$ \\
$6$&$-$&$7$     && $87\tpm 2\tpm 5\tpm 12$             & $83\tpm 1\tpm 3\tpm 11$             & $70\tpm 1\tpm 3\tpm 9$ \\
$7$&$-$&$8$     && $48\tpm 1\tpm 2\tpm 6$               & $46\tpm 1\tpm 2\tpm 6$               & $38\tpm 1\tpm 1\tpm 5$ \\
$8$&$-$&$9$     && $29\tpm 1\tpm 1\tpm 4$               & $25\tpm 1\tpm 1\tpm 3$               & $19.8\tpm 0.5\tpm 0.1\tpm 2.6$ \\
$9$&$-$&$10$   && $14.9\tpm 0.5\tpm 0.4\tpm 2.0$   & $14.5\tpm 0.4\tpm 0.2\tpm 1.9$   & $10.8\tpm 0.4\tpm 0.1\tpm 1.4$ \\
$10$&$-$&$11$ && $9.1\tpm 0.4\tpm 0.3\tpm 1.2$     & $8.3\tpm 0.3\tpm 0.1\tpm 1.1$     & $6.7\tpm 0.3\tpm 0.1\tpm 0.9$ \\
$11$&$-$&$12$ && $5.3\tpm 0.3\tpm 0.2\tpm 0.7$     & $5.0\tpm 0.3\tpm 0.1\tpm 0.7$     & $4.0\tpm 0.2\tpm 0.1\tpm 0.5$ \\
$12$&$-$&$13$ && $3.7\tpm 0.2\tpm 0.1\tpm 0.5$     & $3.0\tpm 0.2\tpm 0.1\tpm 0.4$     & $2.5\tpm 0.2\tpm 0.1\tpm 0.4$ \\
$13$&$-$&$14$ && $2.3\tpm 0.2\tpm 0.1\tpm 0.3$     & $2.3\tpm 0.2\tpm 0.1\tpm 0.3$     & $1.2\tpm 0.1\tpm 0.1\tpm 0.2$ \\
\bottomrule
&&&& \multicolumn{7}{c}{$3.5<y<4.0$} & \multicolumn{7}{c}{$4.0<y<4.5$} \\ \midrule
$0$&$-$&$1$     && $468\tpm 4\tpm 21\tpm 62$         & $341\tpm 4\tpm 21\tpm 45$ \\
$1$&$-$&$2$     && $892\tpm 5\tpm 18\tpm 118$       & $667\tpm 6\tpm 23\tpm 88$ \\
$2$&$-$&$3$     && $721\tpm 5\tpm 16\tpm 95$         & $517\tpm 5\tpm 20\tpm 68$ \\
$3$&$-$&$4$     && $415\tpm 3\tpm 12\tpm 55$         & $282\tpm 4\tpm 13\tpm 37$ \\
$4$&$-$&$5$     && $215\tpm 2\tpm 7\tpm 28$           & $148\tpm 2\tpm 6\tpm 20$  \\
$5$&$-$&$6$     && $104\tpm 1\tpm 4\tpm 14$           & $69\tpm 2\tpm 3\tpm 9$ \\
$6$&$-$&$7$     && $51\tpm 1\tpm 2\tpm 7$               & $31\tpm 1\tpm 1\tpm 4$ \\
$7$&$-$&$8$     && $26\tpm 1\tpm 1\tpm 3$               & $15.8\tpm 0.6\tpm 0.7\tpm 2.1$ \\
$8$&$-$&$9$     && $13.9\tpm 0.5\tpm 0.1\tpm 1.8$   & $7.6\tpm 0.4\tpm 0.3\tpm 1.0$ \\
$9$&$-$&$10$   && $7.1\tpm 0.3\tpm 0.1\tpm 0.9$     & $3.6\tpm 0.3\tpm 0.2\tpm 0.5$ \\
$10$&$-$&$11$ && $4.3\tpm 0.2\tpm 0.1\tpm 0.6$     & $2.0\tpm 0.2\tpm 0.1\tpm 0.3$ \\
$11$&$-$&$12$ && $2.3\tpm 0.2\tpm 0.1\tpm 0.3$     & \\
$12$&$-$&$13$ && $1.0\tpm 0.1\tpm 0.1\tpm 0.1$     & \\
\bottomrule
\end{tabular}
}
\end{center}
\end{table}
\renewcommand{\arraystretch}{1.2}

\renewcommand{\arraystretch}{1.44}
\begin{table}[!ht]
\begin{center}
\caption{\small \label{tab:bfraction}Fraction of \fromb\ (in \%) in bins of the \jpsi\ transverse momentum and rapidity. 
The first uncertainty is statistical, the second systematic (uncorrelated between bins) and the third is the uncertainty 
due to the unknown polarisation of the \prompt; the central values are for unpolarised \jpsi.}
\vskip 0.2cm
\scalebox{0.92}{
\begin{tabular}{@{}r@{}c@{}rrr@{}c@{}r@{}c@{}r@{}rr@{}c@{}r@{}c@{}r@{}rr@{}c@{}r@{}c@{}r@{}r@{}}
\toprule
\multicolumn{4}{c}{$\ptrans\,(\gevc)$} & \multicolumn{6}{c}{$2.0<y<2.5$} & \multicolumn{6}{c}{$2.5<y<3.0$} & \multicolumn{6}{c}{$3.0<y<3.5$} \\
\midrule
$0$&$-$&$1$     && $8.9\tpm 1.7 \tpm 0.3$&$_{-2.4}^{+1.2}$   & $8.2\tpm 0.4\tpm 0.3$&$_{-2.9}^{+1.5}$     & $7.4\tpm 0.3\tpm 0.3$&$_{-2.5}^{+1.3}$  \\
$1$&$-$&$2$     && $9.4\tpm0.7\tpm 0.3$&$_{-2.4}^{+1.3}$     & $9.0\tpm0.2\tpm 0.3$&$_{-2.7}^{+1.4}$      & $8.2\tpm0.2\tpm 0.3$&$_{-2.1}^{+1.1}$   \\
$2$&$-$&$3$     && $11.0\tpm0.5\tpm0.4$&$_{-2.8}^{+1.5}$    &  $10.3\tpm0.2\tpm0.4$&$_{-2.6}^{+1.4}$    & $9.7\tpm0.2\tpm0.3$&$_{-1.9}^{+1.0}$  \\
$3$&$-$&$4$     && $11.9\tpm 0.4\tpm 0.4$&$_{-3.3}^{+1.8}$  & $12.0\tpm 0.2\tpm 0.4$&$_{-2.9}^{+1.5}$   & $11.0\tpm 0.2\tpm 0.4$&$_{-1.8}^{+0.9}$  \\
$4$&$-$&$5$     && $13.6\tpm 0.4\tpm 0.5$&$_{-3.9}^{+2.1}$  & $12.7\tpm 0.3 \tpm 0.5$&$_{-3.0}^{+1.6}$  & $11.9\tpm 0.3 \tpm 0.4$&$_{-1.9}^{+1.0}$ \\
$5$&$-$&$6$     && $15.7\tpm 0.5\tpm 0.6$&$_{-4.3}^{+2.4}$  & $14.6\tpm 0.4\tpm 0.5$&$_{-3.2}^{+1.7}$   & $13.6\tpm 0.4\tpm 0.5$&$_{-2.1}^{+1.1}$  \\
$6$&$-$&$7$     && $18.4\tpm0.7\tpm 0.7$&$_{-4.8}^{+2.6}$   & $17.5\tpm0.5\tpm 0.6$&$_{-3.5}^{+1.9}$    & $15.4\tpm0.5\tpm 0.6$&$_{-2.3}^{+1.2}$  \\
$7$&$-$&$8$     && $20.1\tpm0.8\tpm0.7$&$_{-4.8}^{+2.6}$    & $17.6\tpm0.7\tpm 0.6$&$_{-3.4}^{+1.8}$    & $17.8\tpm0.7\tpm 0.6$&$_{-2.5}^{+1.3}$  \\
$8$&$-$&$9$     && $21.4\tpm 1.0\tpm 0.8$&$_{-4.7}^{+2.6}$  & $21.2\tpm 0.9\tpm 0.8$&$_{-3.5}^{+1.9}$   & $18.6\tpm 1.0\tpm 0.7$&$_{-2.6}^{+1.4}$  \\
$9$&$-$&$10$   && $25.3\tpm 1.4 \tpm 0.9$&$_{-5.1}^{+2.8}$ & $23.1\tpm 1.2\tpm 0.8$&$_{-3.4}^{+1.8}$   & $21.5\tpm 1.3\tpm 0.8$&$_{-2.5}^{+1.3}$  \\
$10$&$-$&$11$ && $27.6\tpm 1.7 \tpm 1.0$&$_{-5.2}^{+2.9}$ & $24.6\tpm 1.5\tpm 0.9$&$_{-3.3}^{+1.8}$   & $20.2\tpm 1.7\tpm 0.7$&$_{-2.5}^{+1.3}$   \\
$11$&$-$&$12$ && $32.8\tpm2.2\tpm 1.2$&$_{-5.2}^{+2.9}$     & $27.0\tpm2.0\tpm 1.0$&$_{-3.3}^{+1.8}$    & $24.7\tpm2.2\tpm  0.9$&$_{-2.4}^{+1.3}$  \\
$12$&$-$&$13$ && $28.9\tpm2.6\tpm 1.0$&$_{-4.7}^{+2.6}$   & $31.3\tpm2.6\tpm 1.1$&$_{-3.5}^{+1.9}$    &  $24.1\tpm2.8\tpm 0.9$&$_{-2.4}^{+1.3}$ \\
$13$&$-$&$14$ && $29.8\tpm 3.6 \tpm 1.1$&$_{-4.8}^{+2.6}$ & $26.5\tpm 2.9\tpm 1.0$&$_{-2.8}^{+1.5}$   & $32.5\tpm 4.1\tpm 1.2$&$_{-2.8}^{+1.5}$ \\
\bottomrule
& & & & \multicolumn{6}{c}{$3.5<y<4.0$} & \multicolumn{6}{c}{$4.0<y<4.5$} \\ \midrule
$0$&$-$&$1$     && $6.3\tpm 0.3\tpm 0.2$&$_{-1.8}^{+0.9}$   & $4.8\tpm 0.4\tpm 0.2$&$_{-1.4}^{+0.7}$  \\
$1$&$-$&$2$     && $6.9\tpm0.2\tpm 0.2$&$_{-1.5}^{+0.8}$    & $6.1\tpm0.2\tpm 0.2$&$_{-1.2}^{+0.6}$  \\
$2$&$-$&$3$     && $7.9\tpm0.2\tpm0.3$&$_{-1.2}^{+0.6}$     & $6.7\tpm0.3\tpm0.2$&$_{-0.9}^{+0.4}$ \\
$3$&$-$&$4$     && $9.3\tpm 0.3\tpm 0.3$&$_{-1.1}^{+0.6}$   & $8.1\tpm 0.4\tpm 0.3$&$_{-0.9}^{+0.4}$ \\
$4$&$-$&$5$     && $10.2\tpm 0.3\tpm 0.4$&$_{-1.2}^{+0.6}$ & $8.4\tpm 0.5\tpm 0.3$&$_{-0.8}^{+0.4}$ \\
$5$&$-$&$6$     && $11.6\tpm 0.5\tpm 0.4$&$_{-1.4}^{+0.7}$ & $10.7\tpm 0.7\tpm 0.4$&$_{-0.9}^{+0.4}$ \\
$6$&$-$&$7$     && $12.7\tpm0.6\tpm 0.5$&$_{-1.6}^{+0.8}$  & $13.3\tpm1.1\tpm 0.5$&$_{-1.1}^{+0.5}$ \\
$7$&$-$&$8$     && $15.7\tpm0.9\tpm0.6$&$_{-1.9}^{+1.0}$   & $13.7\tpm1.4\tpm0.5$&$_{-1.2}^{+0.6}$ \\
$8$&$-$&$9$     && $16.6\tpm 1.2\tpm 0.6$&$_{-2.0}^{+1.0}$ & $15.2\tpm 2.0\tpm 0.5$&$_{-1.4}^{+0.7}$ \\
$9$&$-$&$10$   && $18.0\tpm 1.6\tpm 0.6$&$_{-2.1}^{+1.1}$ & $17.0\tpm 2.9\tpm 0.6$&$_{-1.7}^{+0.9}$ \\
$10$&$-$&$11$ && $19.8\tpm 2.2\tpm 0.7$&$_{-2.1}^{+1.1}$ & $17.7\tpm 3.9\tpm 0.6$&$_{-1.6}^{+0.8}$ \\
$11$&$-$&$12$ && $19.5\tpm2.9\tpm 0.8$&$_{-2.0}^{+1.1}$ \\
$12$&$-$&$13$ && $21.9\tpm4.4\tpm0.8$&$_{-2.4}^{+1.2}$ \\
\bottomrule
\end{tabular}
}
\end{center}
\end{table}
\renewcommand{\arraystretch}{1.2}

\end{document}